\documentclass[twocolumn]{aastex62}
\usepackage{natbib}
\usepackage{topcapt}
\usepackage{booktabs}
\usepackage{subfigure}
\usepackage{amsmath}
\usepackage{multirow}
\usepackage{longtable}

\usepackage{appendix}

\makeatletter
\newcommand*{\centerfloat}{%
  \parindent \z@
  \leftskip \z@ \@plus 1fil \@minus \marginparwidth
  \rightskip \leftskip
  \parfillskip \z@skip}
\makeatother

\usepackage{color,subfigure,lineno} 

\newcommand{\hbeta}{H{$\beta$}}
\newcommand{\halpha}{H{$\alpha$}}

\newcommand{\CIV}{C\,{\sevenrm IV}}

\newcommand{\SiIV}{Si\,{\sevenrm IV}}
\newcommand{\CIII}{C\,{\sevenrm III]}}

\newcommand{\AlIII}{Al\,{\sevenrm III}}

\newcommand{\SiIII}{Si\,{\sevenrm III]}}

\def\FeII{Fe\,{\sc ii}}
\def\MgII{Mg\,{\sc ii}}
\def\HeII{He\,{\sc ii}}

\def \OIIIUV {O\,{\sc iii}]}

\def \OIII {[O\,{\sc iii}]}

\newcommand{\NII}{[N\,{\sevenrm\,II}]}

\newcommand{\SII}{[S{\sevenrm\,II}]}

\newcommand{\OIV}{O\,{\sevenrm IV]}}

   \font\sevenrm=cmr7 scaled 1000
\newcommand{\comments}[1]{}

\def\kms{{\rm km\,s^{-1}}}

\shorttitle{SDSS-RM: BLR Breathing}
\shortauthors{Wang et al.}

\begin{document}

\title{The Sloan Digital Sky Survey Reverberation Mapping Project: How Broad Emission Line Widths Change When Luminosity Changes}

\author{Shu Wang}
\affiliation{Kavli Institute for Astronomy and Astrophysics, Peking University, Beijing 100871, China; wangshukiaa@pku.edu.cn, jiangKIAA@pku.edu.cn}
\affiliation{Department of Astronomy, School of Physics, Peking University, Beijing 100871, China}
\affiliation{Department of Astronomy, University of Illinois at Urbana-Champaign, Urbana, IL 61801, USA}

\author[0000-0002-6893-3742]{Yue Shen}
\altaffiliation{Alfred P. Sloan Research Fellow}
\affiliation{Department of Astronomy, University of Illinois at Urbana-Champaign, Urbana, IL 61801, USA}
\affiliation{National Center for Supercomputing Applications, University of Illinois at Urbana-Champaign, Urbana, IL 61801, USA}

\author[0000-0003-4176-6486]{Linhua Jiang}
\affiliation{Kavli Institute for Astronomy and Astrophysics, Peking University, Beijing 100871, China; wangshukiaa@pku.edu.cn, jiangKIAA@pku.edu.cn}
\affiliation{Department of Astronomy, School of Physics, Peking University, Beijing 100871, China}

\author{C.~J.~Grier}
\affiliation{Department of Astronomy and Astrophysics, Eberly College of Science, The Pennsylvania State University, 525 Davey Laboratory, University Park, PA 16802}
\affiliation{Institute for Gravitation \& the Cosmos, The Pennsylvania State University, University Park, PA 16802}
\affiliation{Steward Observatory, The University of Arizona, 933 North Cherry Avenue, Tucson, AZ 85721, USA} 

\author{Keith~Horne}
\affiliation{SUPA Physics and Astronomy, University of St. Andrews, Fife, KY16 9SS, Scotland, UK} 

\author{Y.~Homayouni}
\affiliation{Department of Physics, University of Connecticut, 2152 Hillside Rd Unit 3046, Storrs, CT 06269, USA}

\author{B.~M.~Peterson}
\affiliation{Department of Astronomy, The Ohio State University, 140 W 18th Avenue, Columbus, OH 43210, USA}
\affiliation{Center for Cosmology and AstroParticle Physics, The Ohio State University, 191 West Woodruff Avenue, Columbus, OH 43210, USA}
\affiliation{Space Telescope Science Institute, 3700 San Martin Drive, Baltimore, MD 21218, USA }

\author{Jonathan R. Trump}
\affiliation{Department of Physics, University of Connecticut, 2152 Hillside Road, Unit 3046, Storrs, CT 06269, USA}

\author{W.~N.~Brandt}
\affiliation{Department of Astronomy and Astrophysics, Eberly College of Science, The Pennsylvania State University, 525 Davey Laboratory, University Park, PA 16802}
\affiliation{Institute for Gravitation \& the Cosmos, The Pennsylvania State University, University Park, PA 16802}
\affiliation{Department of Physics, The Pennsylvania State University, University Park, PA 16802, USA}

\author{P.~B.~Hall}
\affiliation{Department of Physics and Astronomy, York University, Toronto, ON M3J 1P3, Canada}

\author{Luis~C.~Ho}
\affiliation{Kavli Institute for Astronomy and Astrophysics, Peking University, Beijing 100871, China; wangshukiaa@pku.edu.cn, jiangKIAA@pku.edu.cn} 
\affiliation{Department of Astronomy, School of Physics, Peking University, Beijing 100871, China} 

\author{Jennifer~I-Hsiu~Li}
\affiliation{Department of Astronomy, University of Illinois at Urbana-Champaign, Urbana, IL 61801, USA}

\author{J. V. Hernandez Santisteban}
\affiliation{SUPA Physics and Astronomy, University of St. Andrews, Fife, KY16 9SS, Scotland, UK} 

\author{K.~Kinemuchi} 
\affiliation{Apache Point Observatory and New Mexico State University, P.O. Box 59, Sunspot, NM, 88349-0059, USA}

\author{Ian~D.~McGreer}
\affiliation{Steward Observatory, The University of Arizona, 933 North Cherry Avenue, Tucson, AZ 85721, USA}

\author{D.~P.~Schneider}
\affiliation{Department of Astronomy and Astrophysics, Eberly College of Science, The Pennsylvania State University, 525 Davey Laboratory, University Park, PA 16802}
\affiliation{Institute for Gravitation \& the Cosmos, The Pennsylvania State University, University Park, PA 16802}

\begin{abstract}
Quasar broad emission lines are largely powered by photoionization from the accretion continuum. Increased central luminosity will enhance line emissivity in more distant clouds, leading to increased average distance of the broad-line-emitting clouds and decreased averaged line width, known as the broad-line region (BLR) ``breathing''. However, different lines breathe differently, and some high-ionization lines, such as \CIV, can even show ``anti-breathing'' where the line broadens when luminosity increases. Using multi-year photometric and spectroscopic monitoring data from the Sloan Digital Sky Survey Reverberation Mapping project, we quantify the breathing effect ($\Delta\log W=\alpha\Delta\log L$) of broad \halpha, \hbeta, \MgII, \CIV, and \CIII\ for statistical quasar samples over $z\approx 0.1-2.5$. We found that \hbeta\ displays the most consistent normal breathing expected from the virial relation ($\alpha\sim-0.25$), \MgII\ and \halpha\ on average show no breathing ($\alpha\sim 0$), and \CIV\ (and similarly \CIII\ and \SiIV) mostly shows anti-breathing ($\alpha>0$). The anti-breathing of \CIV\ can be well understood by the presence of a non-varying core component in addition to a reverberating broad-base component, consistent with earlier findings. The deviation from canonical breathing introduces extra scatter (a luminosity-dependent bias) in single-epoch virial BH mass estimates due to intrinsic quasar variability, which underlies the long argued caveats of \CIV\ single-epoch masses. Using the line dispersion instead of FWHM leads to less, albeit still substantial, deviations from canonical breathing in most cases. Our results strengthen the need for reverberation mapping to provide reliable quasar BH masses, and quantify the level of variability-induced bias in single-epoch BH masses based on various lines. 

\end{abstract}


\section{Introduction} \label{sec:introduction}

Reverberation mapping studies of broad-line Active Galactic Nuclei (AGN) have revealed a ``breathing'' mode of the broad Balmer lines such that when luminosity increases, the average (i.e., line-emissivity-weighted) BLR size increases and the average line width decreases \citep[e.g.,][]{Gilbert03,Denney09,Park_etal_2012,Barth_etal_2015,Runco_etal_2016}. This observation is consistent with the expectation from photoionization \citep[e.g.,][]{Korista_Goad_2004,Cackett06}, where the responsivity of the line emission depends on the radial distributions of incident ionizing flux and cloud densities. However, most of these results are based on \hbeta\ and \halpha\ targeted in previous RM programs. RM results on other strong broad emission lines such as \CIV\ and \MgII\ are scarce, but recent observations of the variability of broad \CIV\ and \MgII\ for distant quasars revealed different results: \MgII\ appears to have much weaker breathing than broad \hbeta\ \citep[e.g.,][]{Shen_2013,Yang_etal_2019a,Homan_etal_2019}, and the \CIV\ width appears to increase when luminosity increases \citep[e.g.,][]{Richards_etal_2002b,Wilhite06,Shen_etal_2008,Sun_etal_2018}, in the opposite sense of breathing. These different behaviors of breathing for different broad lines suggest different distributions of the line-emitting clouds within the BLR and likely also different kinematic structures from viralized motion such as winds and outflows. For example, previous photoionization calculations have shown that the broad \MgII\ emission comes from clouds that are on average further out than the Balmer lines in the BLR \citep[e.g.,][]{Goad_etal_1993,O'Brien95,Korista00,Goad_etal_2012}. \citet[][]{Guo_etal_2019} further suggested that, based on photoionization calculations, if most of the broad \MgII\ is produced near the physical outer boundary of the BLR, then \MgII\ will show weaker breathing than the Balmer lines. Quantifying the behaviors of breathing for different broad lines therefore offers useful constraints on the structure of the BLR and photoionization models. 

In this work we perform a detailed observational study of the BLR breathing behavior for several strong broad emission lines in quasar spectra. We quantify the general behaviors of \hbeta/\halpha, \MgII, \CIII, and \CIV, in terms of breathing, using well sampled light curves that capture the continuum variability and echoed broad-line variability from the Sloan Digital Sky Survey Reverberation Mapping (SDSS-RM) project \citep[e.g.,][]{Shen_etal_2015a}. With these results, we attempt to establish an observational consensus on BLR breathing to provide valuable inputs for theoretical modeling of photoionization and kinematics of the BLR. 

The structure of the paper is as follows. In \S\ref{sec:data} we describe the data. In \S\ref{sec:result} we present the results for different broad lines. We discuss our results in \S\ref{sec:disc} and summarize in \S\ref{sec:con}. 

\section{Data}\label{sec:data}

The SDSS-RM project is a dedicated multi-object optical reverberation mapping program \citep{Shen_etal_2015a} that has been spectroscopically monitoring a single 7 ${\rm deg}^2$ field, using the SDSS Baryon Oscillation Spectroscopic Survey \citep[BOSS,][]{Eisenstein_etal_2011,Dawson_etal_2013} spectrographs  \citep[][]{Smee_etal_2013} on the 2.5m SDSS telescope \citep{Gunn_etal_2006} at Apache Point Observatory. We also acquired accompanying photometric data in the $g$ and $i$ bands with the 3.6 m Canada-France-Hawaii Telescope (CFHT) and the Steward Observatory 2.3 m Bok telescope during the spectroscopic monitoring. The SDSS-RM program started in 2014 as one of the dark-time ancillary programs in SDSS-III \citep{Eisenstein_etal_2011} and has continued in SDSS-IV \citep{Blanton_etal_2017}.

The SDSS-RM monitoring data used in this work include the multi-epoch spectroscopy taken by the BOSS spectrographs, as well as photometric light curves from the CFHT and Bok telescopes from 2014 to 2017. The spectroscopic data were first re-processed with a custom pipeline to improve flux calibration \citep[][]{Shen_etal_2015a}, followed by another recalibration process called PrepSpec to further improve the spectrophotometric calibration using the flux of the narrow emission lines as a constant reference \citep[e.g.,][]{Shen_etal_2016a,Grier_etal_2017}. 

The SDSS-RM sample includes 849 broad-line quasars flux-limited to $i=21.7$, covering a broad redshift range of $0.1<z<4.5$. In this work we use a subset of the SDSS-RM sample that have well measured broad-line lags to examine the breathing effect. The detection of a broad-line lag ensures there is significant continuum variability already, facilitating the investigation of broad-line width responses. Our initial sample includes 44 SDSS-RM quasars with \hbeta\ lag measurements \citep[][hereafter G17]{Grier_etal_2017}, 57 quasars with \MgII\ lag detections \citep[][hereafter H20]{Homayouni_etal_2020} and 52 quasars with \CIV\ lag measurements \citep[][hereafter G19]{Grier_etal_2019}. The imaging and spectroscopic light curves from the G17 sample cover one season (2014), and those from the G19 and H20 samples cover four seasons (2014--2017). In this work we use the same light curve sets from these three papers, but with several modifications as described in \S\ref{sec:spectral_analysis}.

Assuming a canonical $R-L$ relation \citep[e.g., $R\propto L^{0.5}$,][]{Bentz_etal_2013} and that the broad line width faithfully traces the virial velocity of the BLR, the expected correlation between changes in line width ($\Delta\log W$) and continuum luminosity ($\Delta\log L$) is $\Delta \log W = -0.25\Delta\log L$. Only objects with significant continuum variability can be used to robustly measure the breathing effect. Therefore, we first refine our initial sample using a continuum variability significance threshold. We define a continuum variability signal to noise ratio as ${\rm SNR}_{\rm Var,con} = \sqrt{\frac{1}{N-1} \times \sum_{i=1}^{N}{(F_{\rm i} - \overline{F})^2 / F_{{\rm err},i}^2}}$, where $F_{i}$ is the $g$-band flux in the $i$th epoch, $\overline{F}$ is the average $g$-band flux, and $N$ is the total number of the epochs. Compared to the SNR2 reported by Prepspec (see G17, G19, and H20), our definition is directly related to the study of $\Delta\log W-\Delta\log L$ correlation (see details in \S\ref{sec:lowz}). We restrict our sample to objects with ${\rm SNR}_{\rm Var,con} > 2$, which is a good balance between sample size and ${\rm SNR}_{\rm Var,con}$. Choosing a more stringent criterion, e.g., ${\rm SNR}_{\rm Var,con} > 3$, the results are consistent (see details in \S \ref{sec:lowz} and \ref{sec:highz}). Second, we require that the quality rating of lag measurement reported in G17 and G19 is at least 3, as we will use the measured lag to synchronize the line response with continuum variability in the investigation of breathing. This criterion on lag quality is set to be slightly less stringent than that of the gold sample, which is set to be 4 in G17 and G19. Using this criterion increases the sample size by $\sim$ 30\% relative to the G17 gold sample and by $\sim$ 60\% relative to the G19 gold sample, while removing the least reliable lag measurements reported in G17 and G19. H20 did not report quantitative quality ratings on lag measurements, but defined a gold sample. We adopt the gold sample from H20. We summarize our parent sample in Table \ref{tab:sampleinformation} for further spectral analysis in \S\ref{sec:finalsample}.

\setlength{\tabcolsep}{0.04in}

\begin{longtable*}{cccccccccccc}

    \caption{Sample Summary} \label{tab:sampleinformation} \\

    \toprule
    \toprule
        & &   & $L_{\rm Bol}$ &   & line &  line & $\tau$  & $\tau$ quality  &  &  & $g$-band \\
        RMID & SDSS$-$Identifier & $z$ & (erg s$^{-1}$)  & SNR$_{\rm Var,con}$ & name & SNR & (rest-frame) & ratings &  sample & source & F\_Host  \\
    \midrule
    \endfirsthead

    \multicolumn{12}{c}{\tablename\ \thetable\ -- \textit{Continued}} \\ \\
    \toprule
    \toprule
        & &   & $L_{\rm Bol}$ &   & line &  line & $\tau$  & $\tau$ quality  &  &  & $g$ -band \\
        RMID & SDSS$-$Identifier & $z$ & (erg s$^{-1}$)  & SNR$_{\rm Var,con}$ & name & SNR & (rest-frame) & ratings &  sample & source & F\_Host  \\
    \midrule
    \endhead
    \bottomrule
    \endfoot
    \bottomrule
    \multicolumn{12}{p{\textwidth}}{Notes. The quality ratings of the lags are taken directly from the corresponding papers. H20 only provided the most reliable lags in their Gold sample, indicated by ``G'' in the ``$\tau$ quality ratings'' column.}
    \endlastfoot

\multicolumn{12}{c}{\halpha, \hbeta, and \MgII\ sample} \\ 
\midrule
016 & J141606.95+530929.8 & $0.850$ & $45.563$ & 4.06 & H$\beta$ & 5.44 & $32.0_{-15.5}^{+11.6}$ & 3 &  Lag0 & G17 & ...\\ 
   &   &   &   &   & Mg II & 7.85   &   &   &  Lag1 &  G17 &\\ 
017 & J141324.28+530527.0 & $0.457$ & $44.875$ & 4.30 & H$\alpha$ & 4.99 & $56.6_{-15.1}^{+7.3}$ & 5 &  Lag0 & G17 & 0.34\\ 
   &   &   &   &   & H$\beta$ & 4.99 & $25.5_{-5.8}^{+10.9}$ & 4 &  Lag0 & G17 & 0.34\\ 
   &   &   &   &   & Mg II & 6.90   &   &   &  Lag1 &  G17 &\\ 
044 & J141622.83+531824.3 & $1.233$ & $45.626$ & 2.09 & Mg II & 4.85 & $65.8_{-4.8}^{+18.8}$ & G &   Lag0 & H20 &...\\ 
088 & J141151.78+525344.1 & $0.517$ & $44.963$ & 2.28 & H$\alpha$ & 4.79 & $54.8_{-5.1}^{+2.9}$ & 3 &  Lag0 & G17 & 0.12\\ 
   &   &   &   &   & H$\beta$ & 6.70   &   &   &  Lag1 &  G17 &\\ 
   &   &   &   &   & Mg II & 5.09   &   &   &  Lag1 &  G17 &\\ 
101 & J141214.20+532546.7 & $0.458$ & $45.351$ & 2.41 & H$\beta$ & 15.88 & $21.4_{-6.4}^{+4.2}$ & 5 &  Lag0 & G17 & 0.17\\ 
   &   &   &   &   & H$\alpha$ & 20.62   &   &   &  Lag1 &  G17 &\\ 
   &   &   &   &   & Mg II & 8.50   &   &   &  Lag1 &  G17 &\\ 
160 & J141041.25+531849.0 & $0.360$ & $44.508$ & 3.61 & H$\alpha$ & 18.42 & $21.0_{-2.8}^{+1.4}$ & 4 &  Lag0 & G17 & 0.18\\ 
   &   &   &   &   & H$\beta$ & 18.42 & $21.9_{-2.4}^{+4.2}$ & 3 &  Lag0 & G17 & 0.18\\ 
   &   &   &   &   & Mg II & 8.39   &   &   &  Lag1 &  G17 &\\ 
177 & J141724.59+523024.9 & $0.482$ & $44.983$ & 6.04 & H$\beta$ & 6.66 & $10.1_{-2.7}^{+12.5}$ & 4 &  Lag0 & G17 & 0.21\\ 
   &   &   &   &   & H$\alpha$ & 6.14   &   &   &  Lag1 &  G17 &\\ 
   &   &   &   &   & Mg II & 5.03   &   &   &  Lag1 &  G17 &\\ 
191 & J141645.58+534446.8 & $0.442$ & $44.499$ & 2.63 & H$\alpha$ & 2.71 & $16.7_{-5.5}^{+4.1}$ & 4 &  Lag0 & G17 & 0.40\\ 
   &   &   &   &   & H$\beta$ & 2.71 & $8.5_{-1.4}^{+2.5}$ & 5 &  Lag0 & G17 & 0.40\\ 
229 & J141018.04+532937.5 & $0.470$ & $44.441$ & 2.32 & H$\alpha$ & 3.52 & $22.1_{-7.3}^{+7.7}$ & 3 &  Lag0 & G17 & 0.25\\ 
   &   &   &   &   & H$\beta$ & 3.52 & $16.2_{-4.5}^{+2.9}$ & 5 &  Lag0 & G17 & 0.25\\ 
   &   &   &   &   & Mg II & 2.56   &   &   &  Lag1 &  G17 &\\ 
252 & J141751.14+522311.1 & $0.281$ & $43.826$ & 2.57 & H$\alpha$ & 7.78 & $10.1_{-1.9}^{+2.4}$ & 5 &  Lag0 & G17 & 0.59\\ 
   &   &   &   &   & H$\beta$ & 2.74   &   &   &  Lag1 &  G17 &\\ 
267 & J141112.72+534507.1 & $0.588$ & $45.127$ & 2.07 & H$\beta$ & 5.25 & $20.4_{-2.0}^{+2.5}$ & 5 &  Lag0 & G17 & 0.21\\ 
   &   &   &   &   & Mg II & 9.93   &   &   &  Lag1 &  G17 &\\ 
272 & J141625.71+535438.5 & $0.263$ & $44.850$ & 4.66 & H$\alpha$ & 44.93 & $32.2_{-12.6}^{+15.6}$ & 3 &  Lag0 & G17 & 0.17\\ 
   &   &   &   &   & H$\beta$ & 44.93 & $15.1_{-4.6}^{+3.2}$ & 5 &  Lag0 & G17 & 0.17\\ 
300 & J141941.11+533649.6 & $0.646$ & $45.583$ & 3.86 & H$\beta$ & 4.35 & $30.4_{-8.3}^{+3.9}$ & 4 &  Lag0 & G17 & 0.05\\ 
   &   &   &   &   & Mg II & 5.33   &   &   &  Lag1 &  G17 &\\ 
301 & J142010.25+524029.6 & $0.548$ & $44.947$ & 6.24 & H$\beta$ & 2.75 & $12.8_{-4.5}^{+5.7}$ & 4 &  Lag0 & G17 & 0.19\\ 
   &   &   &   &   & Mg II & 5.19   &   &   &  Lag1 &  G17 &\\ 
303 & J141830.20+522212.4 & $0.821$ & $44.938$ & 6.19 & Mg II & 5.74 & $57.7_{-8.3}^{+10.5}$ & G &   Lag0 & H20 &...\\ 
320 & J142038.52+532416.5 & $0.265$ & $44.584$ & 2.69 & H$\alpha$ & 13.73 & $20.2_{-9.3}^{+10.5}$ & 4 &  Lag0 & G17 & 0.24\\ 
   &   &   &   &   & H$\beta$ & 13.73 & $25.2_{-5.7}^{+4.7}$ & 4 &  Lag0 & G17 & 0.24\\ 
371 & J141123.42+521331.7 & $0.473$ & $44.909$ & 3.19 & H$\alpha$ & 10.79 & $22.6_{-1.5}^{+0.6}$ & 3 &  Lag0 & G17 & 0.17\\ 
   &   &   &   &   & H$\beta$ & 10.79 & $13.0_{-0.8}^{+1.4}$ & 3 &  Lag0 & G17 & 0.17\\ 
   &   &   &   &   & Mg II & 8.55   &   &   &  Lag1 &  G17 &\\ 
373 & J141859.75+521809.7 & $0.882$ & $45.726$ & 3.45 & H$\beta$ & 12.32 & $20.4_{-7.0}^{+5.6}$ & 3 &  Lag0 & G17 & ...\\ 
377 & J142043.53+523611.4 & $0.337$ & $44.324$ & 2.09 & H$\beta$ & 2.65 & $5.9_{-0.6}^{+0.4}$ & 3 &  Lag0 & G17 & 0.52\\ 
   &   &   &   &   & H$\alpha$ & 8.60   &   &   &  Lag1 &  G17 &\\ 
419 & J141201.94+520527.6 & $1.272$ & $45.761$ & 4.07 & Mg II & 4.78 & $95.5_{-15.5}^{+15.2}$ & G &   Lag0 & H20 &...\\ 
422 & J140739.16+525850.7 & $1.074$ & $45.418$ & 4.33 & Mg II & 5.82 & $109.3_{-29.6}^{+25.4}$ & G &   Lag0 & H20 &...\\ 
440 & J142209.13+530559.7 & $0.754$ & $45.646$ & 2.61 & Mg II & 10.61 & $114.6_{-10.8}^{+7.4}$ & G &   Lag0 & H20 &0.04\\ 
   &   &   &   &   & H$\beta$ & 3.34   &   &   &   Lag1 & H20 & \\ 
449 & J141941.75+535618.0 & $1.218$ & $45.684$ & 3.04 & Mg II & 4.63 & $119.8_{-24.4}^{+14.7}$ & G &   Lag0 & H20 &...\\ 
459 & J141206.95+540827.3 & $1.156$ & $45.699$ & 3.01 & Mg II & 4.23 & $122.8_{-5.7}^{+5.1}$ & G &   Lag0 & H20 &...\\ 
589 & J142049.28+521053.3 & $0.751$ & $45.403$ & 4.92 & H$\beta$ & 3.91 & $46.0_{-9.5}^{+9.5}$ & 3 &  Lag0 & G17 & 0.09\\ 
   &   &   &   &   & Mg II & 7.73   &   &   &  Lag1 &  G17 &\\ 
645 & J142039.80+520359.7 & $0.474$ & $44.870$ & 5.03 & H$\alpha$ & 7.78 & $24.2_{-5.3}^{+10.2}$ & 5 &  Lag0 & G17 & 0.11\\ 
   &   &   &   &   & H$\beta$ & 7.78 & $20.7_{-3.0}^{+0.9}$ & 4 &  Lag0 & G17 & 0.11\\ 
   &   &   &   &   & Mg II & 3.81   &   &   &  Lag1 &  G17 &\\ 
651 & J142149.30+521427.8 & $1.486$ & $45.896$ & 3.64 & Mg II & 6.59 & $76.5_{-15.6}^{+18.0}$ & G &   Lag0 & H20 &...\\ 
675 & J140843.79+540751.2 & $0.919$ & $45.791$ & 3.41 & Mg II & 43.32 & $139.8_{-22.6}^{+12.0}$ & G &   Lag0 & H20 &...\\ 
   &   &   &   &   & H$\beta$ & 3.18   &   &   &   Lag1 & H20 & \\ 
694 & J141706.68+514340.1 & $0.532$ & $45.148$ & 2.58 & H$\beta$ & 4.66 & $10.4_{-3.0}^{+6.3}$ & 5 &  Lag0 & G17 & 0.09\\ 
709 & J140855.07+515833.2 & $1.251$ & $45.693$ & 2.77 & Mg II & 4.33 & $85.4_{-19.3}^{+17.7}$ & G &   Lag0 & H20 &...\\ 
714 & J142349.72+523903.6 & $0.921$ & $45.500$ & 2.56 & Mg II & 6.50 & $320.1_{-11.2}^{+11.3}$ & G &   Lag0 & H20 &...\\ 
756 & J140923.42+515120.1 & $0.852$ & $45.113$ & 3.10 & Mg II & 4.08 & $315.3_{-16.4}^{+20.5}$ & G &   Lag0 & H20 &...\\ 
761 & J142412.92+523903.4 & $0.771$ & $45.496$ & 3.09 & Mg II & 12.19 & $102.1_{-7.4}^{+8.2}$ & G &   Lag0 & H20 &0.05\\ 
   &   &   &   &   & H$\beta$ & 4.56   &   &   &   Lag1 & H20 & \\ 
768 & J140915.70+532721.8 & $0.259$ & $44.569$ & 5.61 & H$\alpha$ & 21.62 & $42.1_{-2.1}^{+2.7}$ & 5 &  Lag0 & G17 & 0.57\\ 
   &   &   &   &   & H$\beta$ & 6.52   &   &   &  Lag1 &  G17 &\\ 
772 & J142135.90+523138.9 & $0.249$ & $44.273$ & 2.42 & H$\alpha$ & 23.13 & $5.9_{-1.0}^{+1.6}$ & 5 &  Lag0 & G17 & 0.91\\ 
   &   &   &   &   & H$\beta$ & 23.13 & $3.9_{-0.9}^{+0.9}$ & 5 &  Lag0 & G17 & 0.91\\ 
775 & J140759.07+534759.8 & $0.173$ & $44.785$ & 3.48 & H$\beta$ & 12.58 & $16.3_{-6.6}^{+13.1}$ & 4 &  Lag0 & G17 & ...\\ 
   &   &   &   &   & H$\alpha$ & 31.94   &   &   &  Lag1 &  G17 &\\ 
776 & J140812.09+535303.3 & $0.116$ & $44.374$ & 3.56 & H$\alpha$ & 25.17 & $8.3_{-2.3}^{+4.9}$ & 4 &  Lag0 & G17 & ...\\ 
   &   &   &   &   & H$\beta$ & 25.17 & $10.5_{-2.2}^{+1.0}$ & 4 &  Lag0 & G17 & ...\\ 
779 & J141923.37+542201.7 & $0.152$ & $44.164$ & 5.92 & H$\beta$ & 8.20 & $11.8_{-1.5}^{+0.7}$ & 4 &  Lag0 & G17 & ...\\ 
   &   &   &   &   & H$\alpha$ & 14.95   &   &   &  Lag1 &  G17 &\\ 
782 & J141318.96+543202.4 & $0.363$ & $44.920$ & 3.55 & H$\beta$ & 9.87 & $20.0_{-3.0}^{+1.1}$ & 4 &  Lag0 & G17 & 0.12\\ 
   &   &   &   &   & H$\alpha$ & 20.74   &   &   &  Lag1 &  G17 &\\ 
   &   &   &   &   & Mg II & 3.13   &   &   &  Lag1 &  G17 &\\ 
790 & J141729.27+531826.5 & $0.238$ & $44.465$ & 5.15 & H$\beta$ & 8.43 & $5.5_{-2.1}^{+5.7}$ & 3 &  Lag0 & G17 & 0.58\\ 
   &   &   &   &   & H$\alpha$ & 55.41   &   &   &  Lag1 &  G17 &\\ 
840 & J141645.15+542540.8 & $0.244$ & $44.206$ & 4.44 & H$\alpha$ & 12.59 & $10.6_{-2.4}^{+2.3}$ & 5 &  Lag0 & G17 & 0.62\\ 
   &   &   &   &   & H$\beta$ & 12.59 & $5.0_{-1.4}^{+1.5}$ & 5 &  Lag0 & G17 & 0.62\\ 
\midrule
\multicolumn{12}{c}{\CIII, \CIV, and \SiIV\ sample}  \\ 
\midrule
032 & J141313.52+525550.2 & $1.715$ & $45.475$ & 7.05 & \CIV & 3.55 & $22.8_{-3.6}^{+3.5}$ & 5 &   Lag0  &  G19 & ...\\ 
145 & J141818.45+524356.0 & $2.137$ & $45.713$ & 4.29 & \CIV & 3.83 & $180.9_{-4.7}^{+4.7}$ & 3 &   Lag0  &  G19 & ...\\ 
201 & J141215.24+534312.1 & $1.812$ & $46.799$ & 6.00 & \CIV & 16.47 & $41.3_{-19.5}^{+32.0}$ & 3 &   Lag0  &  G19 & ...\\ 
   &   &   &   &   & \CIII & 10.07   &   &   & Lag1 & G19 & \\ 
231 & J142005.59+530036.7 & $1.645$ & $46.232$ & 4.23 & \CIV & 9.68 & $80.4_{-7.5}^{+6.3}$ & 3 &   Lag0  &  G19 & ...\\ 
   &   &   &   &   & \CIII & 7.03   &   &   & Lag1 & G19 & \\ 
275 & J140951.81+533133.7 & $1.577$ & $46.181$ & 8.45 & \CIV & 9.78 & $81.0_{-24.4}^{+8.2}$ & 5 &   Lag0  &  G19 & ...\\ 
295 & J141347.87+521204.9 & $2.352$ & $45.909$ & 4.98 & \CIV & 4.22 & $163.8_{-5.3}^{+8.2}$ & 3 &   Lag0  &  G19 & ...\\ 
298 & J141155.56+521802.9 & $1.635$ & $46.112$ & 2.51 & \CIV & 5.60 & $106.1_{-31.7}^{+18.7}$ & 4 &   Lag0  &  G19 & ...\\ 
   &   &   &   &   & \CIII & 3.78   &   &   & Lag1 & G19 & \\ 
387 & J141905.24+535354.1 & $2.426$ & $46.390$ & 3.48 & \CIV & 13.05 & $30.3_{-3.4}^{+19.6}$ & 4 &   Lag0  &  G19 & ...\\ 
401 & J140957.28+535047.0 & $1.822$ & $46.112$ & 3.98 & \CIV & 11.55 & $47.4_{-8.9}^{+15.2}$ & 4 &   Lag0  &  G19 & ...\\ 
408 & J141409.85+520137.2 & $1.734$ & $46.200$ & 2.02 & \CIV & 5.72 & $178.5_{-4.4}^{+8.2}$ & 3 &   Lag0  &  G19 & ...\\ 
   &   &   &   &   & \CIII & 4.82   &   &   & Lag1 & G19 & \\ 
485 & J141912.47+520818.0 & $2.563$ & $46.700$ & 2.13 & \CIV & 6.06 & $133.4_{-5.2}^{+22.6}$ & 3 &   Lag0  &  G19 & ...\\ 
   &   &   &   &   & \CIII & 4.32   &   &   & Lag1 & G19 & \\ 
   &   &   &   &   & \SiIV & 3.17   &   &   & Lag1 & G19 & \\ 
549 & J141631.45+541719.7 & $2.275$ & $45.846$ & 3.78 & \CIV & 4.04 & $69.8_{-7.2}^{+5.3}$ & 4 &   Lag0  &  G19 & ...\\ 
827 & J141218.03+541817.1 & $1.965$ & $45.748$ & 6.43 & \CIV & 4.27 & $137.7_{-19.4}^{+18.3}$ & 3 &   Lag0  &  G19 & ...\\ 

\end{longtable*}

\setlength{\tabcolsep}{0.06in}

To synchronize the continuum and broad line variations, we utilize the measured broad-line lags. If the specific line has a measured lag, we simply shift the time series of the line response using the lag; for broad lines covered in the same spectrum but without reported lags, we use the lags of other lines if available, e.g., \hbeta\ lag for \MgII, \CIV\ lag for \CIII. We designate the sample with the corresponding lags the ``Lag0'' sample and those with substituted lags the ``Lag1'' sample.

\section{Quantifying the breathing behaviors}\label{sec:result}

\subsection{Spectral fitting procedure}

To perform spectral fits on the multi-epoch spectra of our sample to measure broad-line properties, we adopt the publicly available quasar spectral fitting package {\tt QSOFIT} \citep{Shen_etal_2019b}. {\tt QSOFIT} deploys a multi-component functional fitting approach similar to earlier work \citep[e.g.,][]{Shen_etal_2008, shen_etal_2011}. Its continuum model consists of a power-law,  an optical \FeII\ template \citep{Boroson_Green_1992}, an UV \FeII\ template\footnote{The \FeII\ template is the same as used in \citet{Shen_Liu_2012}. See their \S 3.1 for more details.} \citep{Vestergaard_Wilkes_2001, Tsuzuki_etal_2006, Salviander_etal_2007} and a 5th-order polynomial to account for any possible reddening. The emission line flux from the continuum and \FeII\ subtracted spectrum is then fitted with multiple Gaussians. The fits were performed in the rest-frame of the quasar using systemic redshifts from \citet{Shen_etal_2019b}.

{\tt QSOFIT} allows the user to specify the fitting range and switch on/off individual continuum and line components. To mitigate the difficulty of fitting the global continuum, we fit each spectrum locally in several line complexes by specifying the {\tt fitting range} parameter. Table \ref{tab:fittingpara} summarizes the fitting ranges we use for each line complex, the continuum and line components included in the fitting, as well as the number of broad-line and narrow-line Gaussians used for each line component.

We now describe the multi-epoch spectral fitting procedure in detail. For each object, we first measure the median S/N over the spectrum for all spectroscopic epochs. Then the mean and standard deviation $\sigma$ of individual median S/N are computed in each observing season, and epochs that are 2$\sigma$ below the mean S/N of each season are rejected. This step excludes 2$\sim$6 epochs in the 2014 season, and 0$\sim$3 epochs in each of other three seasons. The remaining good S/N epochs are used to generate the mean spectrum using the SDSS-III spectroscopic pipeline {\tt idlspec2d}. Next, we fit the mean spectra using {\tt QSOFIT}. The high S/N mean spectrum is used to obtain robust estimates of the parameters describing the narrow line emission and the \FeII\ emission, which are held fixed in the multi-epoch spectra fitting. The velocity division between narrow-line and broad-line Gaussian components is set to be 450 km/s in Gaussian $\sigma$. We tie the widths of the narrow lines within the same line complex (e.g., \hbeta) but allow different narrow-line widths across different line complexes (see Table \ref{tab:fittingpara}). Fixing the \FeII\ emission is a simplification, since broad \FeII\ does vary, albeit with smaller amplitude compared to other major broad lines \citep[e.g.,][]{Barth_etal_2013,Hu_etal_2015}. However, the individual epochs of spectra do not have sufficient S/N to allow reliable \FeII\ subtraction, and therefore we use the average \FeII\ emission constrained from the mean spectrum as a template to subtract this component. 

\begin{figure*}
    \centering
    \includegraphics[width=0.95\textwidth]{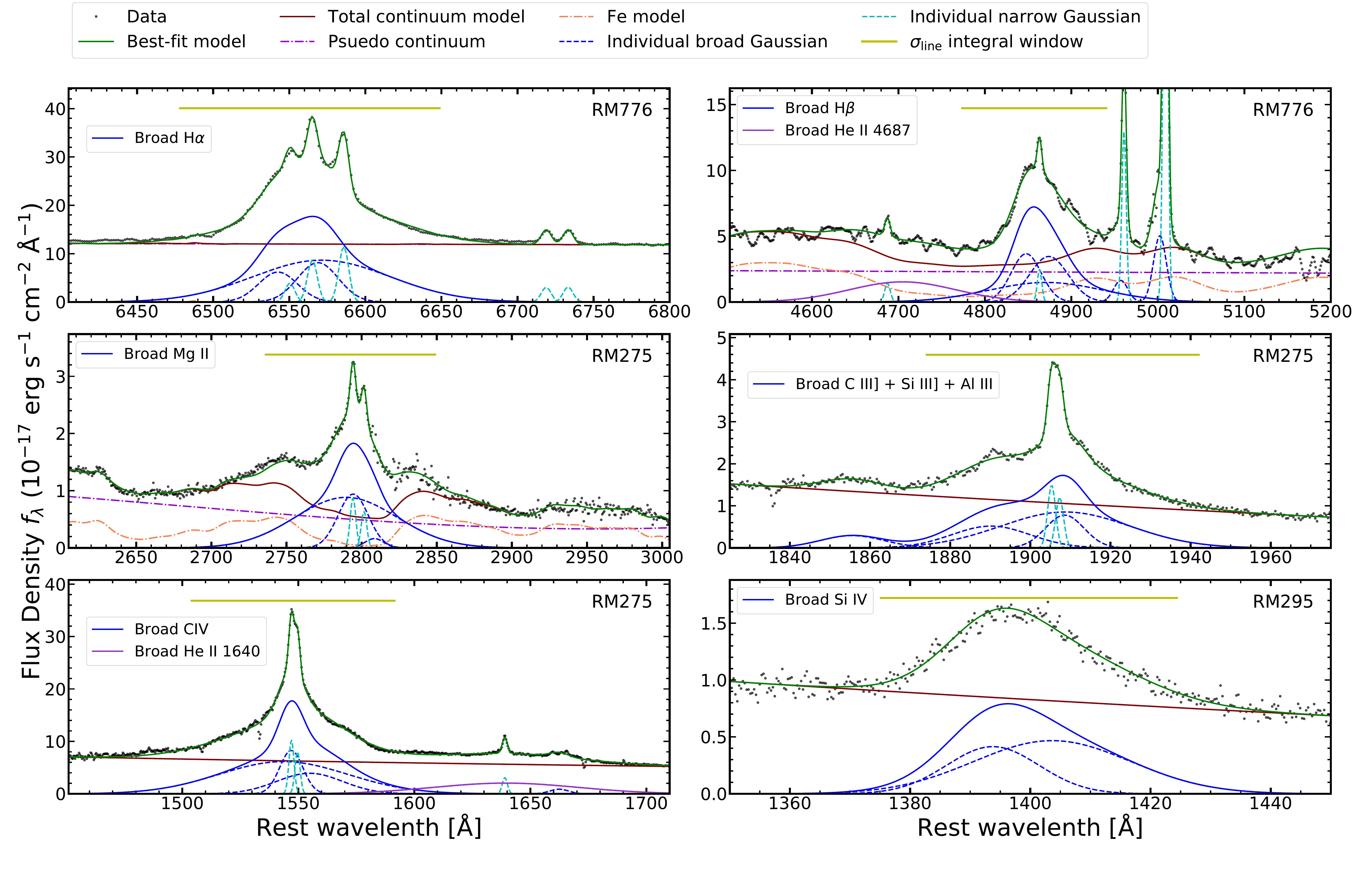}
    \caption{Examples of our multi-components functional fitting result. Each panel displays fitting result of a line complex obtained from the coadded spectra fitting. The RMID of the object used to present is labeled on the top right. The continuum is re-scaled to better show the line and the continuum simultaneously. Notably for RM275, there is strong evidence for a narrow-line component in \MgII, \CIII\ and \CIV. The horizontal line segment indicates the adaptive window we use to calculate the line dispersion $\sigma_{\rm line}$. }
    \label{fig:fittingexample}
\end{figure*}

\begin{figure*}
    \centering
    \includegraphics[width=0.8\textwidth]{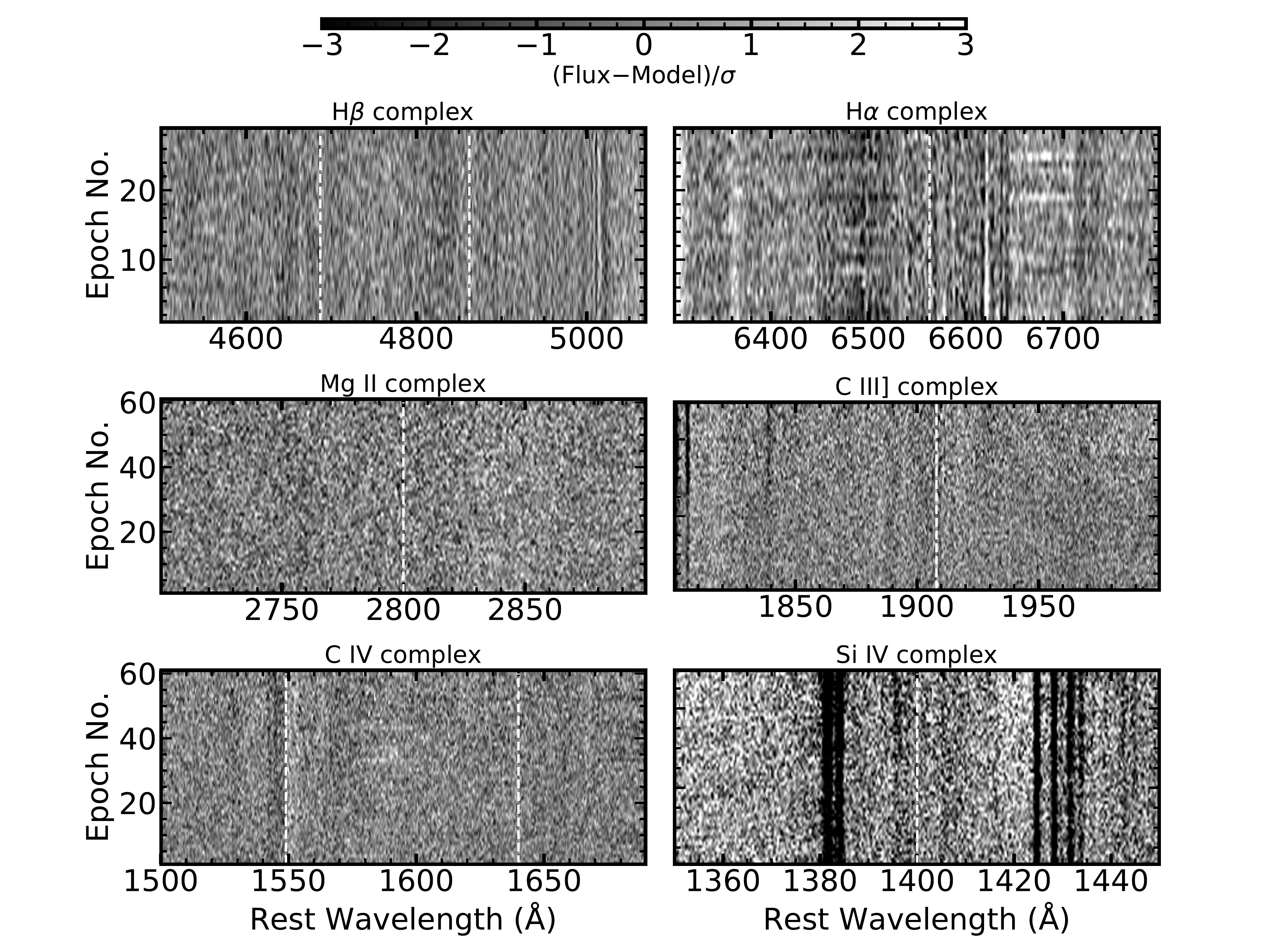}
    \caption{Examples of 2-D normalized residual maps of 6 line complexes in individual objects. The white dashed lines mark the locations of the primary broad emission line in each complex. These residual maps exhibit the quality of our multi-epoch fitting procedure. The dark vertical lines in the \SiIV\ panel are due to strong absorption in this object. }
    \label{fig:fittingresidual}
\end{figure*}

As a demonstration, Figure \ref{fig:fittingexample} shows the multi-component functional fitting results of different line complexes from the mean spectra. Figure \ref{fig:fittingresidual} shows the 2-D residual map to illustrate the quality of the multi-epoch fitting.

\begin{table}[htbp]
   \centering
   \topcaption{Fitting parameters} 
   \label{tab:fittingpara}
   \begin{tabular}{p{1.3cm}p{2cm}p{0.7cm}p{0.8cm}cp{0.8cm} }  
      \toprule
      \toprule
     Line   &  & Line  & Line & &  \\
     Complex   &  Fitting Range  &  Name & Center  & ${N_{gau, B}}^{a}$ & ${N_{gau, N}}^{b}$ \\
      \midrule
     \multirow{5}*{\halpha}  & \multirow{5}*{[6300, 6800]} 	& \halpha              &  6564.61   &  3    & 1 \\
                             &                              & \multirow{2}*{\NII} &  6549.85   &  0    & 1 \\
                             &                         		&  		               &  6585.28   &  0    & 1 \\ 
                             &                              & \multirow{2}*{\SII} &  6718.29   &  0	& 1 \\
                             &                         		&  		               &  6732.67   &  0    & 1 \\ \\
     \multirow{4}*{\hbeta}   & \multirow{4}*{[4670, 5070]}  & \hbeta               &  4862.68   &  3 	& 1 \\
                             & 						        & \HeII                & 4687.02    &  1 	& 1 \\
                             &   							& \multirow{2}*{\OIII} &  4960.30  	&  1    & 1 \\
                             &  				            &                      & 5008.24    &  1    & 1 \\ \\
     \multirow{2}*{\MgII}    &  \multirow{2}*{[2710, 2890]} &  \multirow{2}*{\MgII}&  2795.50 & \multirow{2}*{3$^c$}   & 1 \\
                             &                  		    &  				       & 2802.75    &       & 1 \\ \\
     \multirow{4}*{\CIII}    & \multirow{4}*{[1800, 2000]}  & \multirow{2}*{\CIII} & 1907       & \multirow{2}*{2}		& 1\\
                             &                  			& 				       & 1909       &      	& 1 \\
                             &                              & \AlIII            & 1857.40    & 1     & 0 \\
                             & 							    & \SiIII               &  1892.03  	& 1     & 0  \\ \\
     \multirow{4}*{\CIV}     & \multirow{4}*{[1500, 1703]}  & \multirow{2}*{\CIV}  &  1548.20   & \multirow{2}*{3}  	& 1 \\
                             &                  			&  				       & 1550.77    &      	&1 \\ 
                             &                              & \HeII                & 1640.42   & 1 	& 1 \\
                             &  							& \OIIIUV             & 1663.48  	& 1     & 1  \\ \\
      \multirow{2}*{\SiIV}   &  \multirow{2}*{[1365, 1445]} & \SiIV       		   &  1402.06   & \multirow{2}*{3}  & 1 \\
                             &  							& \OIV                 &  1396.76   &       &1 \\
      \bottomrule
\multicolumn{6}{p{0.48\textwidth}}{Notes. a. The numbers of broad-line Gaussians used in the fit. b. The number of narrow-line Gaussians used in the fit. c. We do not set broad Gaussians separately for the \MgII\ doublets. The three Gaussians are for whole \MgII\ profile, similarly for \CIII, \CIV\ and \SiIV. d. We only include the 5th-order polynomial in the \MgII\ complex fitting because the combination of only power-law and \FeII\ template often provides poor fitting results. e. We do not include the \FeII\ component in \CIV, \halpha, and \SiIV\ as the \FeII\ contribution in these wavelength ranges is low and often difficult to constrain.}
\end{tabular}
\end{table}

\subsection{Spectral measurements}\label{sec:spectral_analysis}

To quantify the line width, we calculate the Full-Width-at-Half-Maximum (FWHM) and the line dispersion $\sigma_{\rm line}$ as defined in \citet{Peterson_etal_2004}, from the model sum of all broad-line Gaussian components. Following the method described in \citet{Wang_etal_2019}, a [$-2.5\times$MAD, 2.5$\times$MAD] window is used for the calculation of $\sigma_{\rm line}$ to mitigate the adverse effects of noise and blending in the line wings. Here MAD is the Median Absolute Deviation, calculated by Equation (\ref{eqn:mad}) below: 
\begin{equation}
    MAD = \int{\left|\lambda - MED \right| F(\lambda)d\lambda} / {\int{F(\lambda)d\lambda}}\ , \label{eqn:mad}
\end{equation}
where MED is the median wavelength, defined as the location dividing the line in equal flux on both sides. As discussed in \citet{Wang_etal_2019}, this latter approach produces $\sigma_{\rm line}$ values that are reproducible, less affected by blending issues and noisy line wings, and is adaptive to the actual line width.

We employ a Monte Carlo approach to estimate uncertainties in the line width measurements \citep[e.g.,][]{Shen_etal_2008,  shen_etal_2011}. We perturb the original spectrum at each pixel by a random value drawn from a zero-mean Gaussian distribution whose $\sigma$ is set to the flux density uncertainty at that pixel and obtain a mock spectrum. We apply the same fitting procedure to the mock spectrum to obtain spectral measurements. We generate 30 mock spectra per epoch of each object, and estimate the measurement uncertainty of a spectral quantity as the semi-amplitude of the range enclosing the 16th and 84th percentiles of the distribution from the 30 trials.

\begin{figure*}[htbp]
    \subfigure{
        \begin{minipage}[t]{0.22\textwidth}
        
        \includegraphics[width=1.8in]{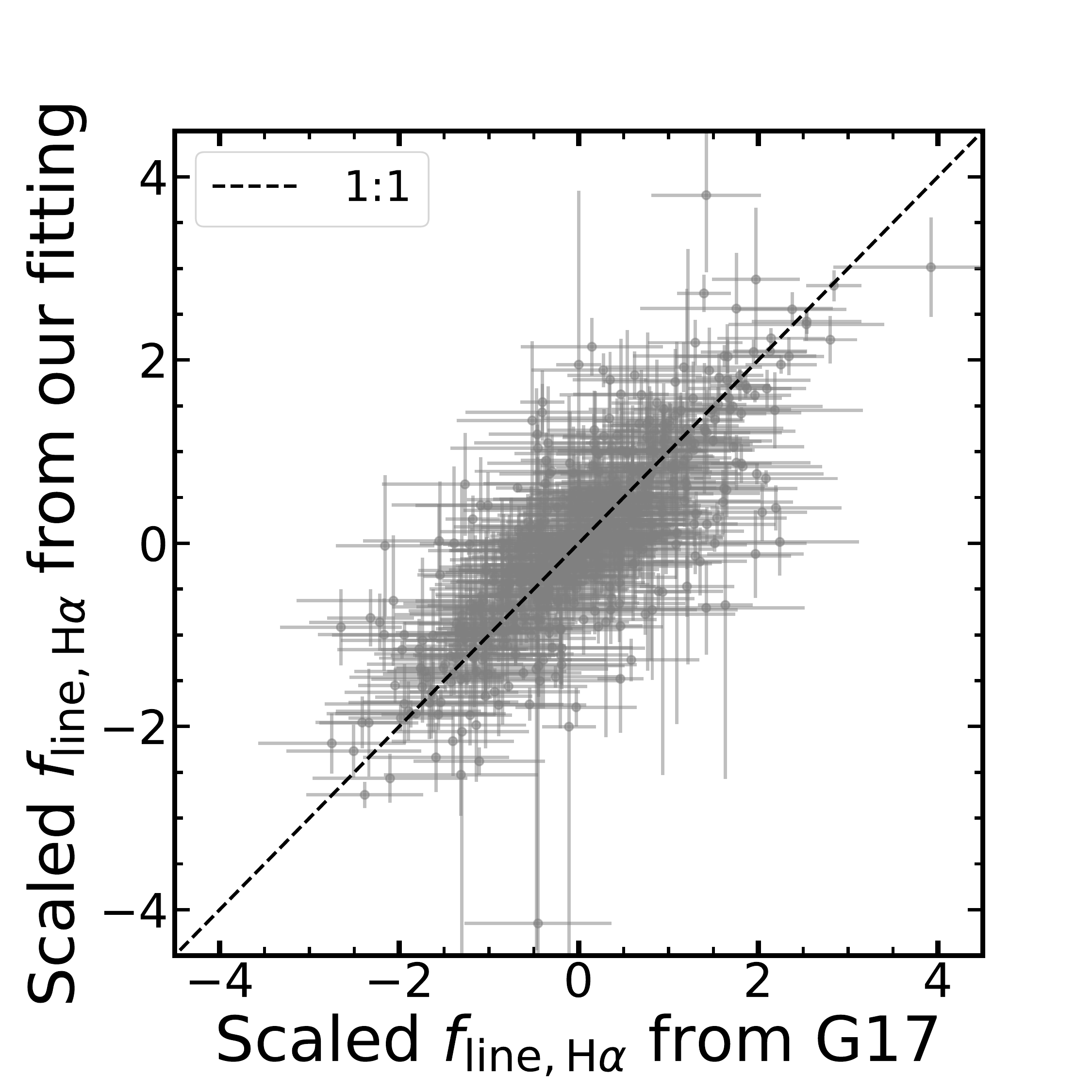}
        \end{minipage}
        }
        \subfigure{
        \begin{minipage}[t]{0.22\textwidth}
        
        \includegraphics[width=1.8in]{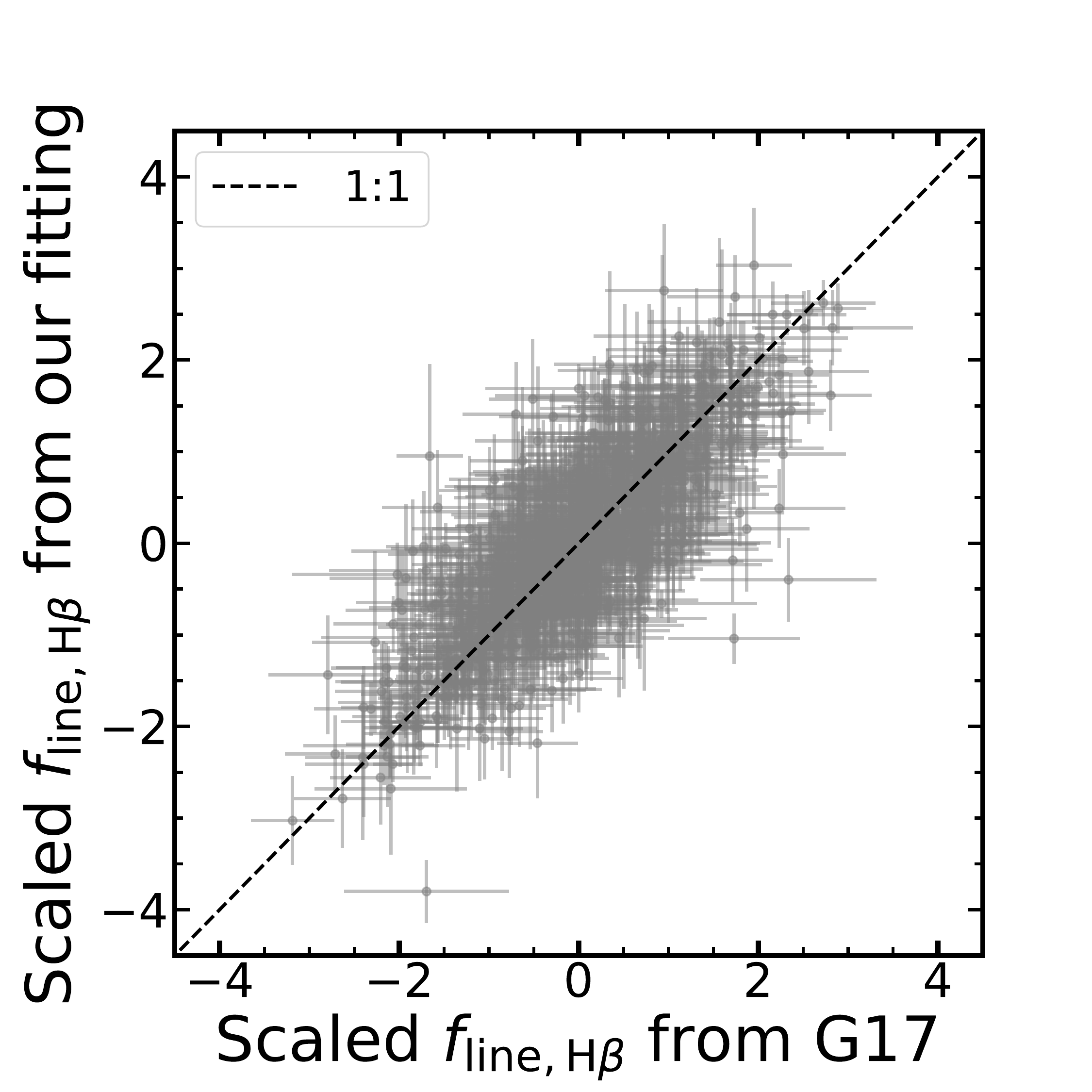}
        \end{minipage}
        }
    \subfigure{
        \begin{minipage}[t]{0.22\textwidth}
        \centering
        \includegraphics[width=1.8in]{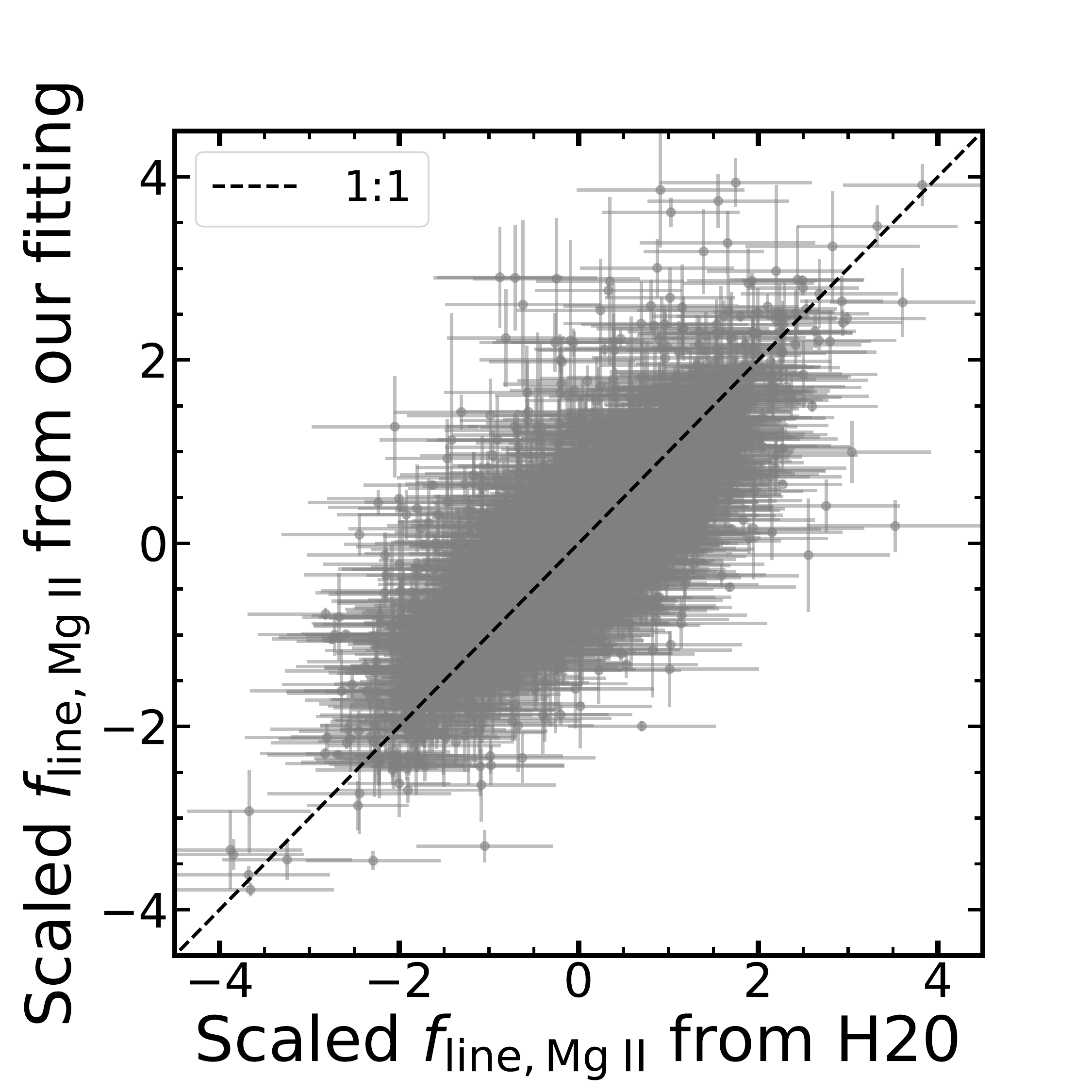}
        \end{minipage}
        
        }
    \subfigure{
        \begin{minipage}[t]{0.22\textwidth}
        
        \includegraphics[width=1.8in]{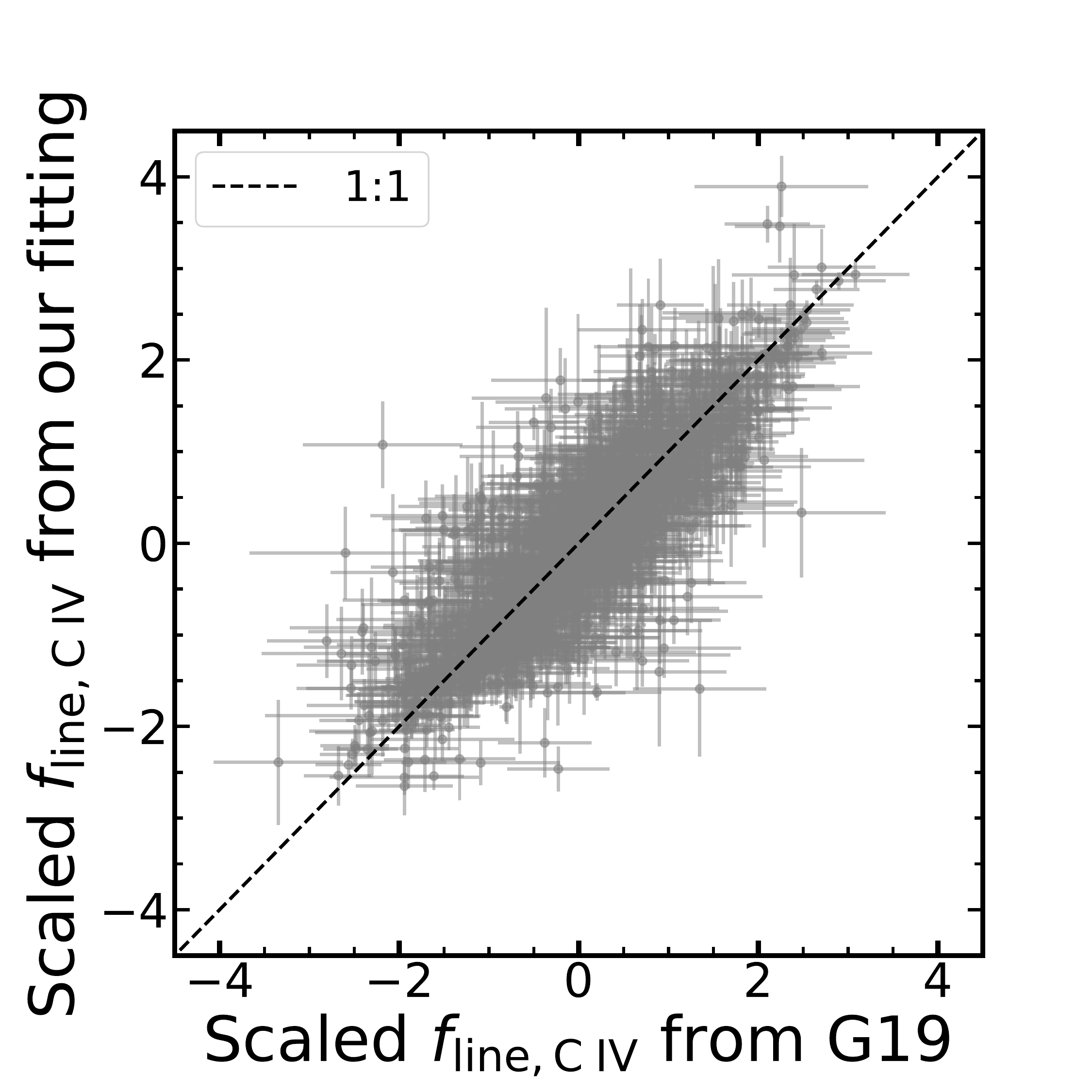}
        \end{minipage}
        }
     \caption{Comparison between our \halpha, \hbeta, \MgII, and \CIV\ line flux measurements and those from PrepSpec outputs reported in G17, H20, and G19. Each grey dot with error bars is a measurement from one epoch of one object. The black dashed line represents the 1:1 correlation. We have scaled the variable line fluxes in each light curve to have zero mean and a unity standard deviation. Our normalized line flux measurements are in good agreement with PrepSpec results.}
    \label{fig:lineflux}
\end{figure*}

In addition to the broad-line widths, we also measure the broad-line flux from our spectral fitting. We compare our flux measurements with those reported in G17, H20, and G19, which are based on PrepSpec outputs, and found reasonable agreement (Figure \ref{fig:lineflux}).

In our following analysis we use the merged $g$-band light curves as our baseline continuum flux, taken from G17\footnote{The units of the light curves in G17 are erg s$^{-1}$ cm$^{-2}$ \AA$^{-1}$ and the flux scales are slightly adjusted in the merging process (see G17 for details). We first calculate the $g$-band flux of each spectroscopic epoch by convolving the PrepSpec-processed spectrum with the SDSS $g$-band filter curve \citep{Doi_etal_2010} using the equation $\int{f_{\lambda}Rd\lambda}/\int{R d\lambda}$, where $R$ is the response curve and $f_{\lambda}$ is the flux density. Next, we scale the G17 $g$-band synthetic light curves to have the same mean and standard deviation as ours. We found good matches between the two sets of light curves. Finally we perform the same re-scaling procedure on the Bok and CFHT $g$-band light curves.}, H20 and G19. These continuum fluxes are not yet host subtracted. \citet{Park_etal_2012} found that in Mrk 40 (Arp 151), it can result in much steeper slopes than the virial expectation if the host contribution is not removed. The reason is that the extra contribution from host in the continuum flux will suppress the dynamic range in luminosity variations, making the slope of the line width -- continuum variability relation in log-log space steeper than the intrinsic value.

To remove the contribution of host contamination in the continuum light curves, we use the host fraction measured in \citet{Yue_etal_2018} (data obtained by private communication)\footnote{We have tested alternative estimation of the host fraction from spectral decomposition performed in \citet{Shen_etal_2015b}. We found that only a few individual objects show noticeable differences, and the median values of the measured breathing effects remain more or less the same. Therefore we conclude our statistical results are insensitive to the detailed scheme of host correction in the continuum light curves.}. \citet{Yue_etal_2018} decomposed the host and nuclear light using CFHT $g$ and $i$ band coadded images and measured the host fraction for 103 SDSS-RM quasars at $z<0.8$. The host fraction is calculated within the 2\arcsec\ diameter BOSS fiber and could be directly applied to spectroscopy. As listed in Table \ref{tab:sampleinformation}, the $g$-band host fraction $f_{{\rm host}, g}$ is moderate or low for most of our objects, but could reach as high as 0.9 (RM772 at $z=0.249$) for low redshift objects. For objects at $z<0.8$, we measured the $g$-band host flux from the mean spectrum. We convolve the mean spectrum with the $g$-band response curve \citep{Doi_etal_2010} to derive the total flux, which is multiplied by the $g$-band host fraction to derive the host flux. The obtained $g$-band host flux is subtracted from the light curves as a constant offset. For objects at $z>0.8$, we assume the host fraction is negligible in $g$-band.

\begin{table}[htbp]
    \centering
    \caption{Sample Statistics}
    \label{tab:Result1}
    \begin{tabular}{p{1.0cm}p{1.0cm}p{1.0cm}p{1.0cm}p{1.0cm}p{1.0cm}}
    \toprule \toprule
        Line  &  Lag0  & Lag1 & Lag1 & Lag1  & Full\\
        complex & sample & sample in G17 & sample in H20 & sample in G19 & sample\\
    \midrule
        \halpha &  14  & 7  & 0 & 0 & 21\\
        \hbeta &  25 & 3   & 3 & 0 & 31\\
        \MgII  & 13 & 15   & 0 & 0  & 28\\
        \CIII & 0 & 0  &  0   & 5 & 5 \\
        \CIV  & 13 & 0 &  0  & 0  & 13 \\
        \SiIV & 0  & 0  &  0  & 1  & 1 \\
    \bottomrule
    \end{tabular}

\end{table}

\subsection{Final Sample} \label{sec:finalsample}

We define a line S/N to restrict the breathing analysis to the subset of quasars with reliable spectral measurements. The line S/N is calculated from the continuum model subtracted residual spectrum and defined as the ratio between the broad-line model flux and the flux uncertainty. The broad-line model flux is calculated within the velocity range of [$-2.5 \times$ MAD, $2.5 \times$MAD] (same as the range for $\sigma_{\rm line}$ calculation). This is a better line S/N criterion than using the continuum-unsubtracted spectral S/N since we could lose some objects with large equivalent widths (EWs) but low continuum levels. 

We first fit all epochs of spectra for all objects in the parent sample, and calculate the median line S/N from all epochs. We select our final sample by requiring the median line S/N$>$2 for the G17 sample, and $>$3 for G19 and H20 samples. The more stringent criterion for the G19 and H20 samples is because the multi-year observations in these two papers have a wider range of S/N variation than that for the single-season observations in G17. 

\begin{figure}
    \centering
    \includegraphics[width=0.48\textwidth]{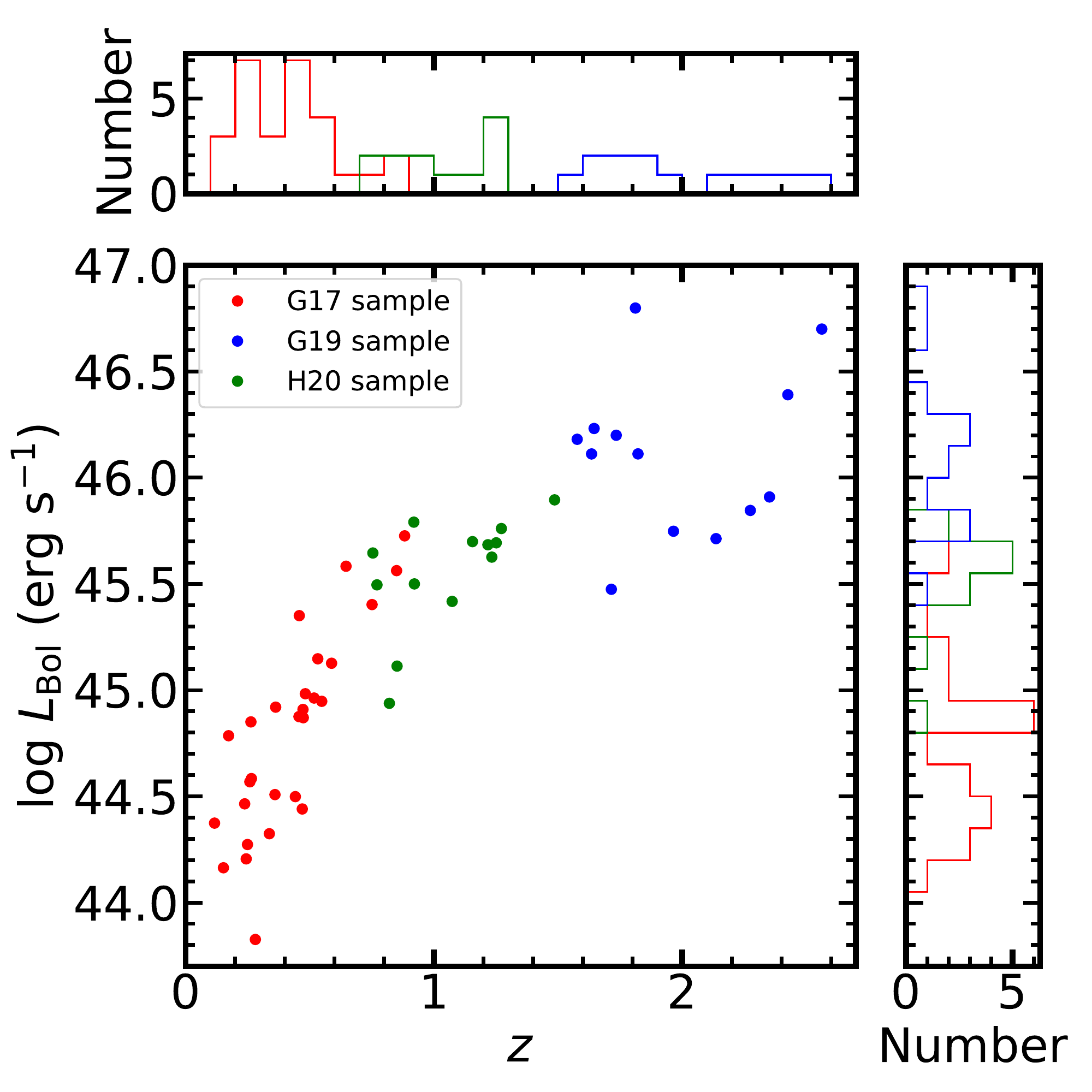}
    \caption{Distribution of quasars in our final sample for the breathing study from the G17, G19, H20 samples in the $L_{\rm Bol}-z$ plane. }
    \label{fig:sampleLzdistribution}
\end{figure}

In addition to the restriction on the median line S/N, we also exclude individual epochs with line S/N less than 2 in our following analysis. Table \ref{tab:Result1} summarizes the statistics of different lines covered by our final sample, and Figure \ref{fig:sampleLzdistribution} displays their redshift and luminosity distribution.

\subsection{\halpha, \hbeta, and \MgII\ breathing}\label{sec:lowz}

We now examine the three major broad lines accessible in optical spectroscopy for low-redshift quasars, \halpha, \hbeta, and \MgII. For the normal breathing effect we are seeking anti-correlations between variations of the continuum flux and line width. If the broad line flux responds to continuum flux variations, we can use the broad-line flux as a surrogate to track luminosity variations. The advantage of using the broad-line flux is that there is no time delay in the anti-correlation between line flux and width variations. However, the line flux light curve is much less sampled than the continuum light curve and generally has worse S/N. Therefore, we chose to use the better sampled continuum light curve as the reference light curve, and take into account the time lag between the continuum variability and broad-line response.

\begin{figure*}
    \centering
    \includegraphics[width=0.49\textwidth]{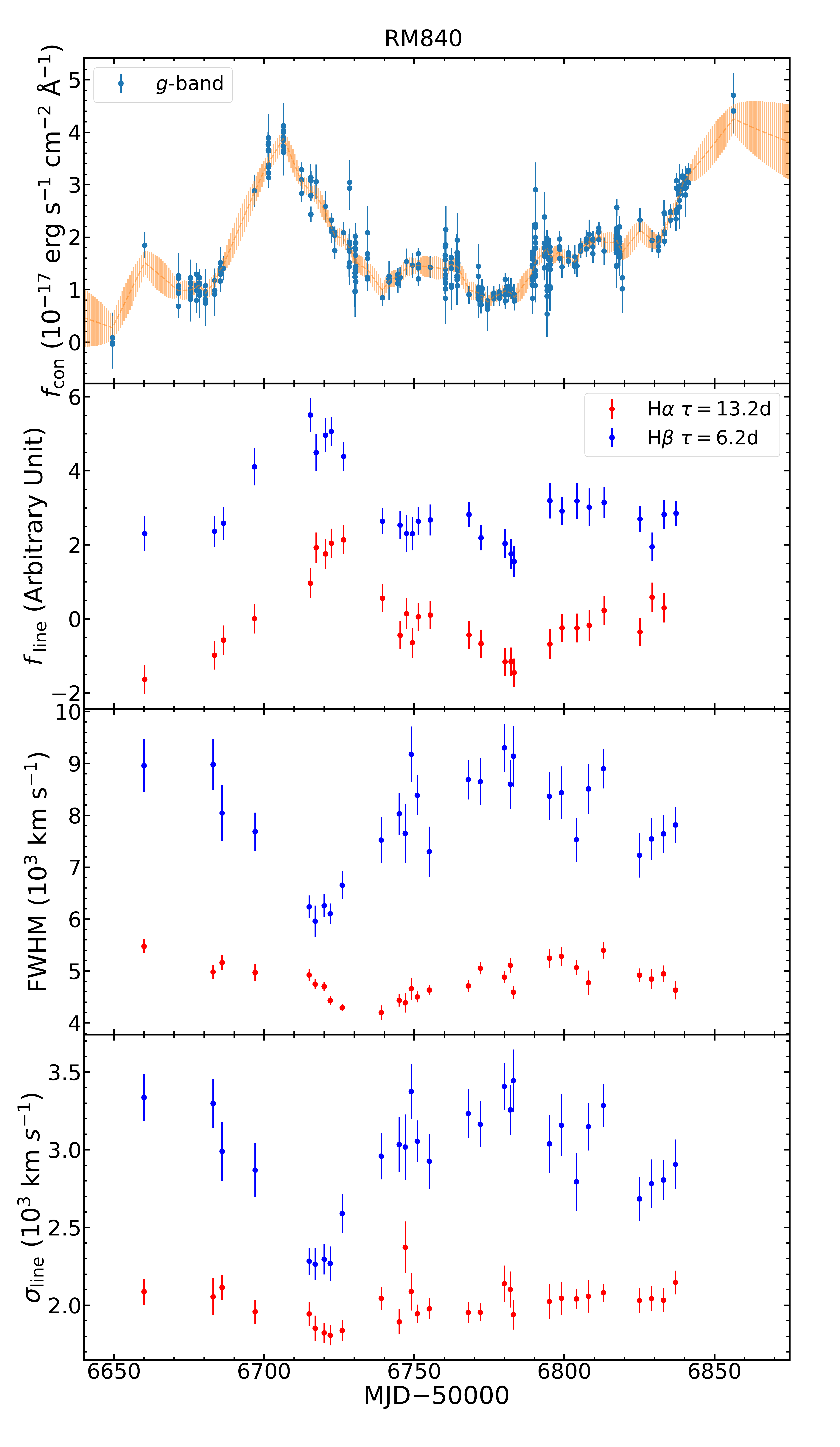}
    \includegraphics[width=0.49\textwidth]{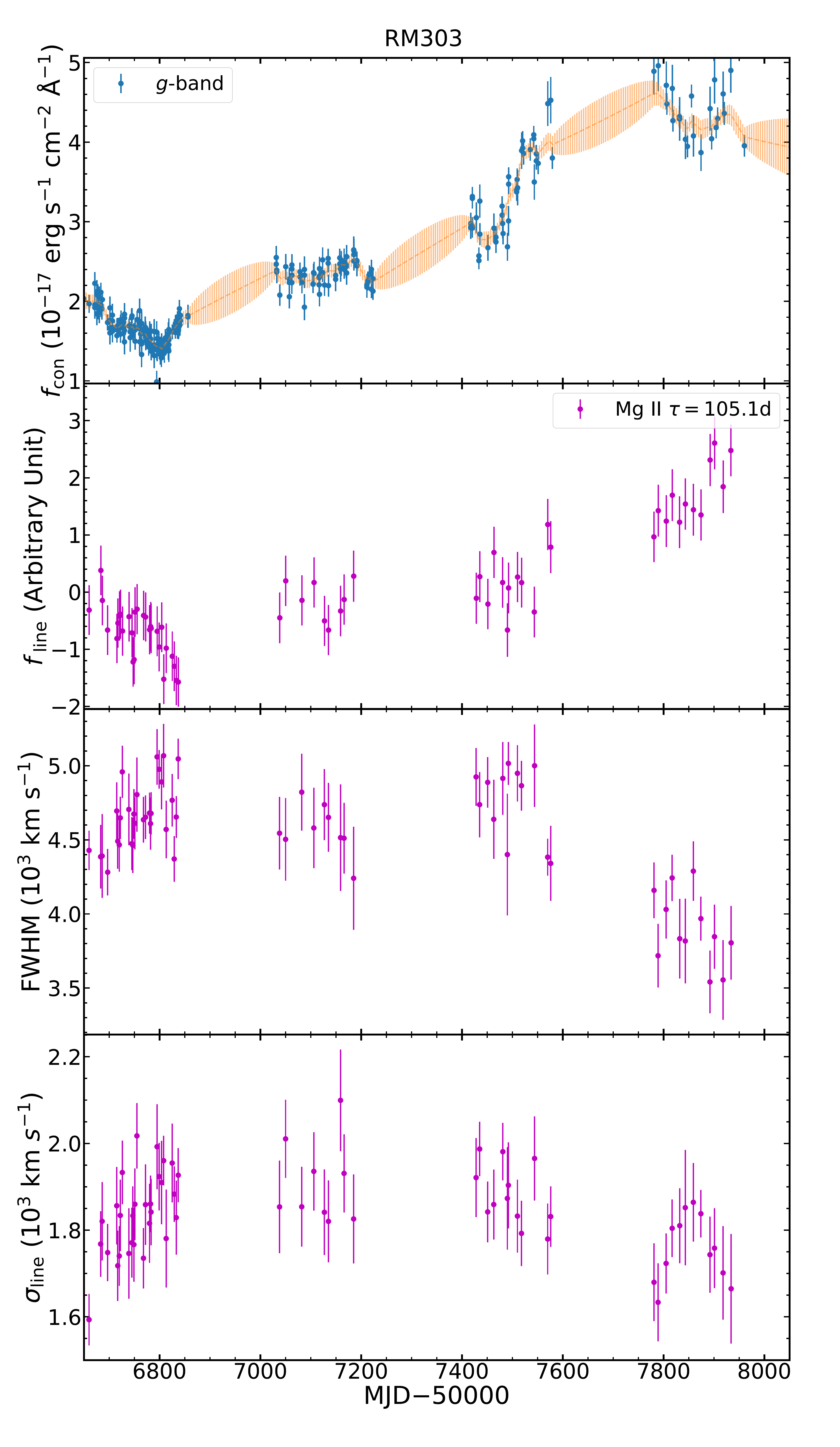}
    \caption{Two examples of time variations of continuum flux $f_{\rm con}$ (first panel), line flux (second panel), line width FWHM (third panel), and $\sigma_{\rm line}$ (fourth panel) of \hbeta\ and \halpha\ in the first observing season (left, RM840), and of \MgII\ in 2014-2017 (right, RM303). Different lines are in different colors and the observed-frame lag of each line (if reported) is also marked. $f_{\rm con}$ is the host-subtracted (if applicable) $g$-band flux. The line fluxes display the lag from the continuum light curves. The line fluxes are in arbitrary units and are vertically offset for clarity. There is clear evidence for normal breathing for all three low-ionization broad lines in these two quasars.}
    \label{fig:L1}
\end{figure*}

Figure \ref{fig:L1} presents two examples of time variations of continuum flux (host subtracted, if applicable), line flux, and broad-line widths of \halpha, \hbeta, and \MgII. There is clear evidence of anti-correlated variations between the continuum flux $f_{\rm con}$ and the line widths (both FWHM and $\sigma_{\rm line}$) for all three lines in these two quasars, i.e., the normal breathing effect. 

Next, following \citet{Shen_2013}, we study the changes in line width as a function of changes in continuum flux in log-log space using the following equation:

\begin{equation}\label{eqn:breathing}
    \Delta({\rm \log} W) = \alpha \Delta({\rm \log}f_{\rm con}) + \beta\ ,
\end{equation}

\noindent where $W$ is the line width, either FWHM or $\sigma_{\rm line}$; $f_{\rm con}$ is the continuum flux; $\alpha$ and $\beta$ are the slope and intercept. We use subscripts $F$ and $S$ to denote the $\alpha$ and $\beta$ results using FWHM and $\sigma_{\rm line}$, respectively.

To synchronize each continuum light curve and the time series of line width variations, we interpolate the continuum light curve using JAVELIN \citep{Zu_etal_2011}. JAVELIN models the continuum variability of quasars assuming a Damped Random Walk (DRW) model \citep[e.g.,][]{Kelly_etal_2009,Kozlowski_etal_2010}, which provides empirical and more accurate interpolations of the light curves than using simple linear interpolations. The interpolated continuum light curves are plotted in the top panel of Figure \ref{fig:L1}. For the uncertainties of the interpolated flux, we take into account both the uncertainties reported in the JAVELIN interpolated light curves and the uncertainties of the lag measurements. For the interpolated flux uncertainty from lag measurements, we assume the true lag is uniformly distributed between $[\tau-\tau_{nerr}, \tau+\tau_{perr}]$, where $\tau$, $\tau_{nerr}$ and $\tau_{perr}$ are the measured lag, lower and upper uncertainties of the lag, respectively. We generate 50 simulated lags within the range above and obtain the interpolated flux and flux uncertainty for each simulated lag. The final interpolated flux and uncertainty are calculated as follows: 
\begin{gather}
    f_{\rm final} = \frac{\sum_{i=1}^{50}{f_{i} * w_{i}}}{\sum_{i=1}^{50}{w_{i}}} \label{eqn:averagef}\\
    f_{\rm err,final} = \left( \frac{1}{50-1}{\sum_{i=1}^{50}{w_{i}}} \right)^{-0.5} \label{eqn:averageferr} 
\end{gather}
where $w_{i}$ is the weight calculated by $\frac{1}{f_{{\rm err},i}^2}$. If an interpolated flux point is located in a seasonal gap, it will have larger uncertainties as showed in Figure \ref{fig:L1}.

We then pair the synchronized continuum flux and line widths at two different epochs. For each pair of epochs with synchronized line width and continuum flux, we calculate two symmetric sets of differences: $[\log f_{\rm con,1} - \log f_{\rm con,2}$, $\log W_{1} - \log W_{2}]$ and $[\log f_{\rm con,2} - \log f_{\rm con,1}$, $\log W_{2} - \log W_{1}]$, where subscripts 1 and 2 denote the two epochs. For an object with $N$ pairs, we have $N(N-1)$ sets of $[\Delta(\log f_{\rm con})$, $\Delta(\log W)]$. This redundancy of data points enforces a symmetric distribution of the flux and line width changes and hence a zero intercept $\beta$. Therefore, the slope $\alpha$ measures the breathing effect\footnote{If we do not use duplicated points in the regression, the slope $\alpha$ does not change. The slope uncertainty increases by $\sim 40\%$.}. The uncertainty of $\Delta(\log f_{\rm con})$ and $\Delta(\log W)$ are derived using error propagation from the uncertainties in $f_{\rm con}$ and $W$, respectively. As in \citet{Shen_2013}, most points are located around zero and these small changes in continuum flux and line widths are consistent with measurement uncertainties. We remove those points with absolute value of $\Delta(\log f_{\rm con})$ less than 1.5 times their uncertainties $\Delta(\log f_{\rm con,err})$. Our selection criterion of SNR$_{\rm Var,con}>$2 almost guarantees that there are enough points left for slope measurements. There is only one object, RM392 in the H20 sample, that does not have enough points to meaningfully measure the slope after light curve interpolation. We thus exclude this object from further analysis.

To measure the slope $\alpha$ between $\Delta(\log W)$ and $\Delta(\log f_{\rm con})$ we use the Bayesian linear regression package {\tt linmix} from \citet{Kelly_2007}. We also measures the Pearson correlation coefficient $r$ between the variations in line width and continuum flux. The final results are summarized in Table \ref{tab:LLWslopes}.

\begin{figure}
    \centering
    \includegraphics[width=0.5\textwidth]{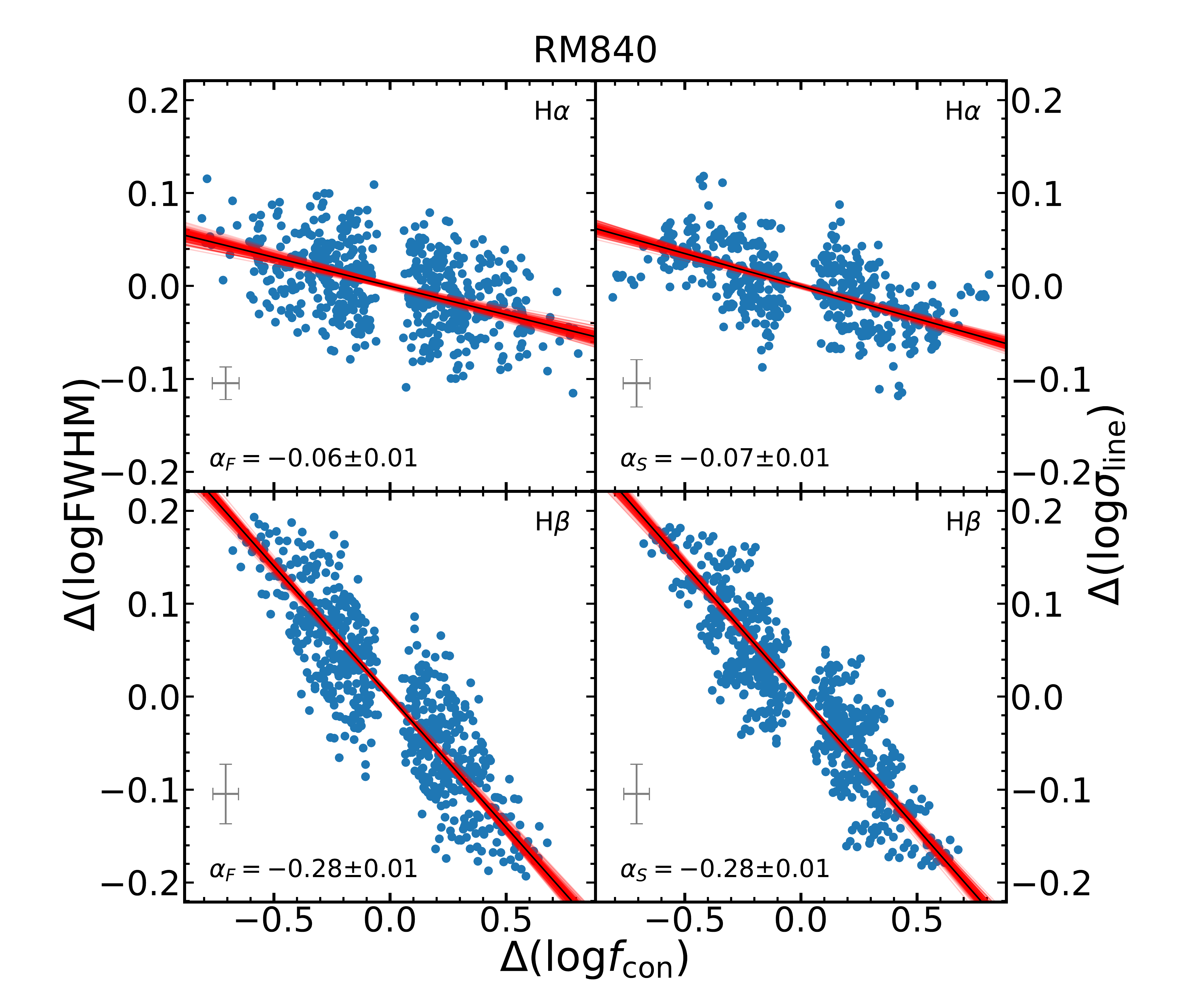}
    \caption{$\Delta\log W-\Delta\log L$ correlations for \halpha\ and \hbeta\ in RM840 in the first observing season of SDSS-RM. For $g$-band continuum flux $f_{\rm con}$, we use the host-subtracted flux. The left two panels present the results of FWHM while the right two panels are for $\sigma_{\rm line}$. The black solid lines and the red shaded regions are the median and 1$\sigma$ confidence ranges of the regression fit, respectively, derived from the Bayesian linear regression package {\tt linmix} \citep{Kelly_2007}. The grey pluses represent the median uncertainties along both axes.}
    \label{fig:LLW1}
\end{figure}

\begin{figure}
    \centering
    \includegraphics[width=0.5\textwidth]{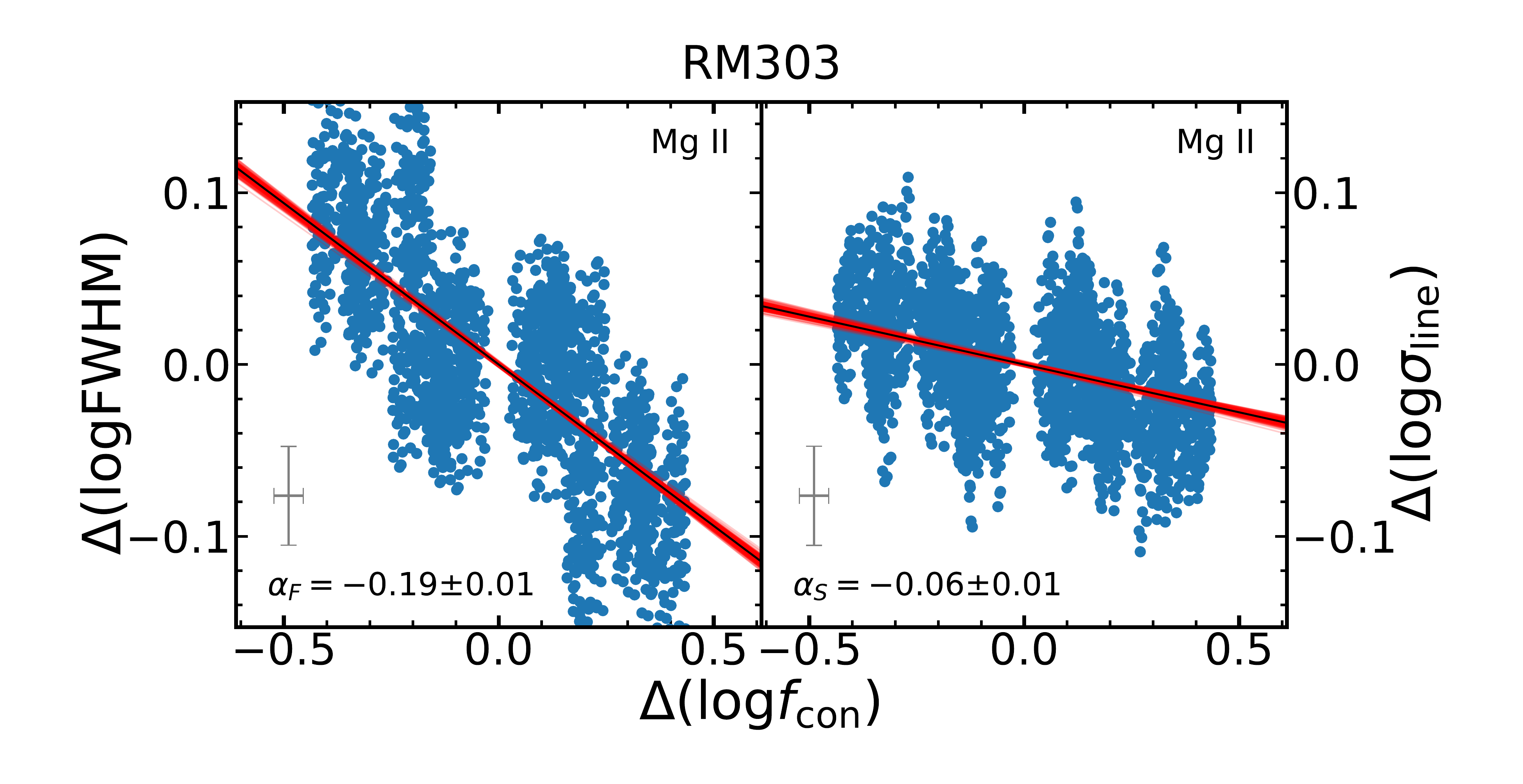}
    \caption{The $\Delta\log W-\Delta\log L$ correlation of RM303 for \MgII\ during the 2014-2017 observations of SDSS-RM.}
    \label{fig:LLW2}
\end{figure}

We present the $\Delta(\log W)-\Delta(\log f_{\rm con})$ correlation of the three low-ionization lines in Figure \ref{fig:LLW1} and \ref{fig:LLW2} for the two examples shown in Figure \ref{fig:L1}. In RM840, the slopes (based on both FWHM and $\sigma_{\rm line}$) of \hbeta\ are consistent with the expected value of $-0.25$ from the virial relation within 3$\sigma$ uncertainties, but the slopes of \halpha\ are shallower. In RM303, the slope based on \MgII\ FWHM is closer to the expected value of $-0.25$ than $\sigma_{\rm line}$.

\begin{figure}
    \centering
    \includegraphics[width=0.5\textwidth]{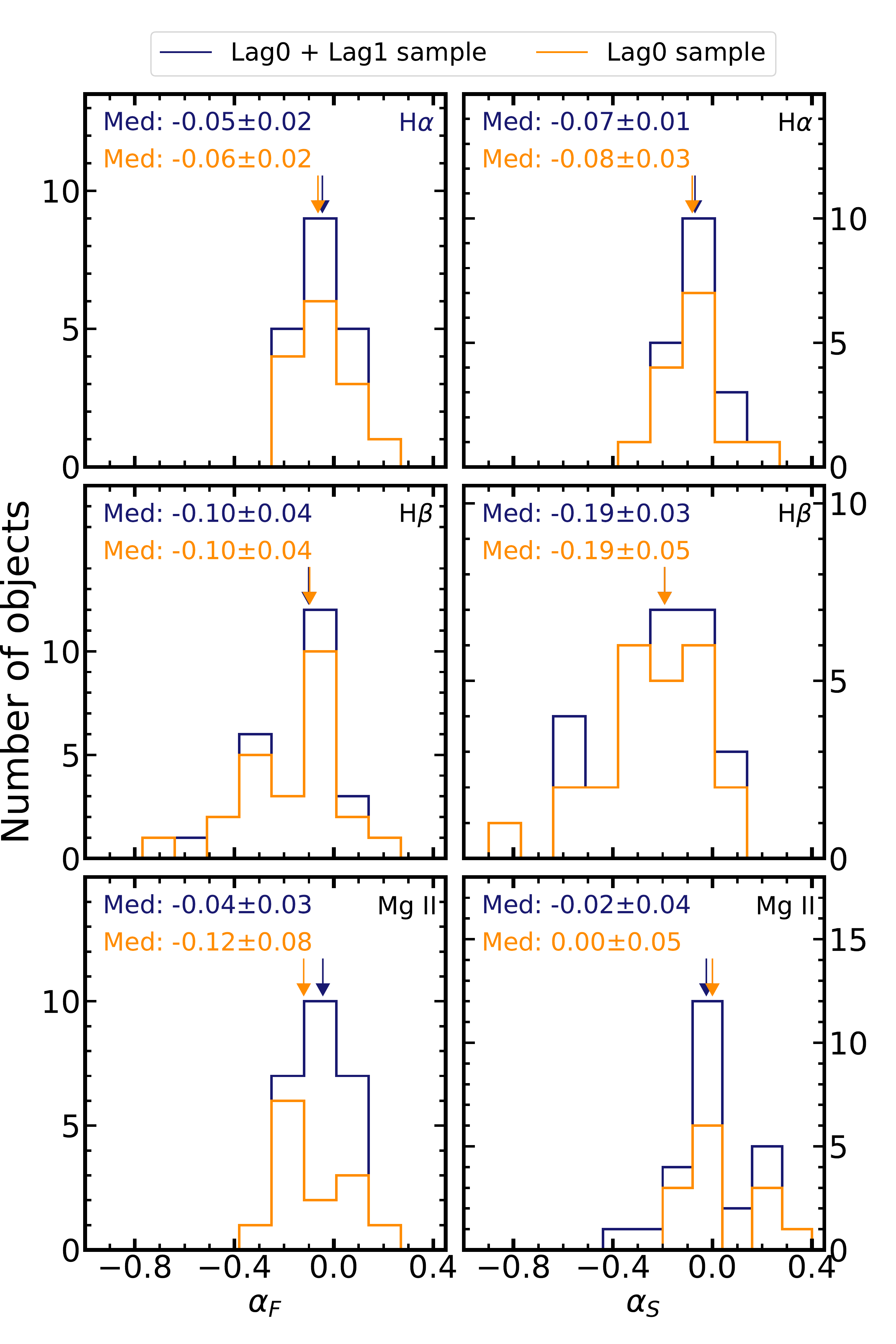}
    \caption{Distributions of the slope in the $\Delta\log W-\Delta\log L$ correlation for broad \halpha, \hbeta, and \MgII, and for $\alpha_{F}$ (left) and $\alpha_{S}$ (right), derived using FWHM and $\sigma_{\rm line}$, respectively. }
    \label{fig:LLWStat}
\end{figure}

We report the values of $\alpha_{F}$ and $\alpha_{S}$ for \hbeta, \halpha, and \MgII, measured for individual objects in our sample, in Table \ref{tab:LLWslopes}. Figure \ref{fig:LLWStat} displays the distributions of the two slopes for the three lines. Median (MED) slopes and their uncertainties are estimated by bootstrap resampling. We adopt the median of the bootstrap distribution and the semi-amplitude of the range enclosing the 16th and 84th percentiles as the measured MEDs and their uncertainties, respectively. In general, the Lag0 sample and the full sample (Lag0 plus Lag1 sample) show consistent MED slopes, suggesting that using the lag from an alternative line to synchronize the continuum light curve with line width changes is a reasonable approximation. We also find that the fractions of the three cases, i.e., breathing, non-detection of breathing, and anti-breathing, are similar in the Lag0 sample and the full sample for \hbeta. For \MgII\ and \halpha, the fraction of non-detections in the full sample is larger than that in the Lag0 sample. This is because objects in the Lag1 sample generally have smaller flux variation, and \hbeta\ is usually more variable than \halpha\ and \MgII. The statistics of the slope distribution are summarized in Table \ref{tab:LLWStat}. The median (MED) values of $\alpha_{F}$ and $\alpha_{S}$ for \hbeta\ are closer to the expected value of $-0.25$ than for \halpha\ and \MgII. On the other hand, for \hbeta, the slope based on $\sigma_{\rm line}$ is closer to the expected value of $-0.25$ than the slope based on FWHM. 

While the median slope is negative, which is indicative of normal breathing to some extent, the distribution of the slope in our sample is broad for all three low-ionization lines, and some objects show a positive slope indicative of very different breathing behaviors. To quantify the different breathing behaviors, we define three breathing modes: normal breathing, no breathing and anti-breathing:

\begin{gather}
    {\rm Normal\;breathing: }\;  \alpha < -3\alpha_{\rm err} \\
    {\rm Non-detection\;of\;breathing:}\;  -3\alpha_{\rm err}\leq \alpha \leq 3\alpha_{\rm err} \\
    {\rm Anti-breathing:}\;  \alpha > 3\alpha_{\rm err}\ ,
\end{gather}
where $\alpha$ is either $\alpha_{F}$ or $\alpha_{S}$, and $\alpha_{\rm err}$ is the slope uncertainty reported by {\tt linmix}. We calculate the fractions of normal breathing, no breathing, and anti-breathing, $q_{-}$, $q_{0}$ and $q_{+}$, and summarize the results for all three lines in Table \ref{tab:LLWStat}. \hbeta\ is the line that is most consistent with normal breathing. On the contrary, both \MgII\ and \halpha\ show a large fraction of no-breathing, consistent with earlier studies \citep[e.g.,][]{Shen_2013}. Anti-breathing is rare for all three low-ionization lines. We discuss the implications of these results further in \S\ref{sec:disc}. 

In our fiducial approach we fixed the \FeII\ parameters in the multi-epoch spectral decomposition. The weak \FeII\ variability may have a non-negligible effect on the wings of the \MgII\ profile, which $\sigma_{\rm line}$ is more sensitive to. To test this effect, we allow the \FeII\ parameters to vary and refit objects included in the \MgII\ Lag0 sample. The median slopes in this case are $-0.11\pm0.08$ and $-0.02\pm0.04$, using FWHM and $\sigma_{\rm line}$, respectively, which are fully consistent with those derived in the fiducial case. Therefore our results are insensitive to the details of treating the \FeII\ emission around \MgII.

Our sample includes a hypervariable quasar SDSS J141324${\rm+}$530527 (RM017). It is identified as a changing-look quasar in \citet{Wang_etal_2018} using spectroscopic data from the SDSS-RM project. \citet{Dexter_etal_2019} studies the response of different emission lines in terms of line flux and width to the dramatic continuum flux variation using monitoring data from 2009 to 2018. In our work, we only include the light curve in the 2014 observing season for this particular object. In 2014, RM017 showed moderate variations in the continuum flux and line width. The derived $\alpha_{S}$ for \halpha, \hbeta, and \MgII\ in this work are $-0.16\pm0.03$, $-0.22\pm0.02$ and $-0.07\pm0.01$, which are smaller but generally consistent with the values measured in \citet{Dexter_etal_2019}: $-0.21\pm0.02$, $-0.26\pm0.03$ and $-0.12\pm0.04$, respectively.

Finally, as mentioned in \S \ref{sec:data}, using a more strict criterion for our sample, e.g., ${\rm SNR}_{\rm Var,con} > 3$, the distributions of breathing slopes are generally consistent. We present the results derived with the samples defined by the new criterion in Figure \ref{fig:appendixllw}. While the sample size of \halpha, \hbeta, and \MgII\ Lag0 sample decreases to 8, 17, 9 (compared to 14, 25, 13 for ${\rm SNR}_{\rm Var,con} > 2$), the median slopes are generally consistent within uncertainties with our fiducial results.

\setlength{\tabcolsep}{0.04in}
\begin{table*}[htbp]
    \scriptsize
    \caption{Breathing effects for \halpha, \hbeta, and \MgII}
    \label{tab:LLWslopes}
    \centering
    \centerfloat
    \begin{tabular}{l p{0.3cm} rrrr rrrr rrrr}
    \toprule \toprule
         &    &  \multicolumn{4}{c}{\halpha\ } & \multicolumn{4}{c}{\hbeta\ } & \multicolumn{4}{c}{\MgII }  \\
        \multicolumn{1}{l}{RMID}  & \multicolumn{1}{c}{$N$} & \multicolumn{1}{c}{$r_{F}$} & \multicolumn{1}{c}{$\alpha_{F}$} & \multicolumn{1}{c}{$r_{S}$} & \multicolumn{1}{c}{$\alpha_{S}$}  & \multicolumn{1}{c}{$r_{F}$} & \multicolumn{1}{c}{$\alpha_{F}$} & \multicolumn{1}{c}{$r_{S}$} & \multicolumn{1}{c}{$\alpha_{S}$} & \multicolumn{1}{c}{$r_{F}$} & \multicolumn{1}{c}{$\alpha_{F}$} & \multicolumn{1}{c}{$r_{S}$} & \multicolumn{1}{c}{$\alpha_{S}$} \\
        \multicolumn{1}{l}{(1)} & \multicolumn{1}{c}{(2)} & \multicolumn{1}{c}{(3)} & \multicolumn{1}{c}{(4)} & \multicolumn{1}{c}{(5)} & \multicolumn{1}{c}{(6)} & \multicolumn{1}{c}{(7)} & \multicolumn{1}{c}{(8)} & \multicolumn{1}{c}{(9)} & \multicolumn{1}{c}{(10)} & \multicolumn{1}{c}{(11)} & \multicolumn{1}{c}{(12)} & \multicolumn{1}{c}{(13)} & \multicolumn{1}{c}{(14)} \\
    \midrule
016* & 28 & ... & ... & ... & ... & $0.49$ & $0.22\pm0.03$ & $0.15$ & $0.07\pm0.04$ & $0.33$ & $0.12\pm0.03$ & $0.45$ & $0.21\pm0.04$ \\ 
017 & 28 & $-0.72$ & $-0.07\pm0.02$ & $-0.83$ & $-0.16\pm0.03$ & $-0.80$ & $-0.30\pm0.03$ & $-0.91$ & $-0.22\pm0.02$ & $-0.12$ & $-0.01\pm0.01$ & $-0.60$ & $-0.07\pm0.01$ \\ 
044* & 62 & ... & ... & ... & ... & ... & ... & ... & ... & $-0.04$ & $0.02\pm0.06$ & $0.56$ & $0.25\pm0.06$ \\ 
088* & 28 & $0.43$ & $0.21\pm0.08$ & $0.50$ & $0.27\pm0.08$ & $0.15$ & $0.06\pm0.06$ & $0.11$ & $0.09\pm0.07$ & $0.14$ & $0.14\pm0.10$ & $-0.16$ & $-0.05\pm0.08$ \\ 
101 & 29 & $-0.71$ & $-0.21\pm0.06$ & $0.04$ & $0.02\pm0.12$ & $-0.45$ & $-0.15\pm0.04$ & $-0.30$ & $-0.07\pm0.04$ & $-0.45$ & $-0.18\pm0.06$ & $-0.46$ & $-0.36\pm0.09$ \\ 
160 & 29 & $-0.78$ & $-0.09\pm0.01$ & $-0.94$ & $-0.16\pm0.01$ & $-0.54$ & $-0.08\pm0.02$ & $-0.57$ & $-0.08\pm0.01$ & $-0.37$ & $-0.06\pm0.01$ & $-0.43$ & $-0.08\pm0.01$ \\ 
177 & 29 & $-0.08$ & $-0.05\pm0.09$ & $0.03$ & $0.01\pm0.02$ & $-0.12$ & $-0.06\pm0.03$ & $-0.65$ & $-0.18\pm0.03$ & $-0.11$ & $-0.05\pm0.04$ & $-0.15$ & $-0.05\pm0.03$ \\ 
191 & 29 & $-0.59$ & $-0.24\pm0.04$ & $-0.25$ & $-0.16\pm0.05$ & $-0.26$ & $-0.06\pm0.05$ & $-0.80$ & $-0.58\pm0.08$ & ... & ... & ... & ... \\ 
229 & 29 & $-0.34$ & $-0.14\pm0.06$ & $-0.19$ & $-0.05\pm0.07$ & $-0.47$ & $-0.50\pm0.08$ & $-0.41$ & $-0.19\pm0.04$ & $0.13$ & $0.09\pm0.06$ & $0.35$ & $0.19\pm0.06$ \\ 
252* & 28 & $-0.07$ & $-0.00\pm0.01$ & $-0.30$ & $-0.02\pm0.01$ & $0.06$ & $-0.02\pm0.02$ & $-0.13$ & $-0.04\pm0.02$ & ... & ... & ... & ... \\ 
267 & 28 & ... & ... & ... & ... & $-0.28$ & $-0.10\pm0.04$ & $0.08$ & $0.06\pm0.06$ & $-0.31$ & $-0.06\pm0.03$ & $0.22$ & $0.10\pm0.05$ \\ 
272 & 29 & $0.12$ & $0.02\pm0.03$ & $-0.05$ & $0.02\pm0.02$ & $0.48$ & $0.10\pm0.02$ & $0.03$ & $-0.00\pm0.02$ & ... & ... & ... & ... \\ 
300 & 29 & ... & ... & ... & ... & $-0.71$ & $-0.74\pm0.09$ & $-0.41$ & $-0.42\pm0.10$ & $0.03$ & $-0.04\pm0.14$ & $-0.21$ & $-0.26\pm0.09$ \\ 
301 & 26 & ... & ... & ... & ... & $-0.65$ & $-0.29\pm0.03$ & $-0.77$ & $-0.28\pm0.02$ & $0.07$ & $0.01\pm0.02$ & $0.19$ & $0.03\pm0.01$ \\ 
303* & 62 & ... & ... & ... & ... & ... & ... & ... & ... & $-0.75$ & $-0.19\pm0.01$ & $-0.43$ & $-0.06\pm0.01$ \\ 
320 & 29 & $-0.09$ & $-0.04\pm0.08$ & $-0.45$ & $-0.15\pm0.09$ & $-0.50$ & $-0.38\pm0.06$ & $-0.66$ & $-0.48\pm0.06$ & ... & ... & ... & ... \\ 
371 & 29 & $0.05$ & $0.07\pm0.02$ & $-0.50$ & $-0.09\pm0.02$ & $-0.03$ & $-0.00\pm0.01$ & $-0.95$ & $-0.30\pm0.01$ & $-0.24$ & $-0.05\pm0.02$ & $0.12$ & $0.02\pm0.02$ \\ 
373* & 26 & ... & ... & ... & ... & $0.77$ & $0.66\pm0.07$ & $0.82$ & $0.82\pm0.10$ & ... & ... & ... & ... \\ 
377 & 29 & $-0.16$ & $-0.05\pm0.05$ & $-0.51$ & $-0.25\pm0.09$ & $-0.10$ & $-0.06\pm0.10$ & $-0.24$ & $-0.19\pm0.08$ & ... & ... & ... & ... \\ 
419* & 56 & ... & ... & ... & ... & ... & ... & ... & ... & $0.65$ & $0.17\pm0.01$ & $0.76$ & $0.16\pm0.01$ \\ 
422* & 63 & ... & ... & ... & ... & ... & ... & ... & ... & $0.12$ &  $0.01\pm0.01$ & $0.20$ & $0.02\pm0.01$ \\
440 & 61 & ... & ... & ... & ... & $-0.88$ & $-0.63\pm0.03$ & $-0.88$ & $-0.58\pm0.03$ & $-0.56$ & $-0.12\pm0.01$ & $0.29$ & $-0.06\pm0.01$ \\ 
449* & 63 & ... & ... & ... & ... & ... & ... & ... & ... & $-0.38$ & $-0.15\pm0.03$ & $0.10$ & $0.04\pm0.03$ \\ 
459* & 60 & ... & ... & ... & ... & ... & ... & ... & ... & $0.05$ & $0.00\pm0.05$ & $0.33$ & $0.18\pm0.04$ \\ 
589 & 28 & ... & ... & ... & ... & $-0.76$ & $-0.22\pm0.03$ & $0.02$ & $-0.03\pm0.04$ & $-0.70$ & $-0.08\pm0.01$ & $-0.76$ & $-0.10\pm0.01$ \\ 
601 & 28 & ... & ... & ... & ... & ... & ... & ... & ... & $-0.43$ & $-0.22\pm0.04$ & $-0.09$ & $-0.03\pm0.03$ \\ 
645 & 28 & $0.51$ & $0.08\pm0.02$ & $-0.07$ & $-0.04\pm0.03$ & $-0.49$ & $-0.13\pm0.02$ & $-0.86$ & $-0.28\pm0.02$ & $0.09$ & $0.01\pm0.02$ & $0.21$ & $0.08\pm0.03$ \\ 
651* & 64 & ... & ... & ... & ... & ... & ... & ... & ... & $-0.01$ & $0.00\pm0.01$ & $0.00$ & $0.00\pm0.01$ \\ 
675* & 58 & ... & ... & ... & ... & $-0.73$ & $-0.93\pm0.06$ & $-0.71$ & $-0.54\pm0.04$ & $-0.48$ & $-0.23\pm0.03$ & $-0.35$ & $-0.14\pm0.03$ \\ 
694 & 29 & ... & ... & ... & ... & $-0.11$ & $-0.01\pm0.04$ & $-0.22$ & $-0.06\pm0.06$ & ... & ... & ... & ... \\ 
709* & 61 & ... & ... & ... & ... & ... & ... & ... & ... & $-0.36$ & $-0.16\pm0.05$ & $-0.37$ & $-0.19\pm0.05$ \\ 
714* & 60 & ... & ... & ... & ... & ... & ... & ... & ... & $-0.56$ & $-0.28\pm0.06$ & $-0.60$ & $-0.12\pm0.04$ \\
756* & 58 & ... & ... & ... & ... & ... & ... & ... & ... & $0.05$ & $0.07\pm0.09$ & $0.78$ & $0.31\pm0.05$ \\ 
761 & 61 & ... & ... & ... & ... & $-0.64$ & $-0.34\pm0.03$ & $-0.46$ & $-0.23\pm0.02$ & $-0.58$ & $-0.18\pm0.01$ & $-0.28$ & $-0.07\pm0.01$ \\ 
768 & 27 & $-0.82$ & $-0.15\pm0.02$ & $-0.71$ & $-0.10\pm0.02$ & $-0.47$ & $-0.10\pm0.03$ & $-0.51$ & $-0.12\pm0.03$ & ... & ... & ... & ... \\ 
772 & 29 & $-0.91$ & $-0.07\pm0.01$ & $-0.89$ & $-0.08\pm0.01$ & $-0.86$ & $-0.31\pm0.02$ & $-0.78$ & $-0.21\pm0.02$ & ... & ... & ... & ... \\ 
775 & 26 & $0.20$ & $0.05\pm0.02$ & $-0.07$ & $-0.07\pm0.03$ & $0.46$ & $0.13\pm0.02$ & $-0.83$ & $-0.33\pm0.02$ & ... & ... & ... & ... \\ 
776 & 29 & $-0.37$ & $-0.16\pm0.04$ & $-0.43$ & $-0.36\pm0.06$ & $-0.66$ & $-0.41\pm0.04$ & $-0.80$ & $-0.87\pm0.05$ & ... & ... & ... & ... \\ 
779 & 29 & $0.21$ & $0.02\pm0.01$ & $-0.78$ & $-0.06\pm0.02$ & $-0.08$ & $0.01\pm0.01$ & $-0.84$ & $-0.19\pm0.01$ & ... & ... & ... & ... \\ 
782 & 29 & $-0.01$ & $-0.01\pm0.08$ & $0.53$ & $0.40\pm0.22$ & $-0.09$ & $-0.02\pm0.10$ & $-0.76$ & $-0.60\pm0.09$ & ... & ... & ... & ... \\ 
790 & 29 & $0.21$ & $0.42\pm0.18$ & $-0.27$ & $-0.06\pm0.21$ & $-0.52$ & $-0.10\pm0.02$ & $-0.50$ & $-0.08\pm0.01$ & ... & ... & ... & ... \\ 
840 & 29 & $-0.50$ & $-0.06\pm0.01$ & $-0.59$ & $-0.07\pm0.01$ & $-0.86$ & $-0.28\pm0.01$ & $-0.89$ & $-0.28\pm0.01$ & ... & ... & ... & ... \\ 
\bottomrule

    \multicolumn{14}{p{\textwidth}}{Notes. Column (1): SDSS-RM identifier RMID; * indicates an object where host measurements are unavailable; (2): Number of epochs used for spectral analysis; (3)-(4): Pearson correlation coefficient $r$ and slopes in the $\Delta\log W-\Delta\log L$ relation derived using FWHM for \halpha; (5)-(6): Pearson correlation coefficient $r$ and slopes in the $\Delta\log W-\Delta\log L$ relation derived using $\sigma_{\rm line}$; (7)-(10): same as (3)-(6) but for \hbeta; (11)-(14): same as (3)-(6) but for \MgII.}
\end{tabular} 
\end{table*}

\setlength{\tabcolsep}{0.06in}

\begin{table*}[]
    \centering
    \scriptsize
    \topcaption{Slope statistics of \halpha, \hbeta, and \MgII\ breathing} 
    \label{tab:LLWStat}
    \begin{tabular}{c c c ccc c ccc }
        \toprule
        \toprule
         sample     &    & \multicolumn{4}{c}{Lag0}  &  \multicolumn{4}{c}{Lag0 + Lag1} \\
         \midrule
        line  &    &  MED   & $q_{-}$ & $q_{0}$ & $q_{+}$  & MED & $q_{-}$ & $q_{0}$ & $q_{+}$ \\
        \midrule
        \multirow{2}*{\halpha}   &  $\alpha_{F}$ & $-0.06\pm0.02$ & 0.43 & 0.43 & 0.14                            & $-0.05\pm0.02$ & 0.33 & 0.57 & 0.10 \\
                                 &  $\alpha_{S}$ & $-0.08\pm0.03$ & 0.57 & 0.36 & 0.07  & $-0.07\pm0.01$ & 0.43 & 0.52 & 0.05 \\
       \multirow{2}*{\hbeta}   & $\alpha_{F}$ & $-0.10\pm0.04$ & 0.52 & 0.32 & 0.16  & 
                                $-0.10\pm0.04$ & 0.55 & 0.32 & 0.12 \\
                                 &  $\alpha_{S}$    &  $-0.19\pm0.05$ &   0.68 & 0.28 & 0.04  & $-0.19\pm0.03$ & 0.68 & 0.29 & 0.03 \\
        \multirow{2}*{\MgII}   &  $\alpha_{F}$ &   $-0.12\pm0.08$ &   0.46 & 0.46 & 0.08  & 
                               $-0.04\pm0.03$ & 0.31 & 0.62 & 0.07 \\
                                 &  $\alpha_{S}$    &   $-0.00\pm0.05$ &   0.46 & 0.15 & 0.38  &
                                 $-0.02\pm0.04$ & 0.38 & 0.35 & 0.29 \\
        \bottomrule
        \end{tabular}

\end{table*}

\subsection{\CIV, \CIII\ and \SiIV}\label{sec:highz}

We now examine the three broad lines, \CIV, \CIII\ and \SiIV, covered by optical spectroscopy for high-redshift quasars. Since the timescales of continuum variability of these high-redshift quasars are longer than those in the low-redshift sample due to their higher luminosity and redshift, we use the multi-year $g$-band light curves compiled in H20 and G19 for this subset of SDSS-RM quasars. No host correction is available for these high-redshift ($z>1$) quasars, but the host contamination in $g$-band is much smaller than their low-$z$ counterparts and therefore does not impact our results much.

\begin{figure}
    \centering
    \includegraphics[width=0.49\textwidth]{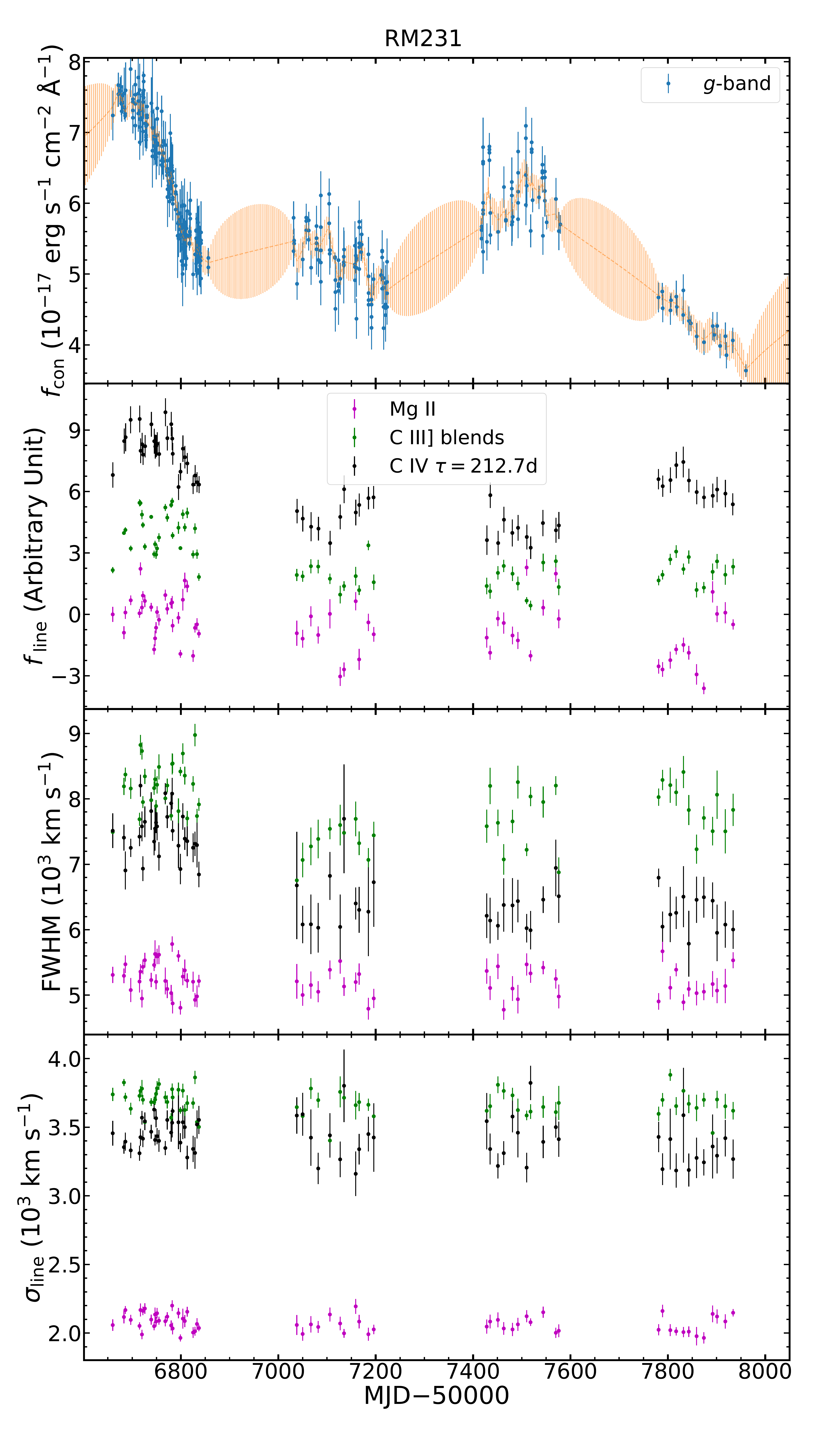}
    \caption{An example (RM231) of time variations of continuum flux $f_{\rm con}$ (first panel), line flux (second panel), line FWHM (third panel), and $\sigma_{\rm line}$ (fourth panel) of \CIV, \CIII, and \MgII\ during 2014-2017. Notations are the same as Figure \ref{fig:L1}. We observe approximately anti-breathing for \CIV\ (and less so for \CIII), in particular if using FWHM, while \MgII\ shows no obvious response in line width.}
    \label{fig:H3}
\end{figure}

Figure \ref{fig:H3} presents an example with coverage of \CIII, \CIV, and \MgII. Interestingly, \CIII\ and \CIV\ show anti-breathing while \MgII\ shows no breathing, which is expected since the \MgII\ line flux does not seem to respond to the drop in continuum flux either. We find that for many objects in our sample with both \MgII\ and \CIV\ coverage, only \CIV\ shows visible reverberation to continuum variations, a reflection of the observed fact that \MgII\ responds to continuum variability to a lesser extent compared to other broad lines \citep[e.g.,][]{Goad_etal_1993,Yang_etal_2019a,Guo_etal_2019}.  

\begin{figure}
    \centering
    \includegraphics[width=0.5\textwidth]{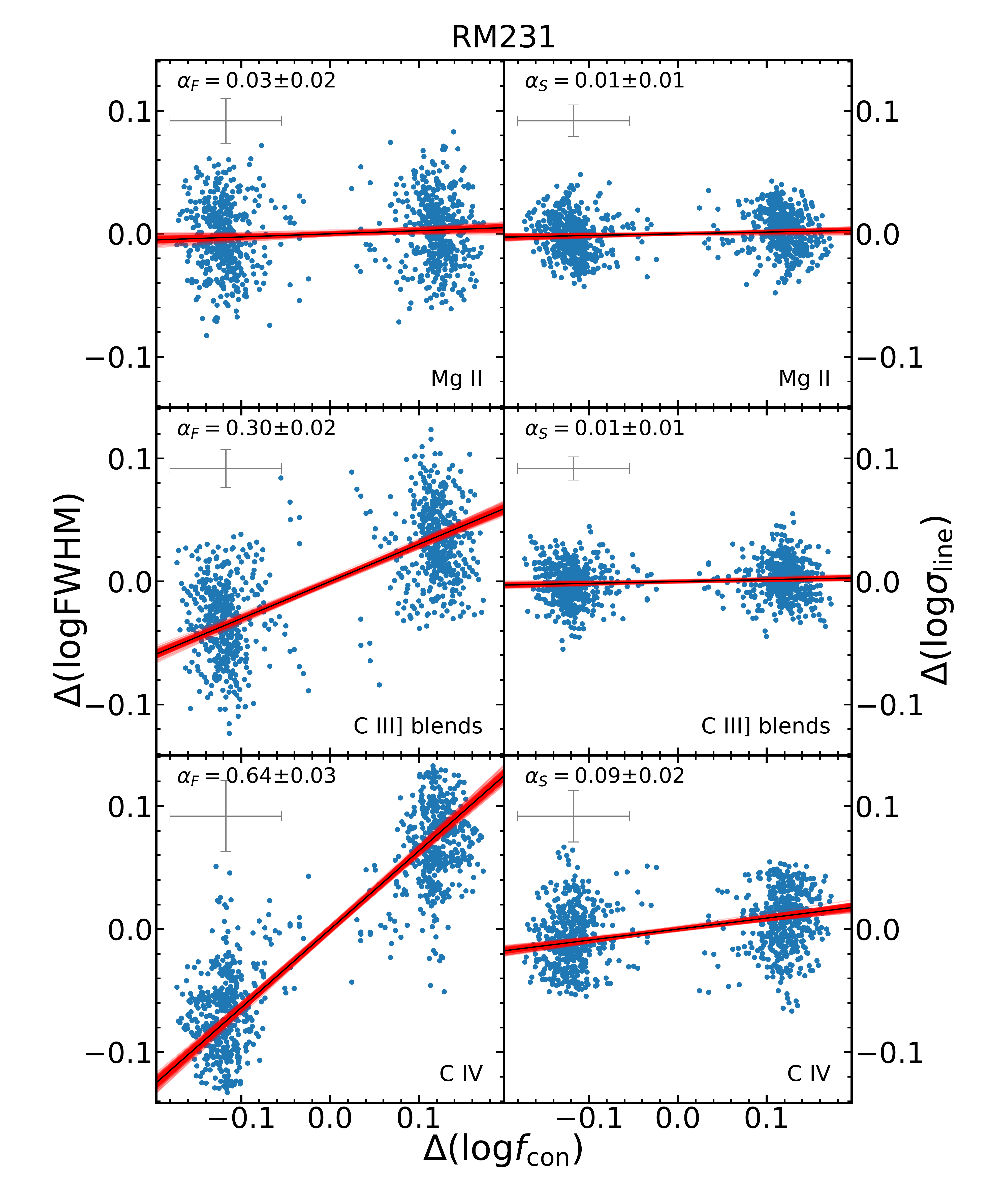}
    \caption{The $\Delta\log W-\Delta\log L$ correlation for \CIII\ and \CIV\ of RM231 during the 2014-2017 observations of SDSS-RM. For continuum flux $f_{\rm con}$, we use the $g$-band flux. \CIV\ and the \CIII\ blends show noticeable anti-breathing, in particular for FWHM, while \MgII\ is consistent with no-breathing (slope close to zero). Notations are the same as Figure \ref{fig:LLW1}.}
    \label{fig:HLW3}
\end{figure}

Next, we follow the same approach in \S\ref{sec:lowz} to analyze the $\Delta(\log W)-\Delta(\log f_{\rm con})$ correlation for the three broad lines in high-$z$ quasars. We present the slope constraints in Figure \ref{fig:HLW3} for the same example shown in Figure \ref{fig:H3}.

\begin{figure}
    \centering
    \includegraphics[width=0.5\textwidth]{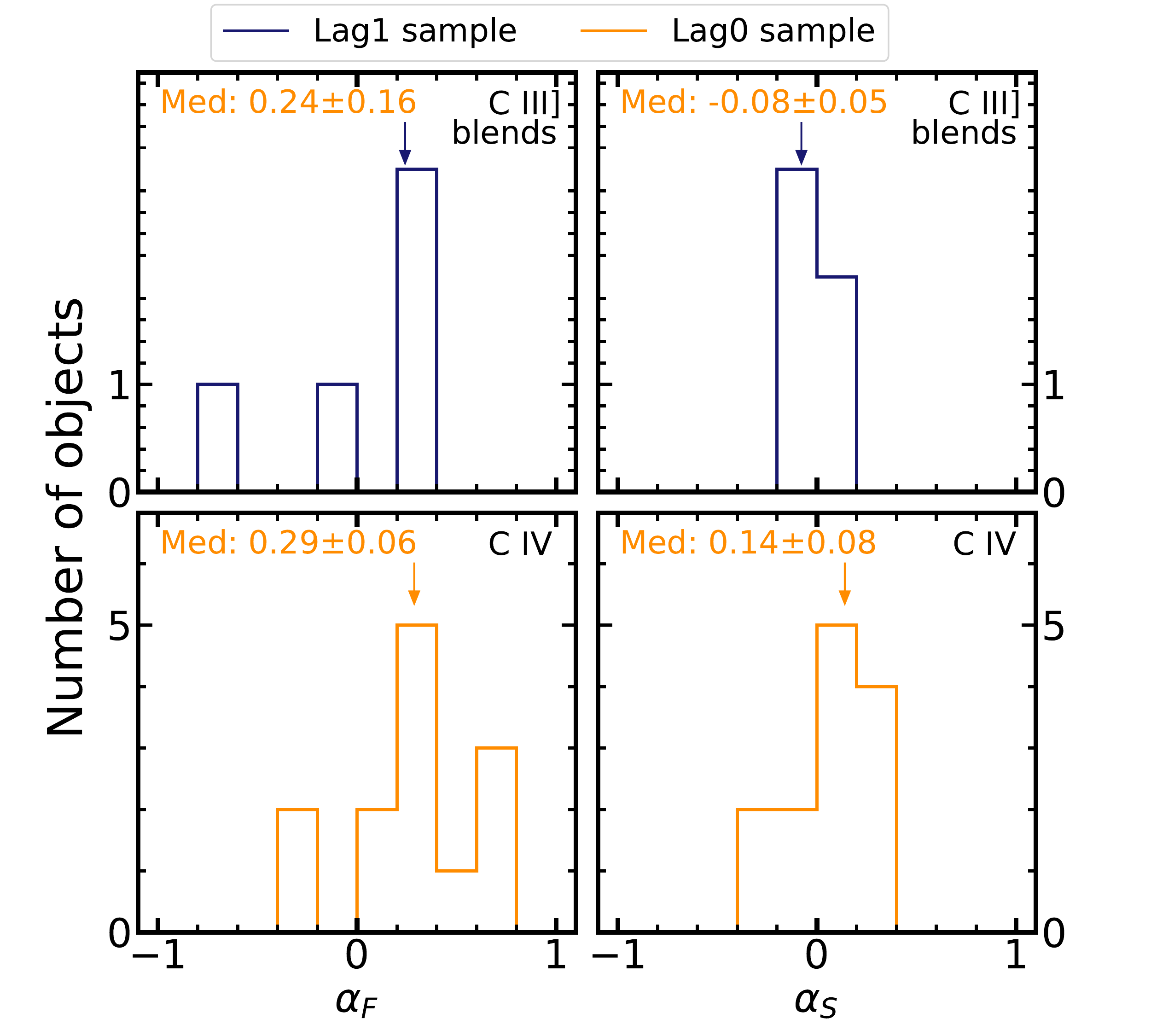}
    \caption{Distributions of the $\Delta\log W-\Delta\log L$ correlation slopes, $\alpha_{F}$ (left) and $\alpha_{S}$ (right), derived using FWHM and $\sigma_{\rm line}$, respectively, for \CIV\ and the \CIII\ blend.}
    \label{fig:HLWStat}
\end{figure}

We summarize the measured $\alpha_{F}$ and $\alpha_{S}$ for these three lines for all objects in our sample in Table \ref{tab:HLWslopes}. Figure \ref{fig:HLWStat} displays the distribution of these slopes. For most objects, $\alpha_{F}$ and $\alpha_{S}$ of \CIV, as well as $\alpha_{F}$ of the \CIII\ blends show anti-breathing. However, there are a few objects showing normal breathing as well. There is only one object in our final sample for which we can measure for \SiIV, and the inferred breathing slope is slightly positive (indicating anti-breathing) but the uncertainty is large. 

Finally, we present the distributions of the breathing slope for \CIII\ and \CIV\ for samples defined by a more stringent criterion, i.e., ${\rm SNR}_{\rm Var,con} > 3$, in Figure \ref{fig:appendixhlw}. The sample size of the \CIV\ Lag0 sample decreases to 10 (compared to 13 for ${\rm SNR}_{\rm Var,con} > 2$). For \CIII, there are only two objects left, which is too small to constrain the statistics. The median breathing slopes for both FWHM and $\sigma_{\rm line}$ for \CIV\ are consistent with our fiducial results. 

\setlength{\tabcolsep}{0.05in}
\begin{table*}[htbp]
    \scriptsize
    \caption{Breathing effects for \CIII, \CIV, and \SiIV}
    \label{tab:HLWslopes}
    \centering
    \centerfloat
    \begin{tabular}{l p{0.3cm} rrrr rrrr rrrr}
    \toprule \toprule
         &    &  \multicolumn{4}{c}{\CIII\ } & \multicolumn{4}{c}{\CIV\ } & \multicolumn{4}{c}{\SiIV }  \\
        \multicolumn{1}{l}{RMID}  & \multicolumn{1}{c}{$N$} & \multicolumn{1}{c}{$r_{F}$} & \multicolumn{1}{c}{$\alpha_{F}$} & \multicolumn{1}{c}{$r_{S}$ } & \multicolumn{1}{c}{$\alpha_{S}$}  & \multicolumn{1}{c}{$r_{F}$} & \multicolumn{1}{c}{$\alpha_{F}$} & \multicolumn{1}{c}{$r_{S}$} & \multicolumn{1}{c}{$\alpha_{S}$} & \multicolumn{1}{c}{$r_{F}$} & \multicolumn{1}{c}{$\alpha_{F}$} & \multicolumn{1}{c}{$r_{S}$} & \multicolumn{1}{c}{$\alpha_{S}$} \\
        \multicolumn{1}{l}{(1)} & \multicolumn{1}{c}{(2)} & \multicolumn{1}{c}{(3)} & \multicolumn{1}{c}{(4)} & \multicolumn{1}{c}{(5)} & \multicolumn{1}{c}{(6)} & \multicolumn{1}{c}{(7)} & \multicolumn{1}{c}{(8)} & \multicolumn{1}{c}{(9)} & \multicolumn{1}{c}{(10)} & \multicolumn{1}{c}{(11)} & \multicolumn{1}{c}{(12)} & \multicolumn{1}{c}{(13)} & \multicolumn{1}{c}{(14)} \\
    \midrule

032 & 62 & ... & ... & ... & ... & $0.97$ & $0.40\pm0.01$ & $0.90$ & $0.19\pm0.01$ & ... & ... & ... & ... \\ 
145 & 63 & ... & ... & ... & ... & $-0.66$ & $-0.24\pm0.04$ & $-0.71$ & $-0.24\pm0.03$ & ... & ... & ... & ... \\ 
201 & 60 & $-0.65$ & $-0.67\pm0.04$ & $-0.49$ & $-0.08\pm0.01$ & $-0.70$ & $-0.36\pm0.02$ & $-0.44$ & $-0.20\pm0.02$ & ... & ... & ... & ... \\ 
231 & 63 & $0.70$ & $0.30\pm0.02$ & $0.10$ & $0.01\pm0.01$ & $0.91$ & $0.64\pm0.03$ & $0.30$ & $0.09\pm0.02$ & ... & ... & ... & ... \\ 
275 & 64 & ... & ... & ... & ... & $0.90$ & $0.37\pm0.01$ & $0.77$ & $0.14\pm0.01$ & ... & ... & ... & ... \\ 
295 & 60 & ... & ... & ... & ... & $0.69$ & $0.29\pm0.04$ & $0.60$ & $0.22\pm0.04$ & ... & ... & ... & ... \\ 
298 & 62 & $0.35$ & $0.24\pm0.05$ & $-0.14$ & $-0.08\pm0.05$ & $0.44$ & $0.25\pm0.05$ & $0.19$ & $0.22\pm0.09$ & ... & ... & ... & ... \\ 
387 & 60 & ... & ... & ... & ... & $0.63$ & $0.40\pm0.03$ & $0.25$ & $0.13\pm0.03$ & ... & ... & ... & ... \\ 
401 & 62 & ... & ... & ... & ... & $0.94$ & $0.71\pm0.01$ & $0.92$ & $0.24\pm0.01$ & ... & ... & ... & ... \\ 
408 & 62 & $-0.06$ & $-0.01\pm0.10$ & $0.15$ & $0.11\pm0.08$ & $0.07$ & $0.04\pm0.13$ & $0.55$ & $0.30\pm0.07$ & ... & ... & ... & ... \\ 
485 & 62 & $0.30$ & $0.32\pm0.13$ & $-0.15$ & $-0.08\pm0.08$ & $0.59$ & $0.73\pm0.13$ & $-0.12$ & $-0.10\pm0.14$ & $-0.06$ & $0.01\pm0.18$ & $0.06$ & $0.13\pm0.14$ \\ 
549 & 63 & ... & ... & ... & ... & $0.25$ & $0.07\pm0.02$ & $0.13$ & $0.04\pm0.02$ & ... & ... & ... & ... \\ 
827 & 63 & ... & ... & ... & ... & $0.74$ & $0.26\pm0.02$ & $-0.04$ & $-0.01\pm0.01$ & ... & ... & ... & ... \\ 

\bottomrule

    \multicolumn{14}{p{\textwidth}}{Notes. Same as Table \ref{tab:LLWslopes}, but for broad \CIII, \CIV\ and \SiIV. }
    \end{tabular} 
\end{table*}

\setlength{\tabcolsep}{0.06in}

\section{Discussion}\label{sec:disc}

\subsection{\CIV\ profile variations}\label{sec:disc1}

\begin{figure*}[htbp]
    \centering
    \includegraphics[width=0.3\textwidth]{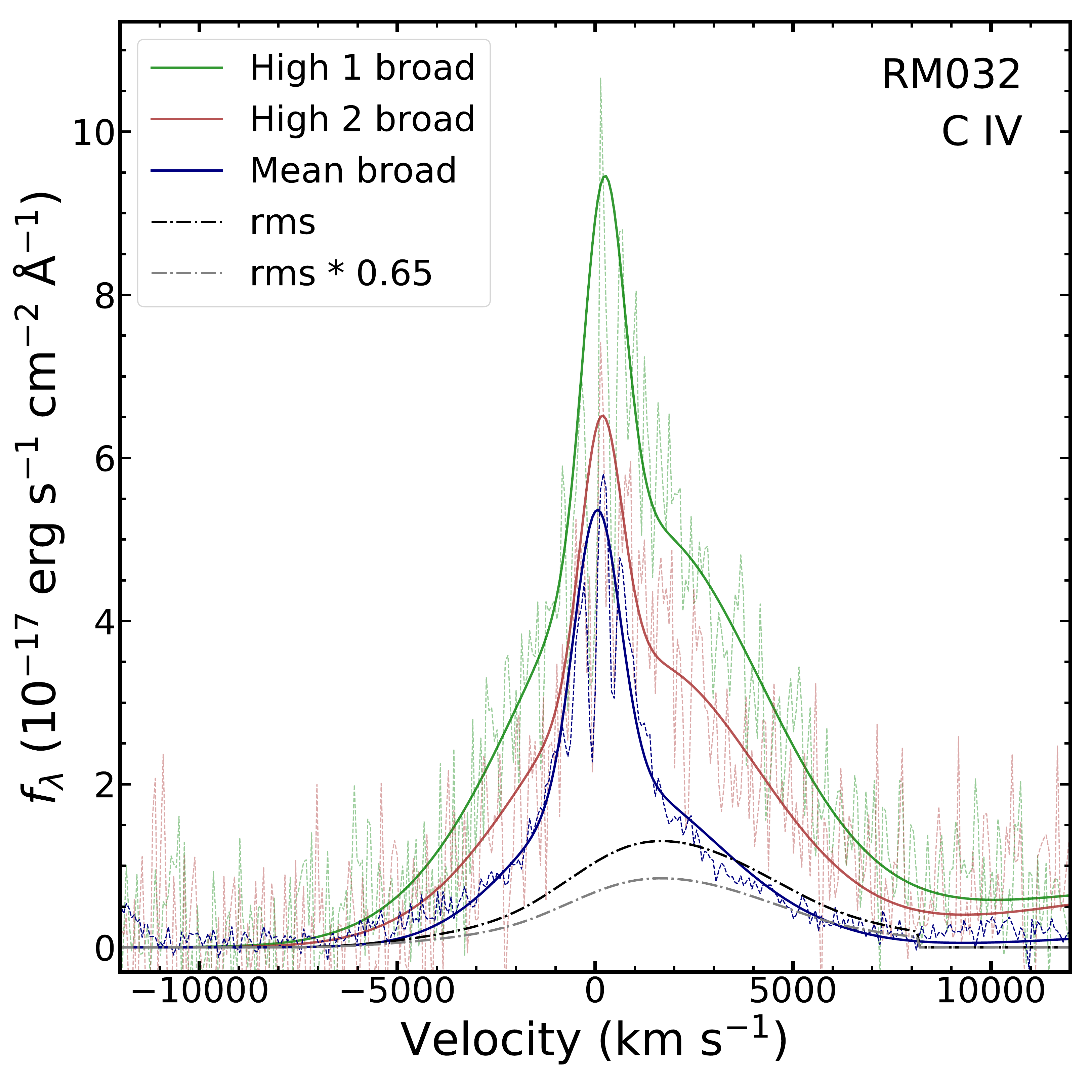}
    \includegraphics[width=0.3\textwidth]{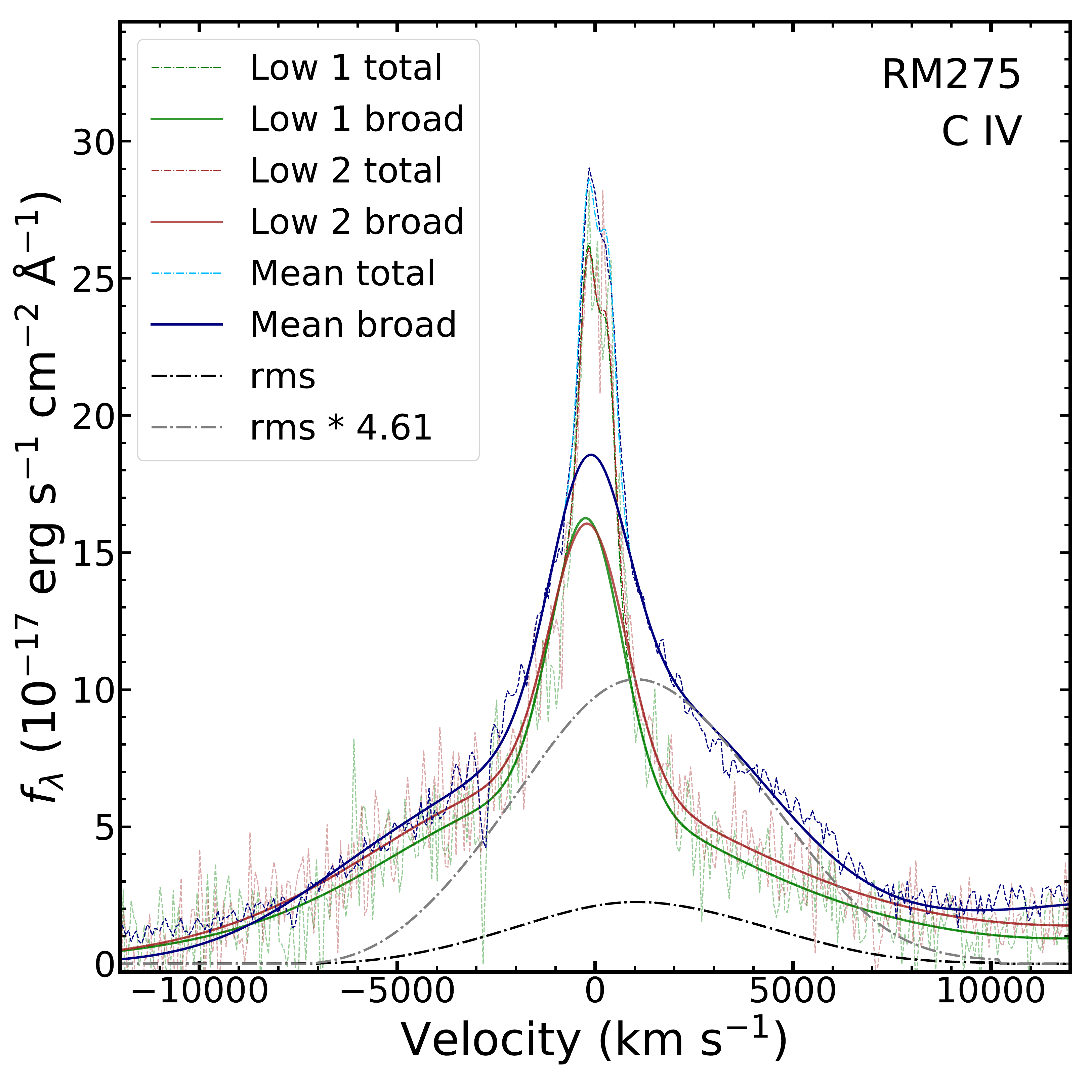}
    \includegraphics[width=0.3\textwidth]{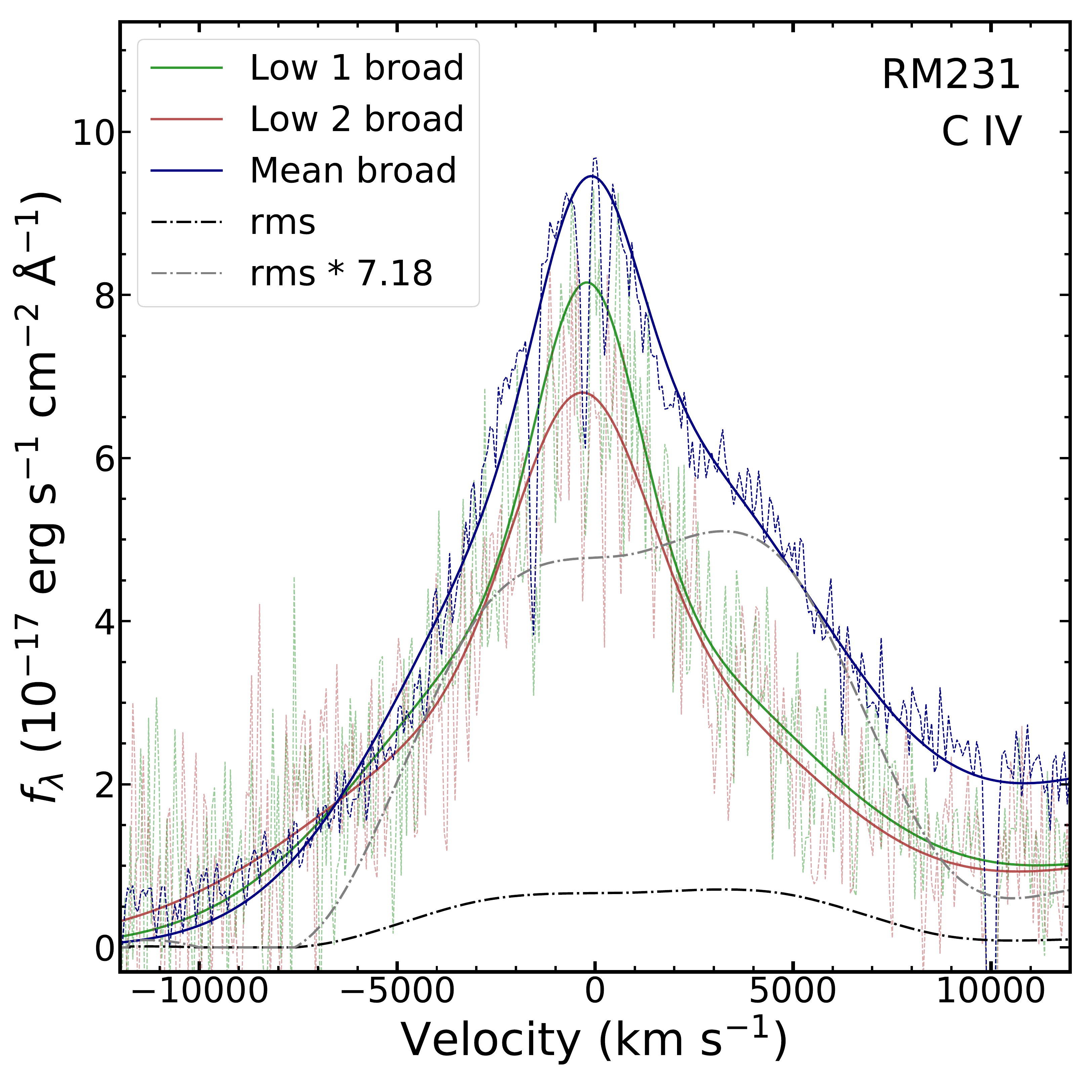}
    \caption{Broad \CIV\ line profile variations of three examples (RMID marked in the top-right corner) showing anti-breathing. We show the mean (blue) and rms spectra (black dash-dotted) of the line, as well as the profiles of two representative epochs (green and red) that show large changes in line width. We also overplot the scaled rms spectrum (grey dash-dotted) that matches at least one side of the mean profile (see text for details). The narrow core of the mean broad-line profile is generally missing from the rms profile. }
    \label{fig:c4profile_egs}
\end{figure*}

\begin{figure*}[htbp]
    \centering
    \includegraphics[width=0.43\textwidth]{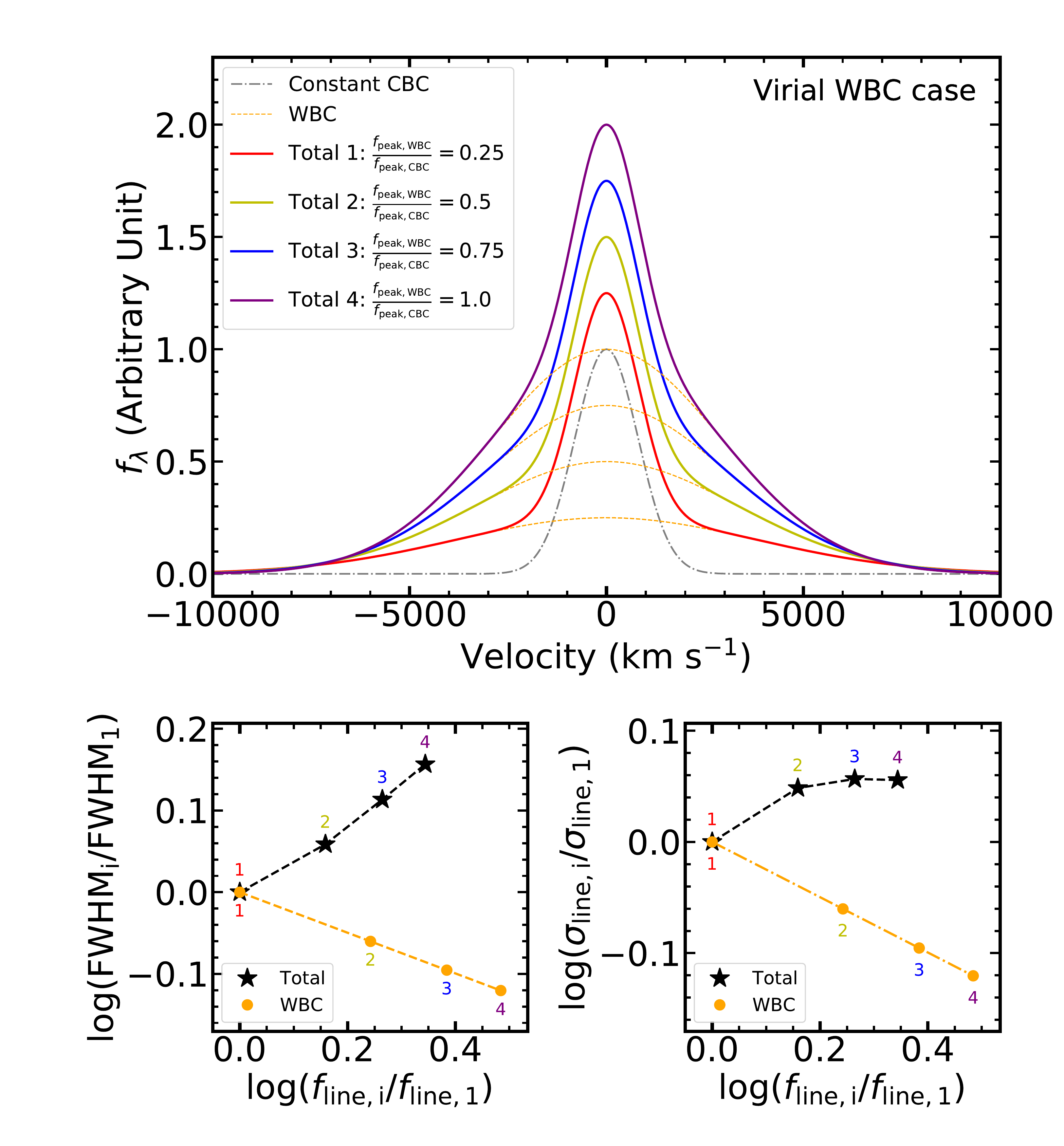}
    \includegraphics[width=0.43\textwidth]{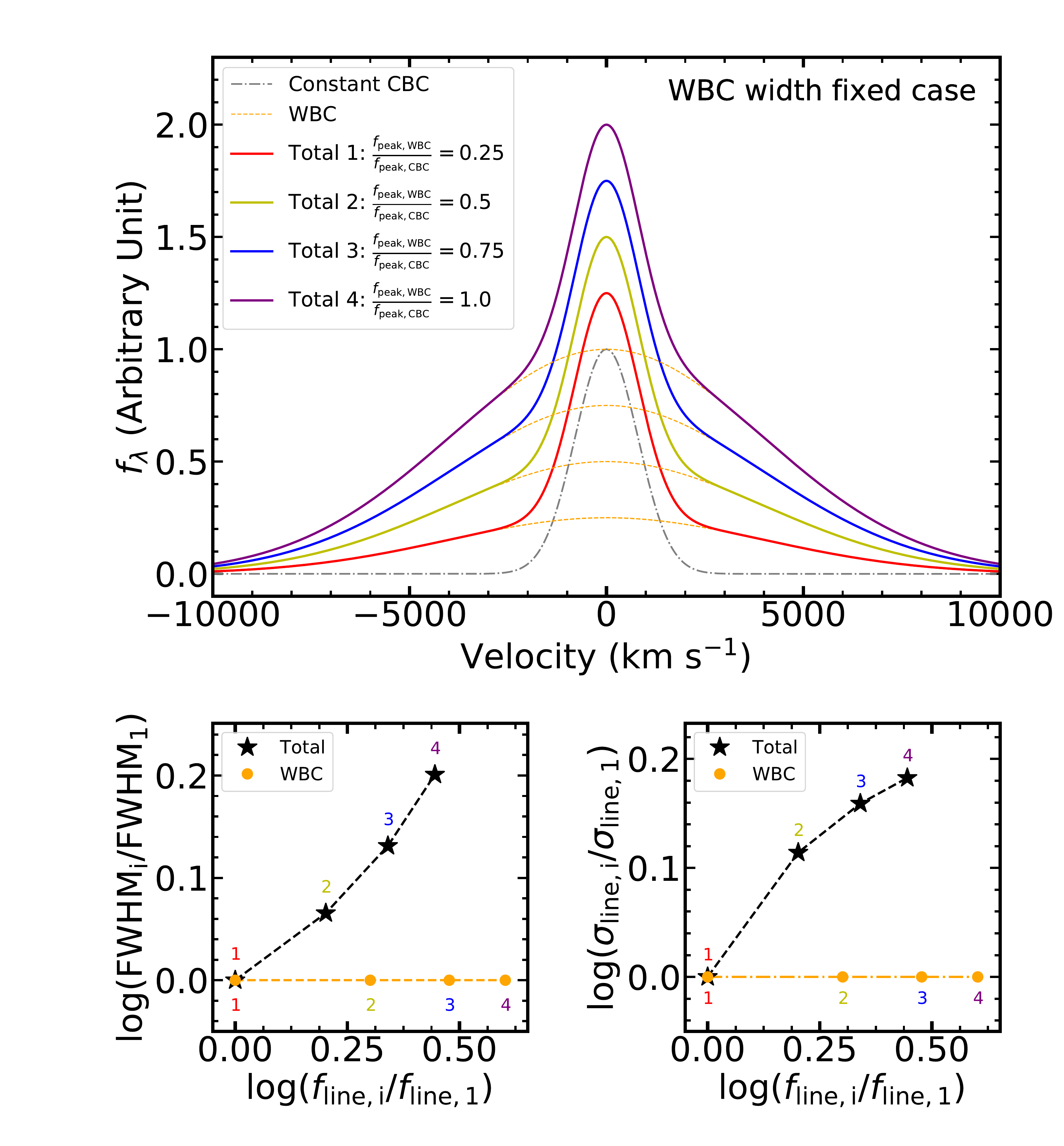}
    \caption{Idealized simulations of the two-components model, a Core Broad Component (CBC) that is non-reverberating and a Wing Broad Component (WBC) that may or may not be reverberating, of \CIV\ profile (top) and the $\Delta\log W-\Delta\log L$ relations (bottom). We test two cases: (1) the WBC follows the expected canonical breathing (left) and (2) the WBC width remains constant (right). In both cases, the CBC (grey) has fixed width and flux, while the WBC (orange) varies in flux. When we increase the flux in the WBC as indicated by the numbers 1 to 4, we find that the line width of the ``total'' profile (CBC + WBC) displays an anti-breathing effect, as observed.}
    \label{fig:simulation1}
\end{figure*}

The observed anti-breathing behavior for \CIV\ is inconsistent with naive expectation from the virial relation. To understand the origin of this anti-breathing behavior, we investigate the profile variations of \CIV. 

We first compare the RMS and mean profiles of the broad \CIV\ line for several examples displaying clear anti-breathing in Figure \ref{fig:c4profile_egs}, where the RMS profile is scaled by a constant factor to best match at least one side of the wings of the line in the mean spectrum \citep{Denney_2012}. If the entire \CIV\ line responds to continuum variations in the same way, the scaled RMS profile should roughly match the mean profile.\footnote{A direct consequence is that the widths measured from the mean and RMS spectra should correlate with each other well, as observed for the low-ionization lines \halpha, \hbeta\ and \MgII\ \citep{Wang_etal_2019}.} There is significant difference between the RMS and mean \CIV\ profiles such that the RMS profile is much broader, while the mean profile likely contains an extra relatively narrower core component that is missing from the RMS profile. This result is consistent with the findings in \citet{Denney_2012} on local RM AGN, where the central component does not seem to respond to continuum variations as much as the broader component during the monitoring period. As pointed out by \citet{Denney_2012}, the likely origin for this central component includes a possible narrow-line region or an optically-thin disk wind component \citep[e.g.,][]{Proga_etal_2000,Proga_Kallman_2004,Waters_etal_2016} that does not respond to photoionization variability on the relevant timescales here. 

It is worth noting that the peak of the \CIV\ line in the mean spectrum is on average blueshifted by hundreds of ${\rm km\,s^{-1}}$ from the systemic velocity of the quasar based on low-ionization lines \citep[][]{Shen_etal_2016b}, commonly known as the \CIV\ blueshift \citep[e.g.,][]{Gaskell_1982,Tytler_Fan_1992,Richards_etal_2002b}. This result may suggest that the non-reverberating core \CIV\ component is more likely associated with a disk wind. On the other hand, the broad-base component revealed in the RMS spectrum is on average redshifted by hundreds of ${\rm km\,s^{-1}}$ from the systemic velocity. The latest calculations \citep{Waters_etal_2016} of reverberation mapping signatures from photoionized disk winds based on the \citet{Proga_Kallman_2004} model reproduce many of the observed \CIV\ properties here, such as the blueshifted non-responding line profile and the shape of the RMS profile. However, the calculations in \citet{Waters_etal_2016} generally predict blueshifted RMS profile from the disk wind; the observed redshifted RMS profile could be explained by the so-called ``hitch-hiking'' gas that falls back toward the disk (T. Waters, private communications). A more thorough comparison between observed \CIV\ line characteristics from reverberation mapping and predictions from disk wind models can further constrain the detailed structure and kinematics of the BLR. 

Under this premise of two components of the broad \CIV\ emission, where only the broader component responds to continuum variations and follows the virial relation between average BLR size and line width, we can qualitatively understand the anti-breathing behavior: when luminosity increases, the virialized broader component increases its flux and decreases its width. However, if the non-reverberating central component is significant, the overall line width may still increase because of the enhanced flux in the wings of the line. To illustrate these effects, we perform the following idealized simulations.

We construct a simple two-component \CIV\ model, where we use two Gaussians to describe the narrower core component and the broader base component. The core component has a fixed Gaussian width (dispersion) of 800\,$\kms$, and the base component has an initial Gaussian width (dispersion) of 4000 $\kms$. The initial peak flux ratio of the two components is ${\rm base}\;:\;{\rm core}=1:4$. We then increase the flux of the base component to mimic the reverberation to continuum changes. For the width of the variable base component, we test two cases, where either the width varies according to the perfect breathing given by the virial relation (left panel of Figure \ref{fig:simulation1}) or is held constant (right panel of Figure \ref{fig:simulation1}). In both cases, we found that the change in the overall line width (core+base) component deviates significantly from the canonical breathing under the virial relation. In particular, the overall line profile always displays anti-breathing if using FWHM, and mostly no breathing (case 1) or anti-breathing (case 2) if using $\sigma_{\rm line}$, in qualitative agreement with our observations shown in Figure \ref{fig:HLWStat}. In this simulation, there is no velocity offset between the core and base components, but the results remain the same when we offset the two components by 1500\,$\kms$ apart.   

A byproduct from this two-component model for \CIV\ is that we expect a somewhat different intrinsic (i.e., due to luminosity variations in a given object) Baldwin effect \citep[e.g.,][]{Baldwin_1977,Kinney_etal_1990,Pogge_Peterson_1992}, compared to the global (i.e., object-by-object) Baldwin effect for quasars with different luminosities. The global Baldwin effect can be reasonably well explained by a systematic change in the ionizing spectral energy distribution (SED) with quasar luminosity and the resulting photoionization response in the broad lines \citep[e.g.,][]{Korista_etal_1998}. The same photonization effects and SED dependence should also apply to the intrinsic Baldwin effect. Additionally, because the central component largely does not respond to (or lag significantly behind) continuum changes on typical RM monitoring timescales, the intrinsic Baldwin effect is expected to have a shallower slope $b$ in the line response to continuum changes ($L_{\rm CIV}\propto L_\lambda^b$) than the global (average) Baldwin effect, consistent with observations \citep[e.g.,][]{Kinney_etal_1990,Pogge_Peterson_1992}.  

We now examine the profile changes in these \CIV\ anti-breathing examples in detail. We use two additional parameters to characterize the shape of the \CIV\ line: the shape parameter $S = {\rm FWHM} / \sigma_{\rm line}$ and the asymmetry parameter $ {\rm Asymm} = (\lambda_{\frac{3}{4}} - \lambda_{\frac{1}{4}}) / {\rm FWHM}$, where $\lambda_{\frac{1}{4}}$ and $\lambda_{\frac{3}{4}}$ are the wavelengths at $\frac{1}{4}$ and $\frac{3}{4}$ of the peak line flux, respectively. We found that when continuum flux increases in anti-breathing examples shown in Figure \ref{fig:c4profile_egs}, indeed the flux in the line wings increases more than in the core. This is consistent with our findings above that most of the line variability is contained in the broader base component and the narrower core component shows less variability. If the broader component has a velocity offset from the core component, increasing its flux will also increase the line asymmetry. The fact that the shape parameter $S$ changes as flux changes also confirms that the narrow core of \CIV\ does not reverberate in the same way as the broader component. On the other hand, we do not believe that we are observing velocity-resolved reverberation of \CIV, since the RMS line profile computed over the entire 4-year monitoring period generally lacks a narrow core RMS component. 

\begin{figure*}[htbp]
    \centering
    \includegraphics[width=0.45\textwidth]{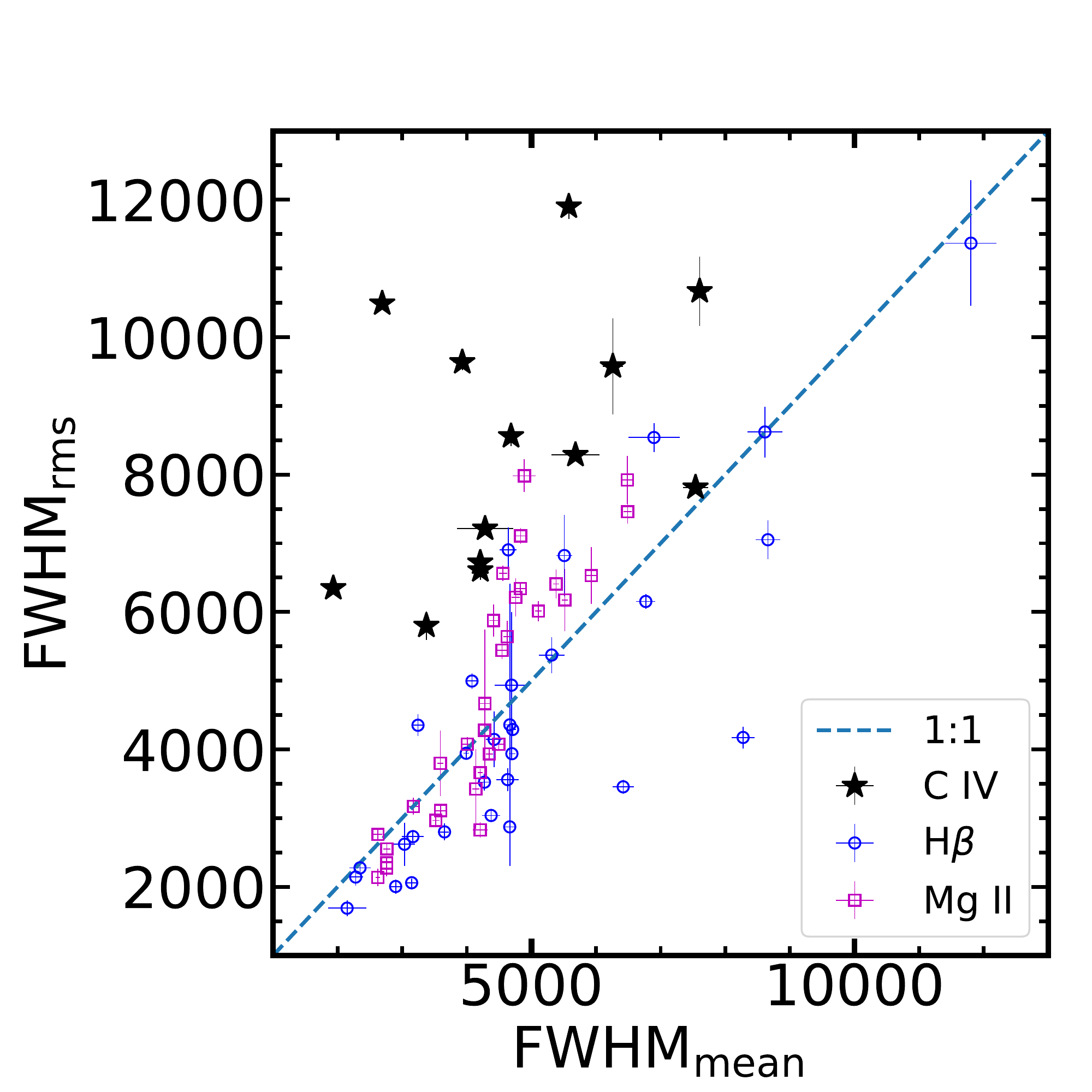}
    \includegraphics[width=0.45\textwidth]{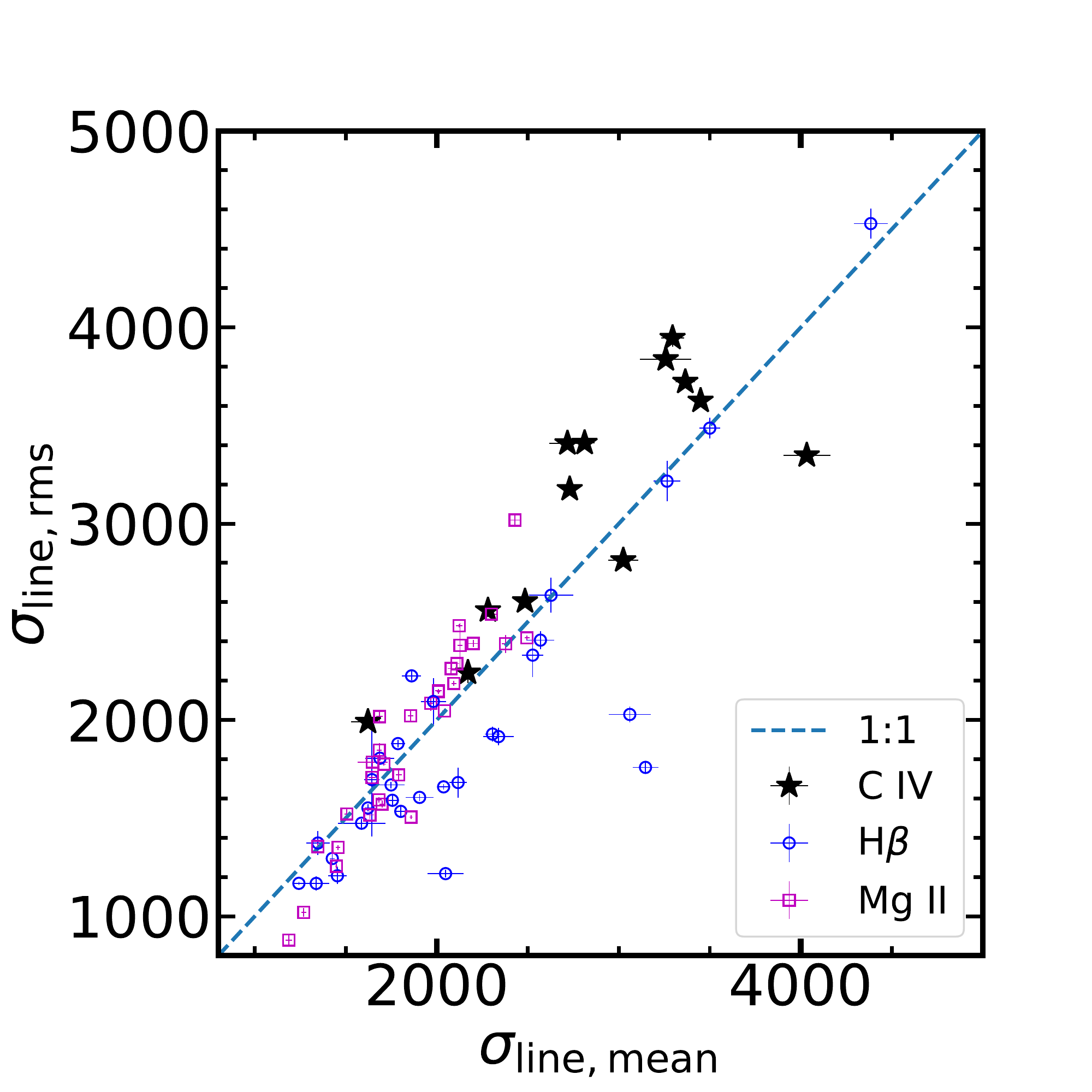}
    \includegraphics[width=0.45\textwidth]{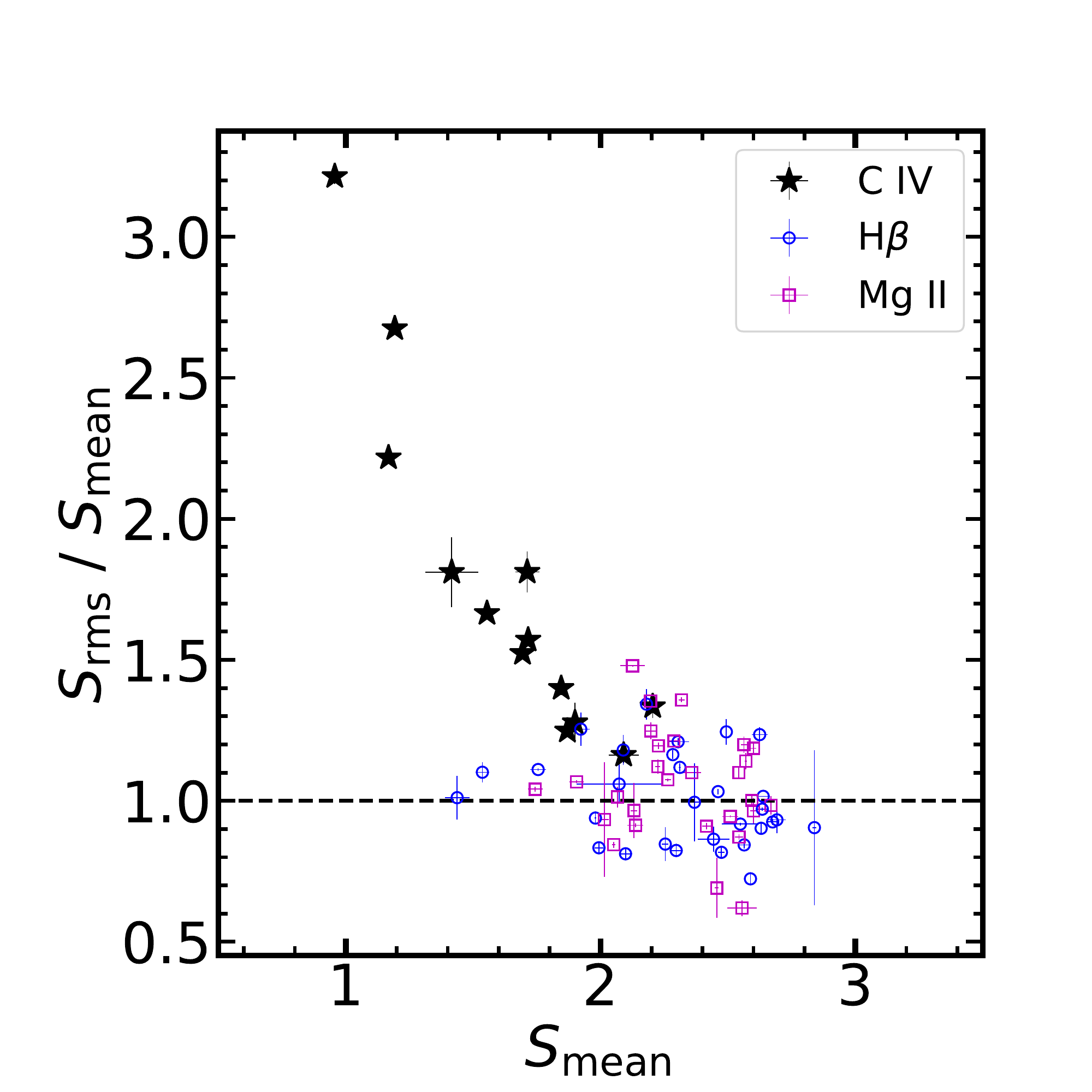}
    \includegraphics[width=0.45\textwidth]{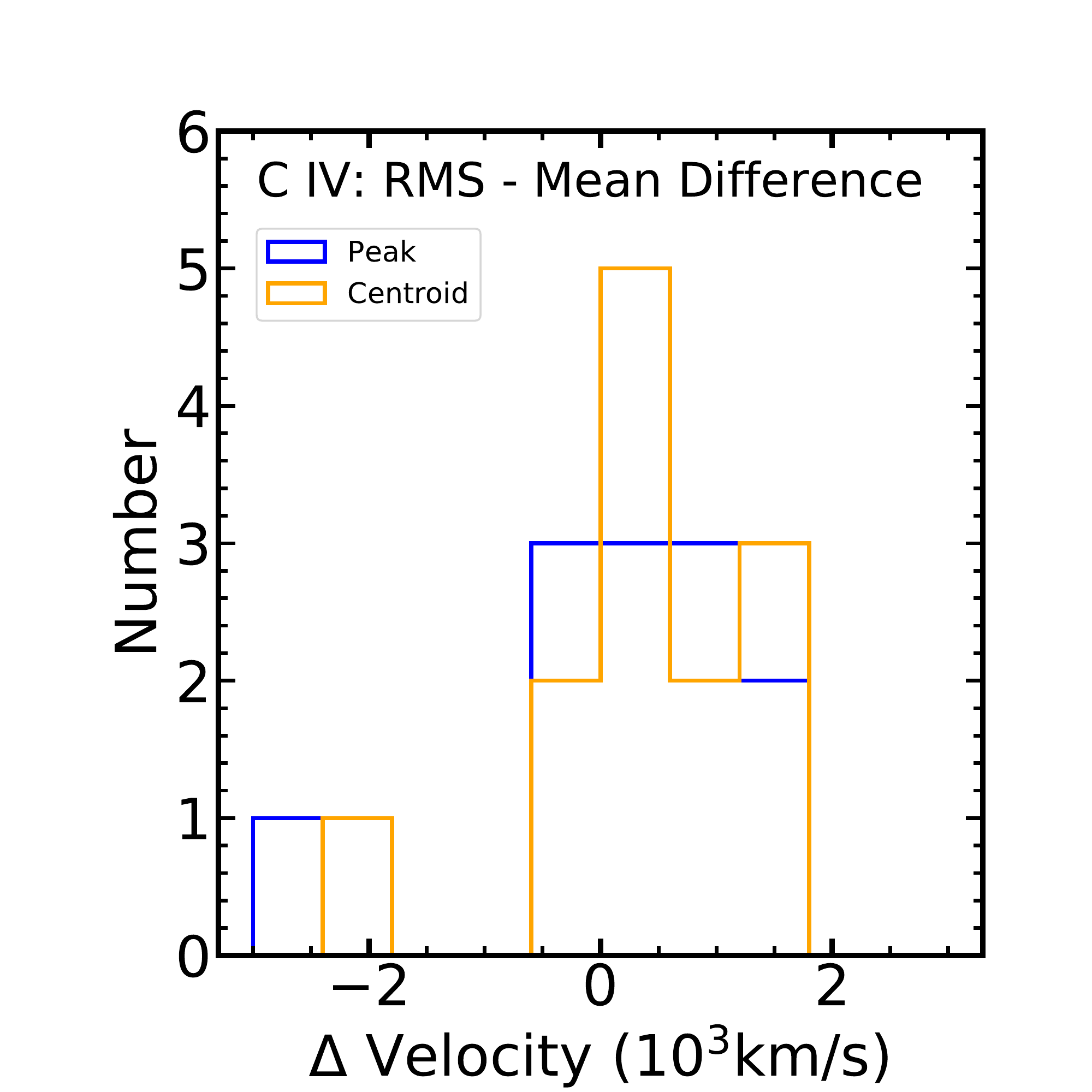}
    \caption{rms and mean profile comparisons for \hbeta, \MgII, and \CIV. The upper two panels are the comparison for FWHM and $\sigma_{\rm line}$, respectively. The lower two panel are line shape ratio comparison and centroid velocity difference distribution, respectively. \CIV\ shows noticeable worse correlation between the rms and mean widths than \hbeta\ and \MgII, mainly due to the non-reverberating core component (see text for details). This non-varying core \CIV\ component has less impact on $\sigma_{\rm line}$ than on FWHM.}
    \label{fig:shape}
\end{figure*}

The existence of a non-varying core component in \CIV\ will lead to enhanced scatter between the line widths measured from the mean and RMS spectra. In Figure \ref{fig:shape}, we compare the correlations between the mean and RMS line widths for different lines. Indeed the correlations are worse for \CIV\ than for the low-ionization broad lines studied in \citet{Wang_etal_2019}. The correlation is particularly poor when we use FWHM. For $\sigma_{\rm line}$, the correlation between the mean and RMS widths for \CIV\ is reasonably good, since the variable line wings contribute significantly to the $\sigma_{\rm line}$ measurements. 

The idea of a two-component model for the \CIV\ line is not new. \citet{Wills_etal_1993} performed a principal components analysis (PCA) on a sample of 15 quasars and found a negative correlation between \CIV\ EW and FWHM in a statistical sense. Considering the \CIV\ Baldwin effect \citep[][]{Baldwin_1977}, this means a positive correlation between continuum luminosity and \CIV\ FWHM, similar to the anti-breathing effect discussed here. \citet{Wills_etal_1993} proposed a two-component model for \CIV, where they attributed the core component to the emission from an intermediate line region \citep[ILR, also see][]{Brotherton_etal_1994} that is the inner extension of the narrow line region (NLR). Variants of the ILR argument and extension to low-ionization lines are further explored in \cite{Hu_etal_2008a,Hu_etal_2008b} and \cite{Hu_etal_2012}. More recently, \citet{Denney_2012} revisited this idea by showing that there is a central non-reverberating part of \CIV\ in local RM AGN, similar to our findings here. \citet{Denney_2012} subtracted the scaled RMS profile from the mean profile and found that the central component often has a blue asymmetry, which is roughly anti-correlated with the EW of the central component. \citet{Denney_2012} argued that this central component could be orientation dependent (face-on objects show more blue asymmetry and less peak value), from either an optically thin (non-reverberating) BLR outflow, or the ILR (much longer time to reverberate). The association of the non-reverberating central component to BLR outflows such as a disk wind is further reasoned in \citet[][and references therein]{Richards_etal_2011}. Our results for high-redshift quasars with \CIV\ RM data are consistent with these earlier studies, and generally support the two-component model for \CIV. 

The average \CIV\ peak blueshift in the mean profile is observed to increase with quasar luminosity over a population of quasars \citep[e.g.,][]{Richards_etal_2011,Shen_etal_2016b}, and can well reach more than 1000\,$\kms$ in the most luminous quasars. If we ascribe the non-reverberating core component to a disk wind, the relative contribution of this wind component then depends on the luminosity (and mostly likely, the Eddington ratio $L/L_{\rm Edd}$ as well) of the quasar. However, for individual objects, the variability of this core component and its blueshift is mild or undetectable within typical monitoring durations of less than a few years, consistent with the findings in e.g., \citet[][]{Sun_etal_2018}. Thus the large dynamic range in \CIV\ blueshift observed over the population of quasars must be driven on much longer timescales, when the long-term-average luminosity state is substantially maintained to establish a new dynamical equilibrium of the wind. Qualitatively, the global dependence of the relative contribution of the non-reverberating component can explain the difference between single-epoch BH masses and RM-based masses as a function of $L/L_{\rm Edd}$ or simply $L$. Fully understanding the physical mechanisms of this wind component is beyond the scope of this work, but is worth exploring in future theoretical studies of disk winds in quasars.

\begin{figure}[htbp]
    \centering
    \includegraphics[width=0.45\textwidth]{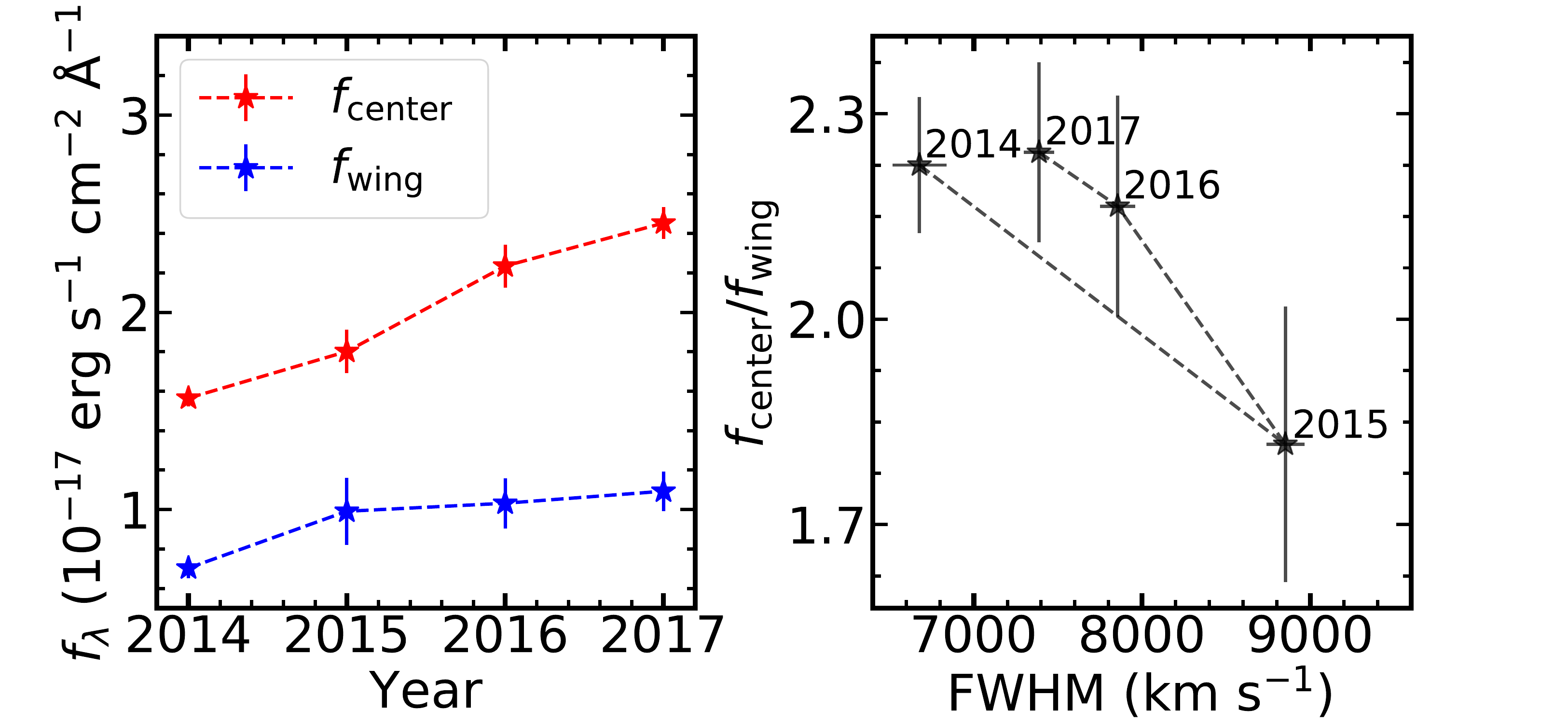}
    \caption{Flux variations of the core (estimated at $v=0$ km s$^{-1}$) and the wing (estimated at $v=5000$ km s$^{-1}$) of the broad \CIV\ profile of RM145 during 2014-2017. }
    \label{fig:RM145_Lcs}
\end{figure}

We conclude this section by noting that some quasars still show normal breathing in \CIV, which is expected if the reverberating component dominates the \CIV\ line \citep[e.g.,][]{Richards_etal_2011}. In our sample, RM201 and RM145 are two examples showing normal breathing. RM145 is a particularly interesting object. The breathing slopes in Table \ref{tab:HLWslopes} suggest that it showed normal breathing overall in the four-year monitoring. However, it actually showed anti-breathing from the first to the second seasons, then went back to normal breathing in the remaining seasons, as shown in Figure \ref{fig:RM145_Lcs}. The changes in breathing slope is more obvious for FWHM. We also examined the red wing asymmetry. We use the flux at $v=5000$ km/s for the red wing, and the flux at $v=0$ km/s for the core. As shown in Figure \ref{fig:RM145_Lcs}, the core and wing fluxes both increased during the four-year monitoring. However, the core to red-wing flux ratio decreased in the first two years then increased in the following years. Since the continuum flux largely monotonically increased during the first three years of monitoring, this transition could result from the lagged response of the core component than the wing component, which could be evidence that the core component is located further away but still reverberates to continuum changes.

\begin{figure}
    \centering
    \includegraphics[width=0.45\textwidth]{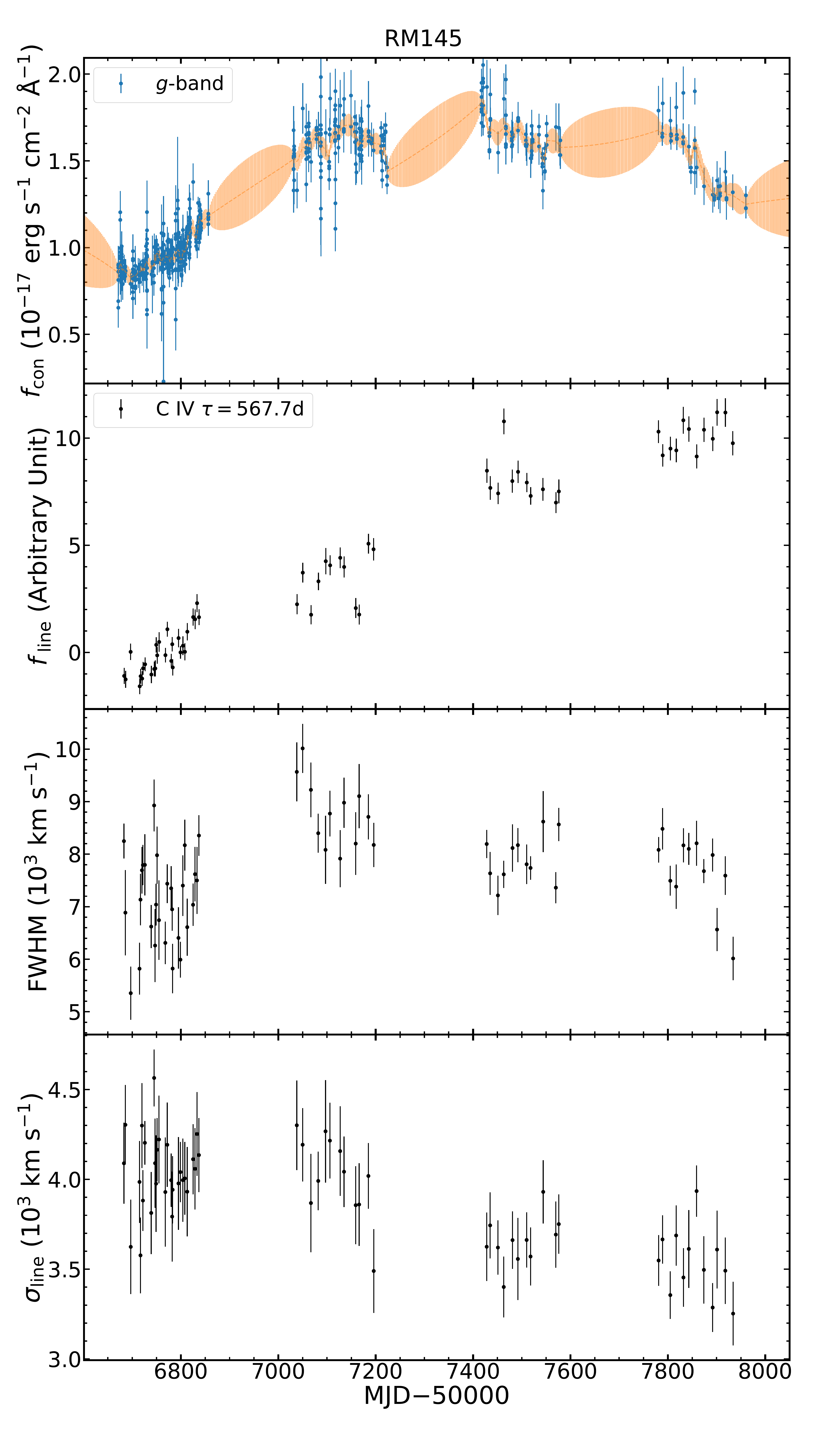}
    \caption{Same as Figure \ref{fig:L1} but of RM145, which shows overall normal breathing for \CIV\ during the four-year monitoring of SDSS-RM. }
    \label{fig:LCs145}
\end{figure}

\begin{figure}
    \centering
    \includegraphics[width=0.49\textwidth]{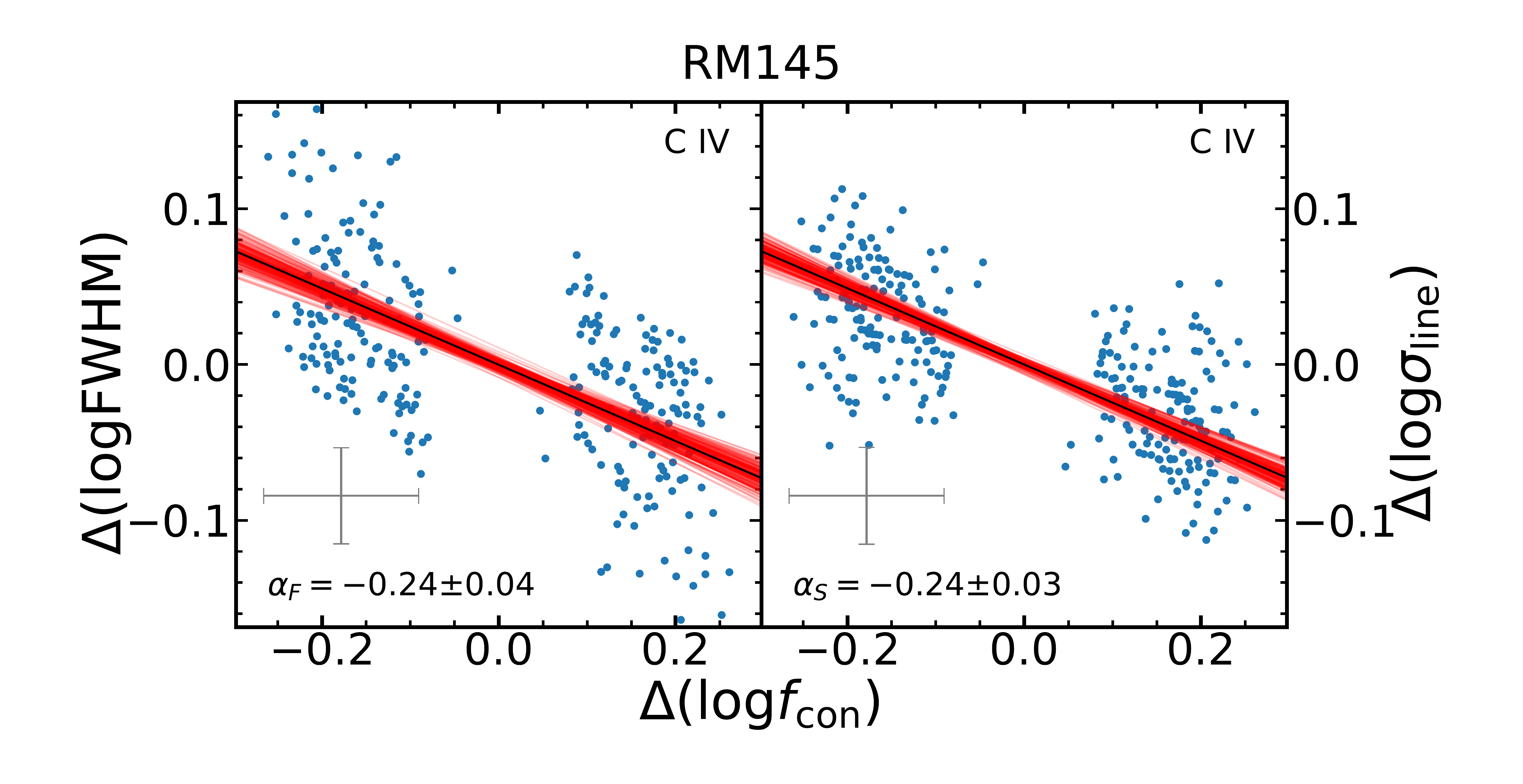}
    \caption{Same as Figure \ref{fig:HLW3} but for RM145, which shows normal breathing for \CIV. Notations are the same as Figure \ref{fig:LLW1}.}
    \label{fig:HLW145 }
\end{figure}

\subsection{\CIII\ and \SiIV\ breathing}

In our high-$z$ sample, we have 5 quasars for which we can measure the breathing effect for the \CIII+\AlIII+\SiIII\ complex. As shown in Figure \ref{fig:HLWStat} and Table \ref{tab:HLWslopes}, \CIII\  displays similar breathing behaviors as \CIV, with most objects showing an anti-breathing. At first glance this may appear unexpected since \CIII\ has a lower ionization potential than \CIV. However, the width of \CIII\ is shown to correlate with the \CIV\ width well \citep[e.g.,][]{Shen_Liu_2012}. Our finding of similar \CIII\ and \CIV\ breathing is consistent with the scenario where \CIV\ and \CIII\ emission originates from similar regions. 

\begin{figure*}
    \centering
    \includegraphics[width=0.9\textwidth]{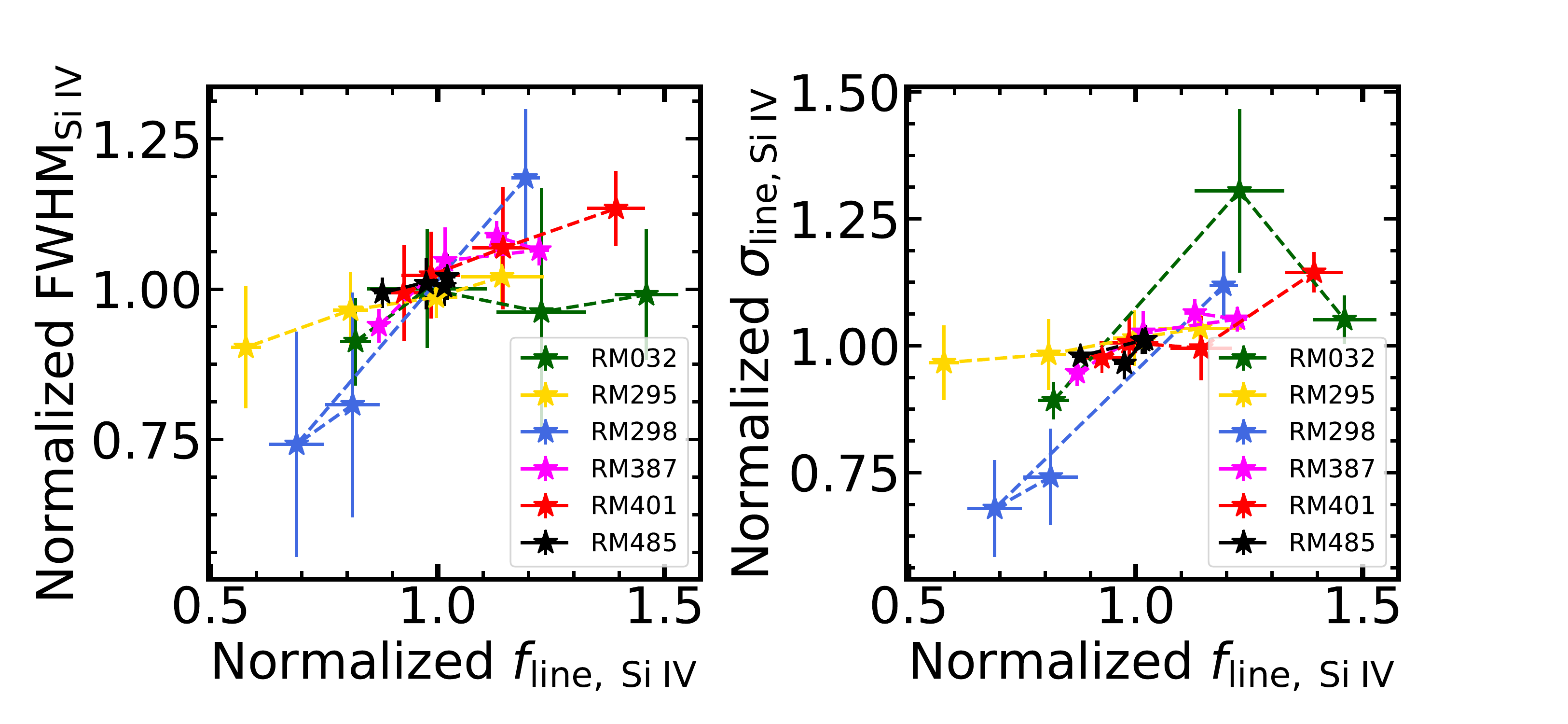}
    \caption{Correlations between line widths and line flux for \SiIV, measured from seasonally coadded spectra for six quasars in our sample. The overall positive correlations suggest \SiIV\ displays an anti-breathing effect, similar to \CIV. }
    \label{fig:siiv}
\end{figure*}

There is only one quasar that passes our S/N criteria for studying \SiIV\ breathing. To boost the S/N of spectral measurements, we use the yearly-coadded spectra on \SiIV. There are six quasars for which we can robustly measure \SiIV\ widths in at least two yearly-coadded spectra. We plot the \SiIV\ ${\rm FWHM}$-$f_{\rm line}$ and $\sigma_{\rm line}$-$f_{\rm line}$ relations in Figure \ref{fig:siiv}. Both line widths increase with line flux, suggesting anti-breathing for \SiIV. The \CIV\ line shows similar anti-breathing in these quasars.   

\subsection{Implications on BLR structure and single-epoch BH masses}

The different breathing behaviors for different broad lines suggest different structures and likely also different emission mechanism of these lines. While \hbeta\ most closely follows the expected normal breathing under the virial assumption, the two other lines with similar ionizing potentials (\halpha\ and \MgII) show less breathing. In particular, \MgII\ displays almost no breathing, along with less response in flux to continuum changes than the Balmer lines, consistent with earlier results \citep[e.g.,][]{Shen_2013,Yang_etal_2019a}. \citet{Guo_etal_2019} performed photoionization calculations of the three low-ionization lines, and were able to qualitatively reproduce their different breathing behaviors, given that the \MgII\ and \halpha\ emitting gas is located further out than the \hbeta\ emitting gas. In particular, if most of the \MgII-emitting gas is near the physical boundary of the BLR, which changes on much longer dynamical timescales, changes in luminosity will not significantly change the average distance of \MgII-emitting gas, leading to mostly no breathing. 

On the other hand, \CIV\ mostly displays an anti-breathing. The fact that the width of \CIV\ does change as luminosity changes suggests that at least part of the \CIV-emitting gas is responding to continuum variations. \S\ref{sec:disc1} further demonstrated that this anti-breathing can be explained if \CIV\ also contains a non-reverberating core component. 

These different breathing behaviors have direct consequences on the single-epoch BH masses \citep[e.g.,][]{Shen_2013} that rely on line widths measured from single-epoch spectroscopy. Any deviations from the expected breathing slope of $\alpha=-0.25$ would introduce a luminosity-dependent bias in the single-epoch BH mass \citep[e.g.,][]{Shen_Kelly_2010,Shen_2013}. While this extra mass bias due to intrinsic quasar variability is still well within the envelope of the systematic uncertainty ($\sim 0.4$ dex) of virial BH mass estimates, given typical amplitude of quasar variability, it may become important when studying flux-limited statistical quasar samples \citep[see detailed discussions in][]{Shen_2013}. Another situation where the lack of breathing may be important is for extreme variability quasars (EVQs) that can vary by more than one magnitude over multi-year timescales \citep[e.g.,][]{Rumbaugh_etal_2018}. Repeated spectroscopy for these EVQs shows that the \MgII\ width remains more or less constant when continuum luminosity varies substantially \citep[e.g.,][]{Yang_etal_2019a}, resulting in significant luminosity-dependent bias in the single-epoch masses. 

In all cases we studied, it appears that the line dispersion $\sigma_{\rm line}$ measured from single-epoch spectra more closely follows the expected virial relation than FWHM, although deviations still exist even with $\sigma_{\rm line}$. This offers support to use $\sigma_{\rm line}$ as a more reliable indicator for the virial velocity in single-epoch spectroscopy, provided that this quantity can be robustly measured. \citet{Wang_etal_2019} provided an empirical recipe to measure $\sigma_{\rm line}$ robustly.  

Based on the breathing results, we therefore conclude \hbeta\ (and to a lesser extent, \halpha) remains the most reliable line to use as a single-epoch virial mass estimator. \MgII\ is not as reliable as \hbeta, given the general behavior of no breathing. \CIV\ is the least reliable line to use as a single-epoch virial mass estimator, given the anti-breathing behavior, although using $\sigma_{\rm line}$ can mitigate the situation to some extent.

However, our analysis does not address the question if for a global population of quasars with different BH masses and luminosities, the measured line widths from single-epoch spectra provide reasonable estimates of the average virial velocity. The breathing effects studied here only provide the extra scatter in single-epoch mass estimates due to intrinsic quasar variability. Following Eqn.~(\ref{eqn:breathing}), the deviation in the single-epoch BH mass estimate for a given quasar scales with the luminosity deviation as $\delta \log M_{\rm BH}=(2\alpha+0.5)\delta \log L$. Therefore for typical quasar RMS variability of 0.1 dex and $\alpha=0$ (no breathing), the intrinsic variability induced scatter in the single-epoch BH mass estimate is only 0.05 dex. However, for extreme quasar variability \citep[more than one magnitude variations, e.g.,][]{Rumbaugh_etal_2018} and an anti-breathing slope of $\alpha=0.25$ (using \CIV\ FWHM), the intrinsic variability induced scatter is 0.4 dex. 

\section{Summary}\label{sec:con}

In this work we have studied the breathing effect, i.e., the changes in broad line width in response to flux variations, for various broad lines using photometric and spectroscopic monitoring data from the SDSS-RM project. Our final sample includes 21 \halpha, 31 \hbeta, 26 \MgII, 5 \CIII, 13 \CIV\ and 1 \SiIV\ quasars for which we can reliably measure breathing; these quasars cover a wide redshift range ($z\approx 0.1-2.5$). While there have been similar studies on a few individual low-$z$ RM AGN focusing on the broad \hbeta\ line \citep[e.g.,][]{Park_etal_2012}, or using few-epoch repeated spectroscopy for large quasar samples \citep[e.g.,][]{Wilhite06,Shen_2013}, our work is the first comprehensive study on the breathing effect with RM monitoring data for statistical quasar samples and for most of the major quasar broad emission lines. Confirming and extending earlier results, our main findings are the following:

\begin{enumerate}
    \item When the lag between line response and continuum variations is taken into account, broad \hbeta\ mainly shows normal breathing (i.e., line width decreases as luminosity increases), and the slope of the breathing is most consistent with the expectation from the virial relation. On the other hand, broad \halpha\ on average shows much less breathing. The delay in the line responses to continuum variations will tend to further flatten the breathing slope if both the continuum luminosity and line width are measured from the same epoch. 
    
    \item Broad \MgII\ on average shows no breathing, i.e., the line width responds less to luminosity changes than broad \hbeta, consistent with earlier results \citep[][]{Shen_2013,Yang_etal_2019a}. This result can be qualitatively understood by the possibility that the \MgII-emitting gas is near the physical boundary of the BLR, and thus will not expand or contract freely when luminosity changes \citep[e.g.,][]{Guo_etal_2019}. 
    
    \item Importantly, the broad \CIV\ line mostly shows an anti-breathing mode, i.e., line width increases with luminosity, consistent with earlier findings based on non-RM data \citep[e.g.,][]{Wills_etal_1993,Wilhite06,Shen_2013}. This anti-breathing can be explained by a two-component model for broad \CIV\ emission, a broad-base reverberating component, and a narrower (but still broader than typical narrow lines), core component that does not respond to luminosity variations as strongly as the reverberating component. This is consistent with the findings in e.g., \citet{Denney_2012} based on local RM results, and follows the general idea of two components for the \CIV\ emission \citep[e.g.,][]{Wills_etal_1993,Richards_etal_2011}. The nature of this non-reverberating component, however, remains unclear. It could originate from an Intermediate Line Region \citep[ILR, e.g.,][]{Wills_etal_1993,Brotherton_etal_1994}, or from a disk wind \citep[e.g.,][]{Proga_etal_2000,Proga_Kallman_2004,Waters_etal_2016}. The existence of a non-reverberating broad \CIV\ component underlies the large scatter between the broad \CIV\ widths measured from the mean and RMS spectra, and the long-argued caveats in single-epoch BH masses based on \CIV\ \citep[e.g.,][]{Baskin_Laor_2005,Sulentic_etal_2007,Shen_etal_2008,Shen_Liu_2012,Coatman_etal_2017}. 

    \item Despite these average behaviors, individual objects can show normal breathing in any of the broad lines studied here. 
    
    \item These diverse breathing behaviors suggest additional uncertainty due to intrinsic quasar variability in the single-epoch virial BH mass estimates that rely on the line width measured from single-epoch spectroscopy. In particular, any deviation from the expected breathing from a perfect virial relation would lead to a luminosity-dependent bias \citep[e.g.,][]{Shen_Kelly_2010,Shen_2013} in the single-epoch mass estimate. Based on the observed breathing behaviors, \CIV\ (and \CIII, \SiIV\ to some extent as well) would have the largest extra scatter due to quasar variability for single-epoch BH mass estimates. \MgII\ and \halpha\ will also induce extra scatter in single-epoch masses due to quasar variability, given their deviations from the expected breathing behavior from the virial relation. 
    
    \item We find evidence for \hbeta that $\sigma_{\rm line}$ could be better than FWHM in preserving the normal breathing behavior, although significant deviations still present in most cases.   
    
\end{enumerate}

The different breathing behaviors are generally consistent with earlier photoionization models \citep[e.g.,][]{Goad_etal_1993,Guo_etal_2019} and qualitative ideas put forward for the \CIV\ BLR \citep[e.g.,][]{Wills_etal_1993,Richards_etal_2011,Denney_2012}. These breathing observations provide valuable information to constrain the different distributions of line-emitting gas and the kinematic structures of the BLR for different line species, in future modeling of the BLR. 

The extra scatter in single-epoch BH mass estimates due to quasar variability scales with the scatter in luminosity as $\delta \log M_{\rm BH}=(2\alpha+0.5)\delta \log L$. This is generally negligible compared with the $\sim 0.4$ dex uncertainty in single-epoch mass estimates. However, for lines with anti-breathing (in particular \CIV) and for large-amplitude quasar variability, this luminosity-dependent bias\footnote{In addition to abnormal breathing, there could be other factors that will lead to similar luminosity-dependent biases in single-epoch BH mass estimates, such as intrinsic scatter in the BLR $R-L$ relation. We recommend the reader to \citet{Shen_2013} for a comprehensive discussion on the systematic biases in single-epoch BH masses.} in single-epoch mass estimate becomes important. \citet{Rumbaugh_etal_2018} has showed that if selection effects are accounted for, quasar variability exceeding one magnitude is not uncommon over multi-year timescales. Thus this luminosity-dependent bias due to abnormal breathing is particularly relevant when using flux-limited quasar samples with single-epoch mass estimates to measure the distribution of BH masses and Eddington ratios \citep[e.g.,][]{Shen_Kelly_2010,Shen_Kelly_2012}. Our work highlights the importance of obtaining direct RM-based BH masses to mitigate the uncertainties in single-epoch BH masses, especially for those based on the \CIV\ line. 

\acknowledgments

We acknowledge support from the National Science Foundation of China (11721303, 11890693, 11991052) and the National Key R\&D Program of China (2016YFA0400702, 2016YFA0400703). YS acknowledges support from an Alfred P. Sloan Research Fellowship and NSF grant AST-1715579. CJG, WNB, JRT, and DPS acknowledge support from NSF grants AST-1517113 and AST-1516784. KH acknowledges support from STFC grant ST/R000824/1. PBH acknowledges support from NSERC grant 2017-05983. YH acknowledges support from NASA grant HST-GO-15650. We thank Tim Waters and Daniel Proga for useful discussion and the anonymous referee for suggestions that have improved the manuscript. 

This work is based on observations obtained with MegaPrime/MegaCam, a joint project of CFHT and CEA/DAPNIA, at the Canada-France-Hawaii Telescope (CFHT) which is operated by the National Research Council (NRC) of Canada, the Institut National des Sciences de l'Univers of the Centre National de la Recherche Scientifique of France, and the University of Hawaii.
The authors recognize the cultural importance of the summit of Maunakea to a broad cross section of the Native Hawaiian community. The astronomical community is most fortunate to have the opportunity to conduct observations from this mountain.

Funding for the Sloan Digital Sky Survey IV has been provided by the Alfred P. Sloan Foundation, the U.S. Department of Energy Office of Science, and the Participating Institutions. SDSS-IV acknowledges support and resources from the Center for High-Performance Computing at the University of Utah. The SDSS web site is www.sdss.org. SDSS-IV is managed by the Astrophysical Research Consortium for the Participating Institutions of the SDSS Collaboration including the Brazilian Participation Group, the Carnegie Institution for Science, Carnegie Mellon University, the Chilean Participation Group, the French Participation Group, Harvard-Smithsonian Center for Astrophysics, Instituto de Astrof\'isica de Canarias, The Johns Hopkins University, Kavli Institute for the Physics and Mathematics of the Universe (IPMU) / University of Tokyo, the Korean Participation Group, Lawrence Berkeley National Laboratory, Leibniz Institut f\"ur Astrophysik Potsdam (AIP),  
Max-Planck-Institut f\"ur Astronomie (MPIA Heidelberg), 
Max-Planck-Institut f\"ur Astrophysik (MPA Garching), 
Max-Planck-Institut f\"ur Extraterrestrische Physik (MPE), 
National Astronomical Observatories of China, New Mexico State University, 
New York University, University of Notre Dame, 
Observat\'ario Nacional / MCTI, The Ohio State University, 
Pennsylvania State University, Shanghai Astronomical Observatory, 
United Kingdom Participation Group,
Universidad Nacional Aut\'onoma de M\'exico, University of Arizona, 
University of Colorado Boulder, University of Oxford, University of Portsmouth, 
University of Utah, University of Virginia, University of Washington, University of Wisconsin, 
Vanderbilt University, and Yale University.

\vspace{5mm}
\facilities{Sloan, CFHT, Bok}


\appendix

\section{Tests on more stringent sample cuts}

\begin{figure}
\centering
\includegraphics[width=0.8\textwidth]{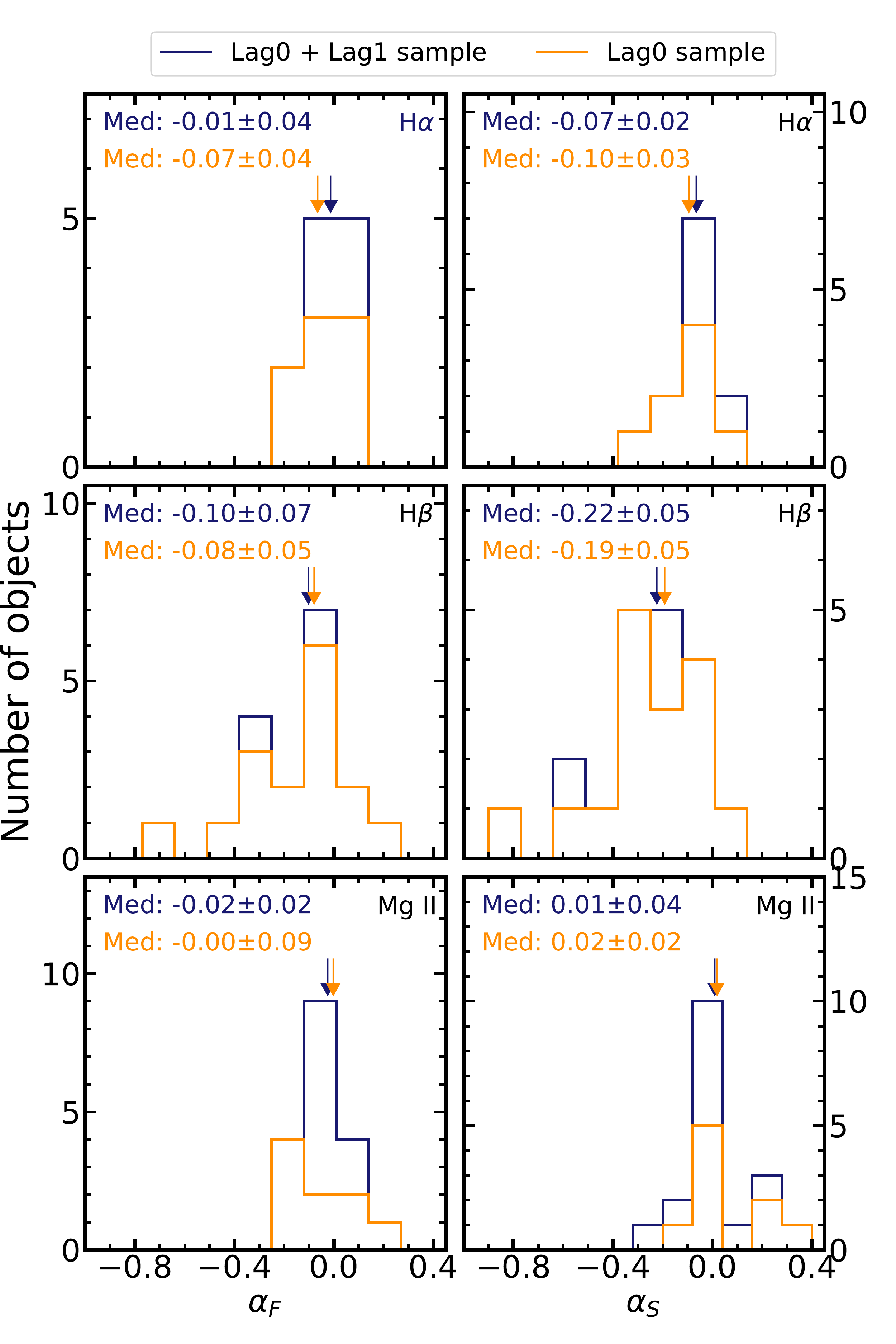}
    \caption{Same as Figure \ref{fig:LLWStat} but for the sample of ${\rm SNR}_{\rm Var,con} > 3$. The median slopes of \halpha, \hbeta, and \MgII\ results are generally consistent with those in Figure \ref{fig:LLWStat}.}
    \label{fig:appendixllw}
\end{figure}

\begin{figure}
\centering
\includegraphics[width=0.8\textwidth]{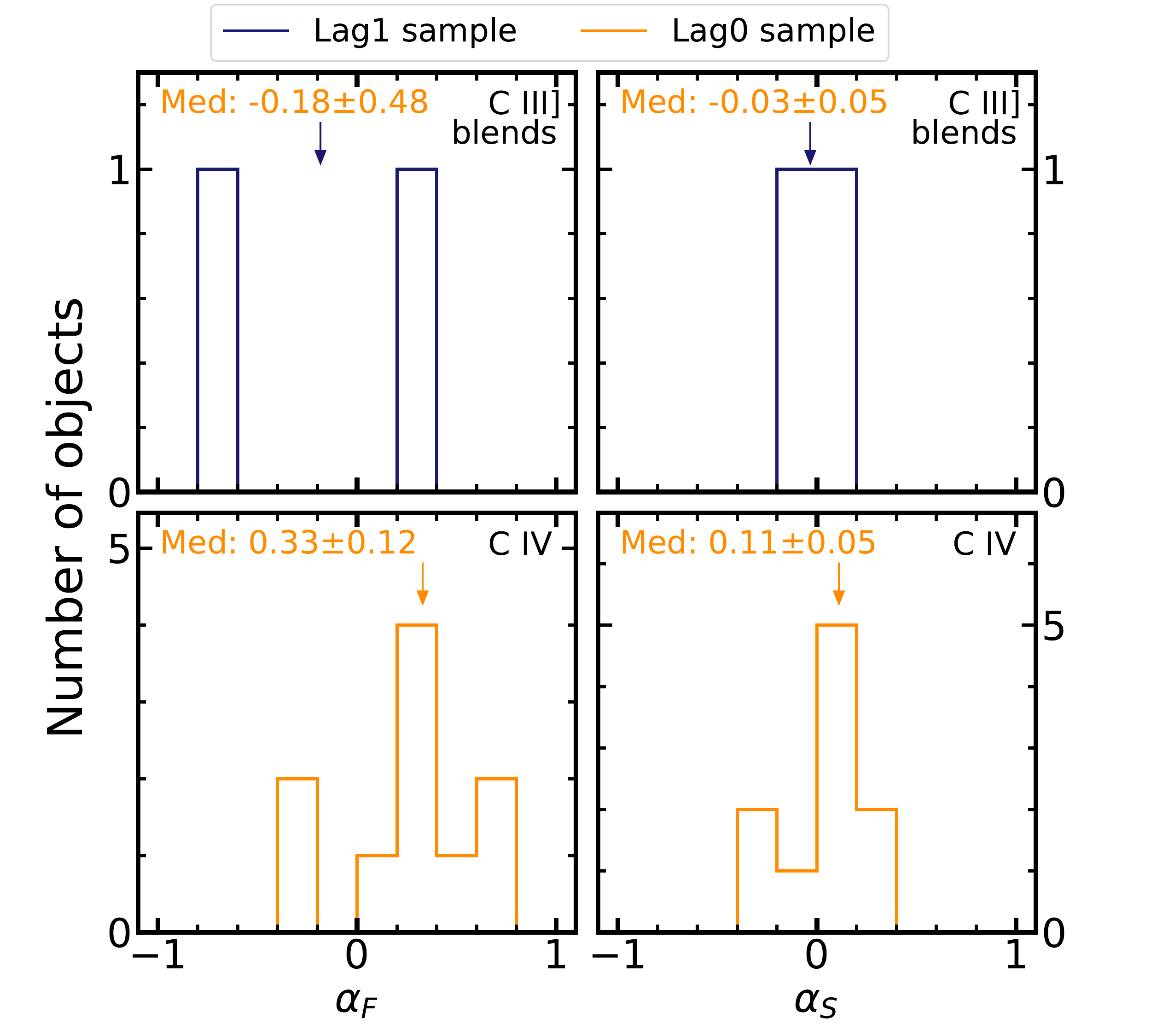}
    \caption{Same as Figure \ref{fig:HLWStat} but for the sample of ${\rm SNR}_{\rm Var,con} > 3$. For \CIII, there are only two objects left and it is difficult to constrain the statistics. For \CIV, the results are consistent with those in Figure \ref{fig:HLWStat}.}
    \label{fig:appendixhlw}
\end{figure}

We provide the distributions of breathing slopes for samples defined by a more stringent continuum S/N cut, ${\rm SNR_{\rm Var,con}}>3$, in Figure~\ref{fig:appendixllw} and Figure~\ref{fig:appendixhlw}.

\section{Mean and rms line profiles}

We provide the mean and rms line profiles for all objects in our final sample below. Figure notations are the same as Figure \ref{fig:c4profile_egs}. For clarity, we only show the model profiles. The mean profiles are from our spectral fits described in \S\ref{sec:spectral_analysis} and the rms profiles are from PrepSpec output. The scaled rms profiles are derived by matching the line flux between the rms model and the mean model in the window of $[-2.5\times{\rm MAD}_{\rm mean},2.5\times{\rm MAD}_{\rm mean}]$.

\newpage

\newpage
\begin{figure*}
    \centering
    \includegraphics[width=\textwidth]{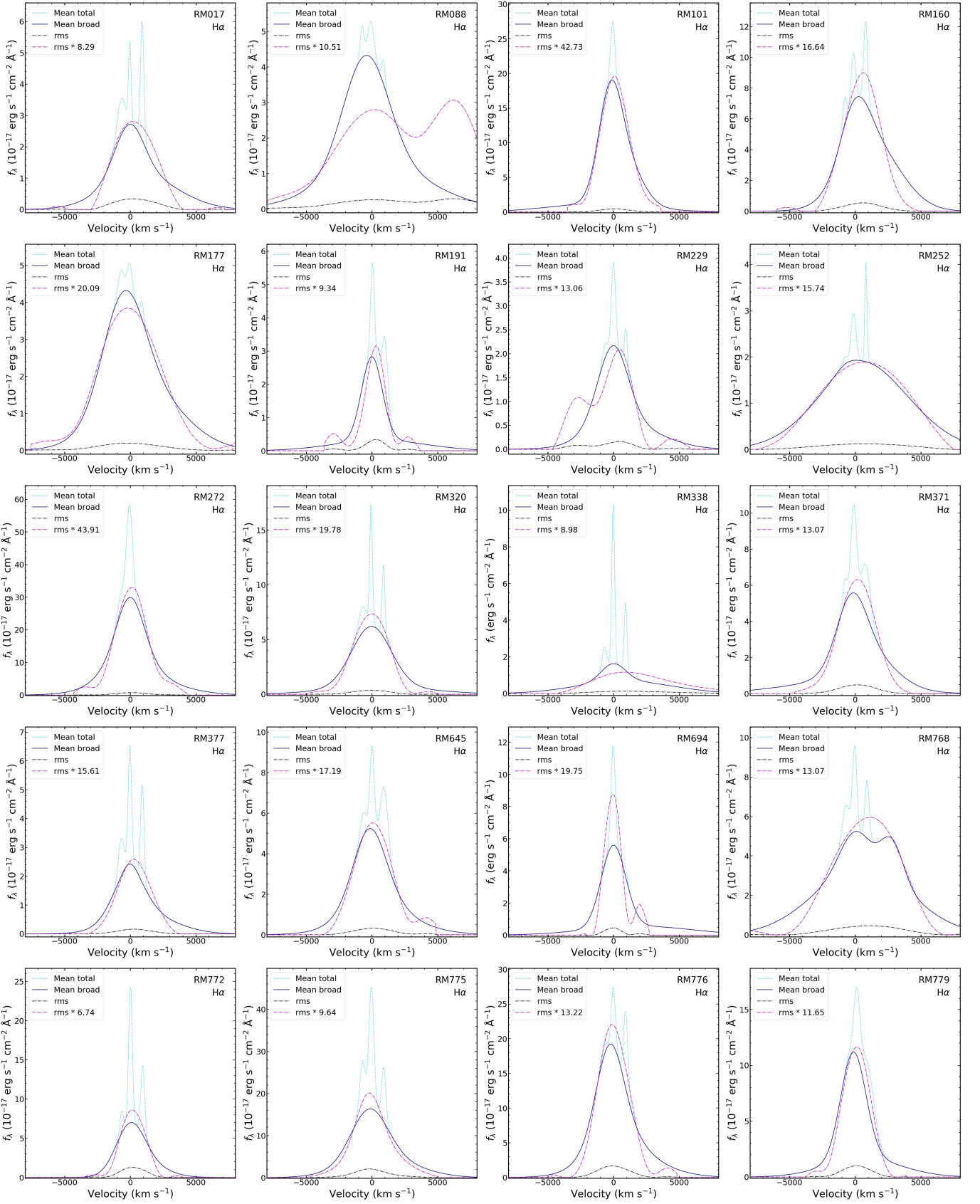}
    \caption{Mean and rms profiles of \halpha. For clarity we only show the model profiles. The mean profiles are from our spectral fits described in \S\ref{sec:spectral_analysis}, and the rms profiles are output by PrepSpec. We also plot a scaled rms profile that matches the integrated flux of the mean broad profile within the window of $\sigma_{\rm line}$ calculation (\S\ref{sec:spectral_analysis}). }
    \label{fig:final1}
\end{figure*}


\addtocounter{figure}{-1}
\begin{figure*}
    \centering
    \includegraphics[width=\textwidth]{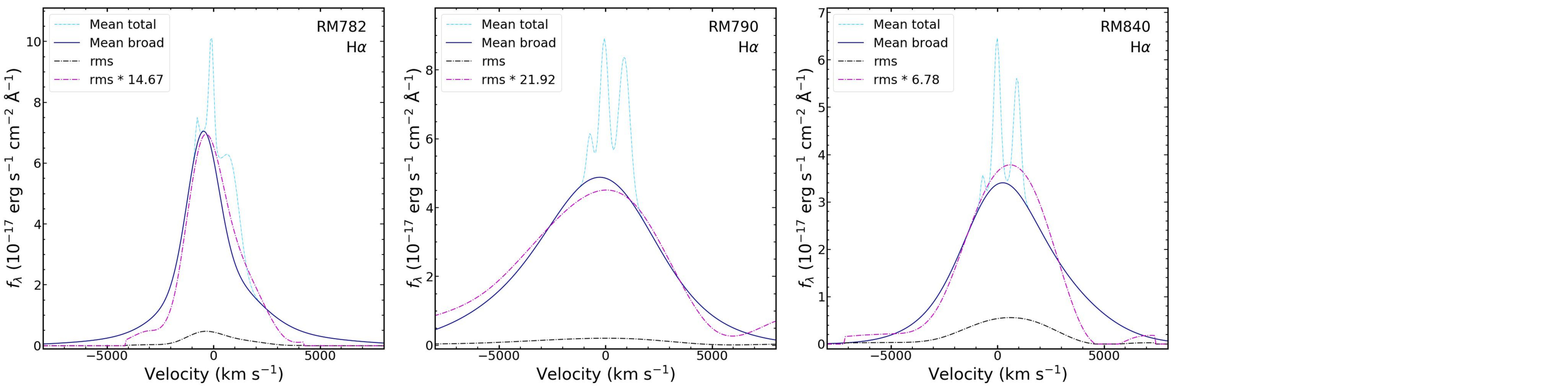}
    \caption{Continued.}
    \label{fig:my_label}
\end{figure*}

\begin{figure*}
    \centering
    \includegraphics[width=\textwidth]{Figure_allprofileHb_central.pdf}
    \caption{Same as Figure \ref{fig:final1} but for \hbeta. PrepSpec failed to produce the correct rms profile for RM301 based on the first-season data.}
    \label{fig:my_label}
\end{figure*}

\addtocounter{figure}{-1}
\begin{figure*}
    \centering
    \includegraphics[width=\textwidth]{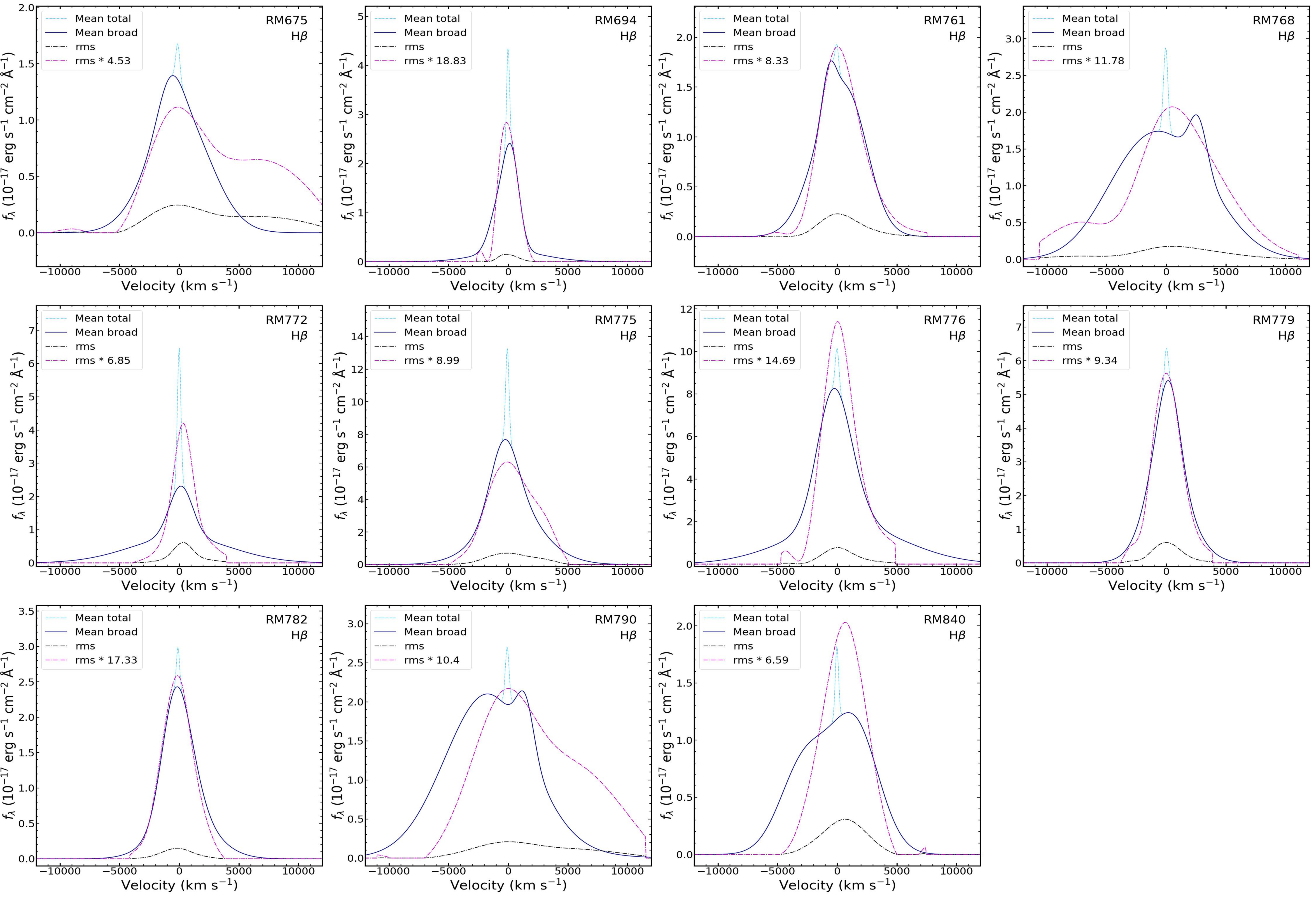}
    \caption{Continued.}
    \label{fig:my_label}
\end{figure*}

\begin{figure*}
    \centering
    \includegraphics[width=\textwidth]{Figure_allprofileMg2_central.pdf}
    \caption{Same as Figure \ref{fig:final1} but for \MgII. }
    \label{fig:my_label}
\end{figure*}

\addtocounter{figure}{-1}
\begin{figure*}
    \centering
    \includegraphics[width=\textwidth]{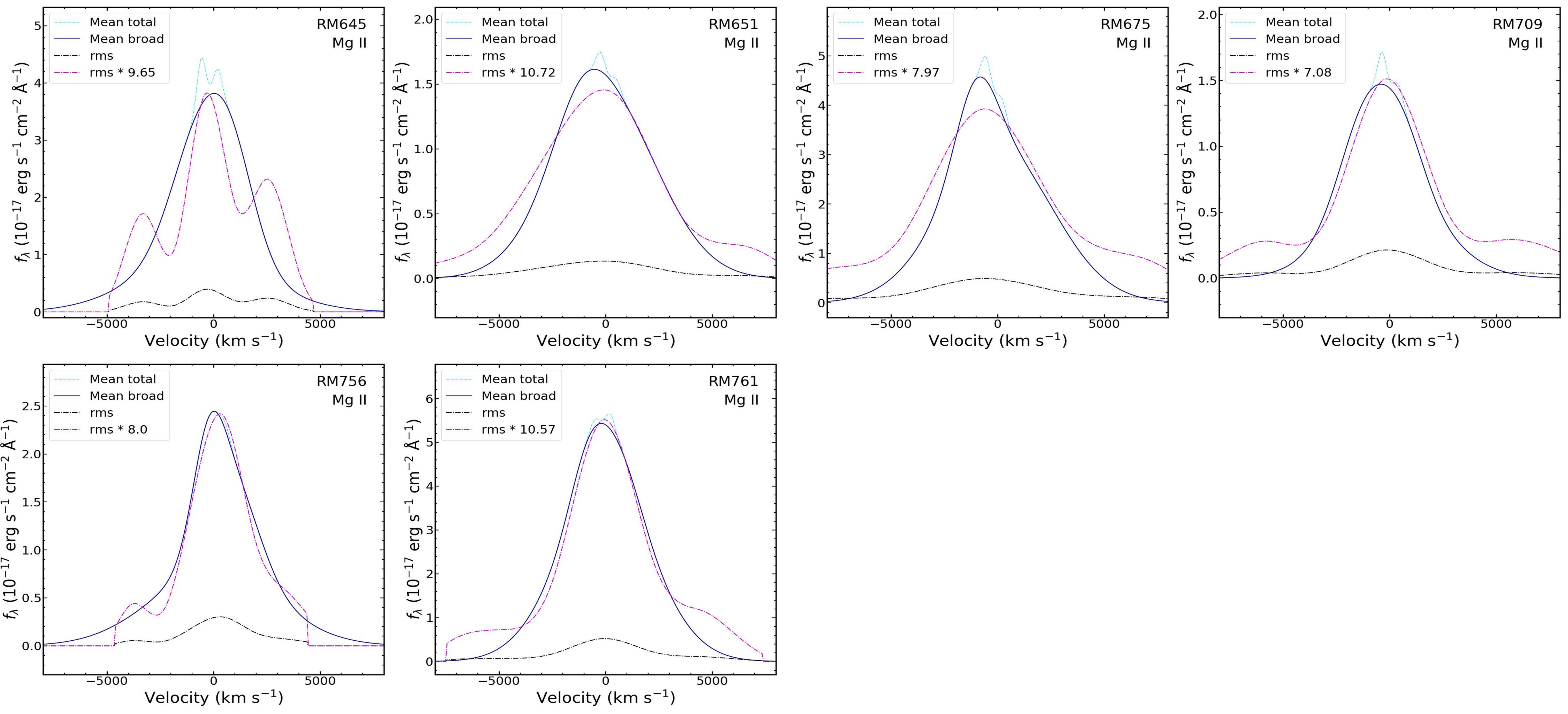}
    \caption{Continued.}
    \label{fig:my_label}
\end{figure*}

\begin{figure*}
    \centering
    \includegraphics[width=\textwidth]{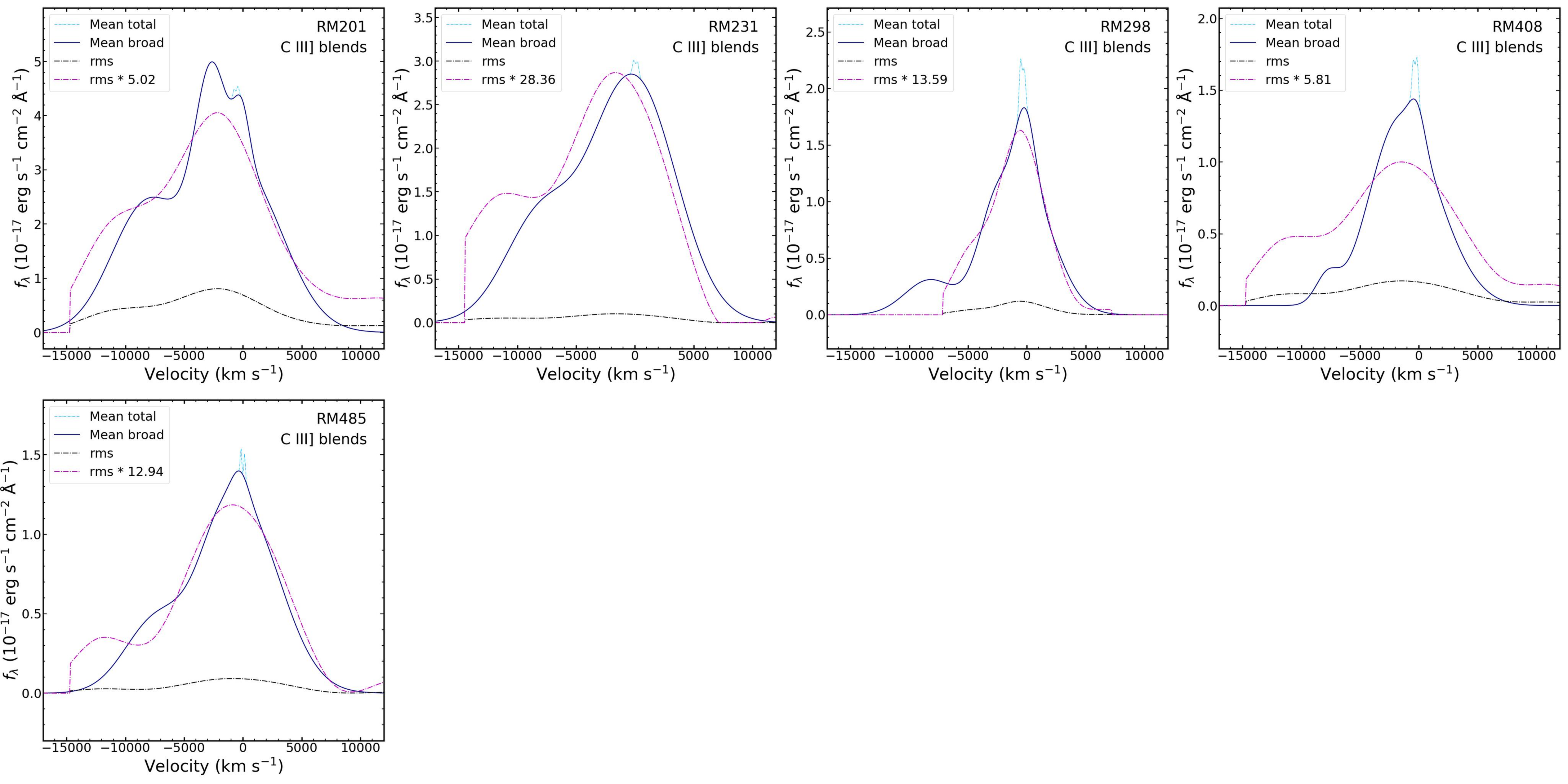}
    \caption{Same as Figure \ref{fig:final1} but for \CIII\ blends.}
    \label{fig:my_label}
\end{figure*}

\begin{figure*}
    \centering
    \includegraphics[width=\textwidth]{Figure_allprofileC4_central.pdf}
    \caption{Same as Figure \ref{fig:final1} but for \CIV. We only show the mean total model if there is evidence for a narrow line component. }
    \label{fig:my_label}
\end{figure*}

\begin{figure*}
    \centering
    \includegraphics[width=\textwidth]{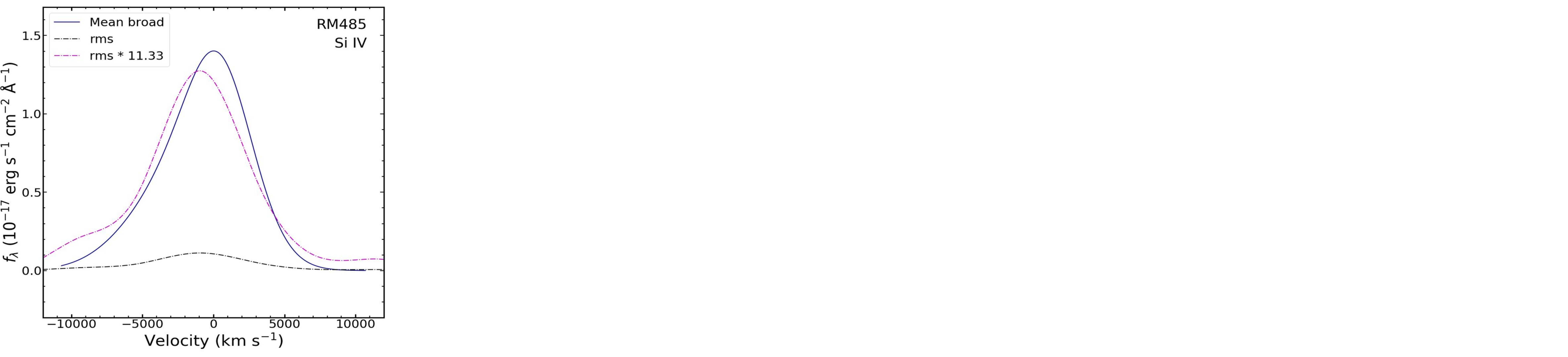}
    \caption{Same as Figure \ref{fig:final1} but for \SiIV.}
    \label{fig:my_label}
\end{figure*}

\bibliography{refs,ref2}

\end{document}